\documentclass[aps,prd,twocolumn,groupedaddress,nofootinbib]{revtex4-2}

\usepackage[utf8]{inputenc}
\usepackage[T1]{fontenc}

\usepackage{graphicx}
\usepackage{booktabs}
\usepackage{afterpage}
\usepackage[colorlinks=true,
            allcolors=blue]{hyperref}
\usepackage{pgfplots}
\pgfplotsset{compat=1.18}
\usetikzlibrary{shadings}
\usepackage[many]{tcolorbox}
\usepackage{placeins}
\usepackage{enumerate}
\usepackage{enumitem}
\usepackage{oubraces}

\usepackage{tikz}
\usetikzlibrary{shapes.geometric}
\usepackage[draft]{tikzit}

\usepackage{multirow}

\usepackage{float}
\usepackage{amsthm}
\usepackage{amsmath,amssymb}
\usepackage{CJKutf8}
\usepackage{xcolor}
\usepackage{pifont}

\newcommand{\cmark}{\text{\ding{51}}} 
\newcommand{\xmark}{\text{\ding{55}}} 

\colorlet{colors01}{red!10}
\colorlet{colors02}{yellow!10}
\colorlet{colors03}{green!10}
\colorlet{colors04}{blue!10}
\colorlet{colors05}{purple!10}

\newcommand{\verticalcenter}[1]{\raisebox{-0.5\height}{\begin{tikzpicture}
    \node at (0,0) {#1};
\end{tikzpicture}}}

\tikzstyle{gauge_gp}=[fill=white, draw=black, shape=circle, minimum size=.4cm]
\tikzstyle{flavour_gp}=[fill=white, draw=black, shape=rectangle, minimum size=.4cm]
\tikzstyle{overb_gauge_gp}=[fill={rgb,255: red,191; green,191; blue,191}, draw=black, shape=circle, minimum size=.4cm]
\tikzstyle{cyan_gauge}=[fill={rgb,255: red,155; green,255; blue,255}, draw=black, shape=circle]
\tikzstyle{blue_gauge_gp}=[fill=white, draw=blue, shape=circle, minimum size=.4cm]
\tikzstyle{red_gauge_gp}=[fill=white, draw=red, shape=circle, minimum size=.4cm]
\tikzstyle{purple_gauge_gp}=[fill=white, draw={rgb,255: red,128; green,0; blue,128}, shape=circle, minimum size=.4cm]
\tikzstyle{pink_gauge_gp}=[draw=magenta, shape=circle, minimum size=.4cm]
\tikzstyle{cyan_gauge_gp}=[draw={rgb,255: red,20; green,221; blue,221}, shape=circle, minimum size=.4cm]
\tikzstyle{yellow_gauge_gp}=[draw={rgb,255: red,168; green,168; blue,0}, shape=circle, minimum size=.4cm]
\tikzstyle{teal_gauge_gp}=[draw={rgb,255: red,0; green,128; blue,128}, shape=circle, minimum size=.4cm]
\tikzstyle{green_gauge_gp}=[draw={rgb,255: red,0; green,200; blue,0}, shape=circle, minimum size=.4cm]
\tikzstyle{grey_flavour}=[draw={rgb,255: red,191; green,191; blue,191}, shape=rectangle, minimum size=.4cm]
\tikzstyle{pink_flavour}=[draw=magenta, shape=rectangle, minimum size=.4cm]
\tikzstyle{green_flavour}=[draw={rgb,255: red,0; green,200; blue,0}, shape=rectangle, minimum size=.4cm]
\tikzstyle{purple_flavour}=[draw={rgb,255: red,128; green,0; blue,128}, shape=rectangle, minimum size=.4cm]
\tikzstyle{orange_flavour}=[draw=orange, shape=rectangle, minimum size=.4cm]
\tikzstyle{teal_flavour}=[draw={rgb,255: red,0; green,128; blue,128}, shape=rectangle, minimum size=.4cm]
\tikzstyle{red_flavour}=[draw=red, shape=rectangle, minimum size=.4cm]
\tikzstyle{blue_flavour}=[draw=blue, shape=rectangle, minimum size=.4cm]
\tikzstyle{cyan_flavour}=[draw={rgb,255: red,20; green,221; blue,221}, shape=rectangle, minimum size=.4cm]
\tikzstyle{yellow_flavour}=[draw={rgb,255: red,168; green,168; blue,0}, shape=rectangle, minimum size=.4cm]
\tikzstyle{red dot}=[fill=red, draw=none, shape=circle, minimum size=0.1cm, inner sep=0pt]
\tikzstyle{blue dot}=[fill=blue, draw=none, shape=circle, minimum size=0.1cm, inner sep=0pt]
\tikzstyle{black dot}=[fill=black, draw=none, shape=circle, minimum size=0.1cm, inner sep=0pt]
\tikzstyle{blank}=[fill=none, draw=none, shape=circle]

\tikzstyle{arrow}=[->, line width=0.5mm]
\tikzstyle{blue}=[-, draw=blue]
\tikzstyle{red}=[-, draw=red]
\tikzstyle{blue arrow}=[draw=blue, ->, line width=0.5mm]
\tikzstyle{blue dashed}=[-, draw=blue, dashed]
\tikzstyle{red dashed}=[-, draw=red, dashed]
\tikzstyle{doubleline}=[-, double distance=1mm]
\tikzstyle{doublelinewide}=[-, double distance=1.5mm]
\tikzstyle{blue double}=[-, double distance=1mm, draw=blue]
\tikzstyle{grey dashed}=[-, draw={rgb,255: red,191; green,191; blue,191}, dashed]
\tikzstyle{grey}=[-, draw={rgb,255: red,191; green,191; blue,191}]
\tikzstyle{yellow}=[-, draw={rgb,255: red,168; green,168; blue,0}]
\tikzstyle{green}=[-, draw={rgb,255: red,0; green,200; blue,0}]
\tikzstyle{pink}=[-, draw=magenta]
\tikzstyle{teal}=[-, draw={rgb,255: red,0; green,128; blue,128}]
\tikzstyle{purple}=[-, draw={rgb,255: red,128; green,0; blue,128}]
\tikzstyle{orange}=[-, draw=orange]
\tikzstyle{cyan}=[-, draw={rgb,255: red,20; green,221; blue,221}]
\tikzstyle{dark grey}=[-, draw={rgb,255: red,64; green,64; blue,64}]
\tikzstyle{wavy}=[-, style={decorate, decoration=snake}]
\tikzstyle{blue wavy}=[-, style={decorate, decoration=snake}, draw=blue]
\tikzstyle{red wavy}=[-, style={decorate, decoration=snake}, draw=red]
\tikzstyle{orange wavy}=[-, style={decorate, decoration=snake}, draw=orange]
\tikzstyle{double arrow}=[<->, line width=0.5mm]

\tcolorboxenvironment{test}{
  enhanced,
  borderline={0.8pt}{0pt}{black},
  borderline={0.4pt}{2pt}{black},
  boxrule=0.4pt,
  colback=white,
  coltitle=black,
  sharp corners,
}

\newcommand{\convexpath}[2]{
  [   
  create hullcoords/.code={
    \global\edef\namelist{#1}
    \foreach [count=\counter] \nodename in \namelist {
      \global\edef\numberofnodes{\counter}
      \coordinate (hullcoord\counter) at (\nodename);
    }
    \coordinate (hullcoord0) at (hullcoord\numberofnodes);
    \pgfmathtruncatemacro\lastnumber{\numberofnodes+1}
    \coordinate (hullcoord\lastnumber) at (hullcoord1);
  },
  create hullcoords
  ]
  ($(hullcoord1)!#2!-90:(hullcoord0)$)
  \foreach [
  evaluate=\currentnode as \previousnode using \currentnode-1,
  evaluate=\currentnode as \nextnode using \currentnode+1
  ] \currentnode in {1,...,\numberofnodes} {
    let \p1 = ($(hullcoord\currentnode) - (hullcoord\previousnode)$),
    \n1 = {atan2(\y1,\x1) + 90}, 
    \p2 = ($(hullcoord\nextnode) - (hullcoord\currentnode)$),
    \n2 = {atan2(\y2,\x2) + 90},
    \n{delta} = {Mod(\n2-\n1,360) - 360}
    in 
    {arc [start angle=\n1, delta angle=\n{delta}, radius=#2]}
    -- ($(hullcoord\nextnode)!#2!-90:(hullcoord\currentnode)$) 
  }
}

\newcommand{\HS}{\text{HS}}
\newcommand{\U}{\mathrm{U}}
\newcommand{\SU}{\mathrm{SU}}
\newcommand{\Sp}{\mathrm{Sp}}
\newcommand{\SO}{\mathrm{SO}}

\begin{document}

\title{Bootstrapping mirror pairs: The beginning of the end}

\affiliation{Mathematical Institute, University of Oxford, Andrew Wiles Building, Woodstock Road, Oxford OX2 6GG, United Kingdom}

\author{Leyi Jiang}
\email{ljiang@mpi-cbg.de}
\altaffiliation[Present address:\\]{Max Planck Institute of Molecular Cell Biology and Genetics (MPI-CBG), Pfotenhauerstraße 108, 01307 Dresden, Germany;\\
Center for Systems Biology Dresden (CSBD), Pfotenhauerstraße 108, 01307 Dresden, Germany;\\
Faculty of Mathematics, Technische Universit\"at Dresden, Helmholtzstraße   \ 10, 01069 Dresden, Germany}

\author{Jazz E. Z. Ooi}
\email{jazzooi21@gmail.com}
\altaffiliation[Present address:\\]{Centre for Quantum Technologies (CQT), National University of Singapore, Block S15, 3 Science Drive 2, Singapore 117543}

\author{Richard Stone}
\email{richstone052@gmail.com}

\author{Zhenghao Zhong}
\email{zhong@simis.cn}
\altaffiliation[Present address:\\]{Shanghai Institute for Mathematics and Interdisciplinary Sciences (SIMIS), Block A, International Innovation Plaza, No.\ 657 Songhu Road, Shanghai, China}

\affiliation{Mathematical Institute, University of Oxford, Andrew Wiles Building, Woodstock Road, Oxford OX2 6GG, United Kingdom}

\begin{abstract}
Three-dimensional supersymmetric gauge theories with eight supercharges possess a unique duality known as 3D mirror symmetry. Under this correspondence, the Coulomb branch of one theory is equivalent to the Higgs branch of its mirror dual, and vice versa. Over the past decades, extensive effort has been devoted to charting the landscape of 3D mirror pairs, though progress has often been constrained by the need to identify suitable brane configurations. In this first installment, we introduce a new quiver-based algorithm, termed growth and fusion, which completes a quartet of Higgsing algorithms alongside decay and fission, quiver subtraction, and quiver addition. Together, these four algorithms provide a systematic framework that circumvents the limitations of brane constructions, enabling us to determine the mirror dual of a given quiver and to systematically bootstrap new 3D mirror pairs. We demonstrate the power of this approach on a new class of circular 3D mirror pairs.
\end{abstract}

\maketitle

\section{Introduction} \label{sec:intro}
In quantum field theories, a theory that is difficult to study can often be approached by studying its dual theory. For example, in string theory, a strongly coupled theory is often impervious to many field theoretic techniques, but by taking the S-dual, one obtains a weakly coupled dual theory that is much more approachable. In three-dimensional supersymmetric field theories with eight supercharges (3D $\mathcal{N}=4$), this S-duality manifests as \emph{3D mirror symmetry}, first introduced in Ref.\ \cite{Intriligator:1996ex}.

A pair of theories is related by 3D mirror symmetry when the Coulomb branch of theory $X$ coincides with the Higgs branch of theory $Y$, and vice versa. Historically, Higgs branches have been relatively easy to study as they are classical objects, while Coulomb branches are notoriously difficult as they receive quantum corrections, parametrized by disorder operators (monopoles) rather than composites of elementary fields. As a result, researchers often turned to the Higgs branch of a mirror theory instead of directly tackling the Coulomb branch. However, in the past decade, significant progress has been made in understanding Coulomb branches, which are now often more computationally efficient to analyze. Ever since the invention of the closely related but slightly different electric/magnetic quiver duality\footnote{A duality that relates the Higgs branch of 3D, 4D, 5D, 6D theories (\emph{electric quivers}) with eight supercharges to the Coulomb branch of 3D $\mathcal{N}=4$ \emph{magnetic quivers} \cite{Bourget:2019aer}.\color{black}}, interest in 3D mirror symmetry has once again soared. 

In Ref.\ \cite{Zhong:2024jxw}, a question was set out regarding how far one can extend the landscape of 3D mirror pairs that both have Lagrangian descriptions. The paper mainly takes Dynkin quivers of classical algebras (i.e.\ quivers shaped like $ABCD$ Dynkin diagrams), putting bouquets of $\U(1)$s extending from them, and identifying the 3D mirror. The construction relies on an extension \cite{USU} of the Hanany-Witten brane system of D3-D5-NS5 branes \cite{Hanany:1996ie}, which allowed the identification of additional $\U(1)$ nodes, as well as special unitary gauge groups. This method successfully generated a large number of new 3D mirror pairs, verified by computing Coulomb/Higgs branch Hilbert series (see Appendix \ref{appx:hs} for an introduction to the concept). However, it remains constrained by its dependence on brane configurations, which are difficult to generalize to highly nonlinear\footnote{Quivers whose gauge groups form a nonlinear diagram.} quivers (beyond the classical Dynkin diagrams) in 3D $\mathcal{N}=4$ theories.

In this paper, we introduce a new method for bootstrapping 3D mirror pairs. The output is quiver gauge theory pairs made of unitary gauge nodes. Our approach combines a known algorithm, \emph{quiver addition}, with a new one, \emph{growth and fusion}. A key advantage of this algorithm is that it does not rely on brane constructions.

Table \ref{tab:algo} shows the four quiver algorithms which are used to generate phase diagrams of the vacua. Quiver subtraction is described in Ref.\ \cite{Bourget:2019aer}, decay and fission in Refs.\ \cite{Decay, Fission}, and quiver addition in Ref.\ \cite{Bourget:2021siw}. Our introduction of the growth and fusion algorithm here will complete this quartet. 

\begin{table}[H]
\label{tab:algo}
\caption{Quiver algorithms.}
\begin{ruledtabular}
\begin{tabular}{lcc}
& Higgs branch & Coulomb branch \\
\hline
Higgsing   & Quiver subtraction & Decay and fission \\
UnHiggsing & Quiver addition    & Growth and fusion
\end{tabular}
\end{ruledtabular}
\end{table}

In Sec.\ \ref{sec:algo}, we introduce the new growth and fusion algorithm and explain why it provides a framework for bootstrapping new 3D mirror pairs. Sec.\ \ref{sec:nonlinear} explores a novel class of circular nonlinear quivers that extend beyond the Dynkin quivers. In Sec.\ \ref{sec:boot}, we demonstrate how the mirrors of these nonlinear quivers can be systematically obtained by utilizing the quartet of quiver algorithms and use them as a basis to bootstrap new mirrors. 

This paper acts as a first installment and sets the stage by introducing the bootstrapping algorithm. In the next installment, we will fully explore its reaches with a myriad of examples.

\section{Growth and Fusion Algorithm}\label{sec:algo}
\subsection{The four quiver algorithms}
The Higgs mechanism is a foundational concept in quantum field theory. In supersymmetric theories, there exist many scalar fields that can acquire vacuum expectation values and thereby trigger Higgsing. For supersymmetric theories with eight supercharges, one can distinguish between \emph{Coulomb branch Higgsing} and \emph{Higgs branch Higgsing}, depending on which scalar fields acquire the expectation values.

In electroweak theory, for example, the gauge group $\SU(2)\times \U(1)$ is Higgsed down to $\U(1)$ once the Higgs boson acquires a vacuum expectation value. In supersymmetric theories, however, Higgsing is far richer. It becomes important to identify not only the immediate daughter theories obtained after Higgsing, but also successive descendants---granddaughter theories, and so on---that arise as gauge symmetries are broken by massive scalars through the Higgs mechanism. To systematically encode these relations, one employs a \emph{Hasse diagram}, which organizes the resulting theories into a partially ordered structure.

The quiver algorithms, as summarized in Table \ref{tab:algo}, are used to implement the Higgs mechanism in supersymmetric theories and to generate the associated Hasse diagrams. The relevant functions are:

\begin{enumerate}[label=(\roman*), noitemsep]
\item \textbf{Quiver subtraction}\\
Higgsing algorithm on the Higgs branch.
\item \textbf{Quiver addition}\\
UnHiggsing algorithm on the Higgs branch.
\item \textbf{Decay and fission}\\
Higgsing algorithm on the Coulomb branch.
\item \textbf{Growth and fusion}\\
UnHiggsing algorithm on the Coulomb branch.
\end{enumerate}

We consider these algorithms only in the context of 3D $\mathcal{N}=4$ supersymmetric theories. It is worth noting that the decay and fission algorithm, when applied to magnetic quivers, acts as a Higgsing algorithm on the Higgs branch of the corresponding electric quivers (which may be 3D, 4D, 5D, or 6D theories). For this reason, decay and fission has often been described as a Higgs branch Higgsing algorithm in the literature. In our case, since we are not dealing with electric and magnetic quivers, and stay within the restricted setting of 3D $\mathcal{N}=4$ theories, it should be more precisely regarded as a Coulomb branch Higgsing algorithm.

The unHiggsing algorithms are central to this work. Their purpose is to generate parent and grandparent theories---that is, possible theories from which a given theory can be obtained through successive Higgsings. These two algorithms provide the framework we use to bootstrap new 3D mirror pairs.

See Appendix \ref{appx:algo} for more details on the quiver algorithms.

\subsection{Quiver subtraction and decay and fission}
Let us consider the first pair of algorithms. Consider the following 3D mirror pair: 
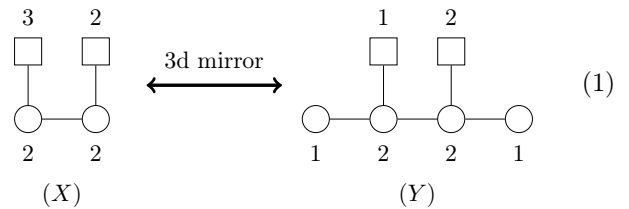
\begin{equation}
    \label{eqn:mirrorpair}
    \scalebox{0.9}{
        \begin{tikzpicture}[baseline=-0.25em,scale=0.5]
    	\begin{pgfonlayer}{nodelayer}
    		\node [style={gauge_gp}] (2) at (-5.5, -1) {};
    		\node [style={gauge_gp}] (3) at (-3.5, -1) {};
    		\node [style={flavour_gp}] (4) at (-5.5, 1) {};
    		\node [style={flavour_gp}] (5) at (-3.5, 1) {};
    		\node [style=none] (6) at (2, 0) {};
    		\node [style=none] (7) at (-2, 0) {};
    		\node [style=none] (8) at (0, 0.75) {3D mirror};
    		\node [style={flavour_gp}] (9) at (5, 1) {};
    		\node [style={flavour_gp}] (10) at (7, 1) {};
    		\node [style={gauge_gp}] (11) at (5, -1) {};
    		\node [style={gauge_gp}] (12) at (7, -1) {};
    		\node [style={gauge_gp}] (13) at (3, -1) {};
    		\node [style={gauge_gp}] (14) at (9, -1) {};
    		\node [style=none] (15) at (5, -2) {2};
    		\node [style=none] (16) at (7, -2) {2};
    		\node [style=none] (17) at (3, -2) {1};
    		\node [style=none] (18) at (9, -2) {1};
    		\node [style=none] (19) at (5, 2) {1};
    		\node [style=none] (20) at (7, 2) {2};
    		\node [style=none] (21) at (-5.5, 2) {3};
    		\node [style=none] (22) at (-3.5, 2) {2};
    		\node [style=none] (23) at (-5.5, -2) {2};
    		\node [style=none] (24) at (-3.5, -2) {2};
    		\node [style=none] (25) at (-4.5, -3.25) {$(X)$};
    		\node [style=none] (26) at (6, -3.25) {$(Y)$};
    	\end{pgfonlayer}
    	\begin{pgfonlayer}{edgelayer}
    		\draw (2) to (3);
    		\draw (3) to (5);
    		\draw (4) to (2);
    		\draw [style=double arrow] (7.center) to (6.center);
    		\draw (12) to (11);
    		\draw (11) to (9);
    		\draw (13) to (11);
    		\draw (14) to (12);
    		\draw (12) to (10);
    	\end{pgfonlayer}
    \end{tikzpicture}
    
    }
\end{equation}

As with all the quivers in this paper, the gauge nodes and flavor nodes represent unitary gauge groups with the ranks specified next to them. By performing the algorithms decay and fission on quiver $X$ and Quiver Subtraction on quiver $Y$, we can construct a Hasse diagram where the resulting quivers on the two sides are 3D mirrors of each other: 
\begin{widetext}
\begin{equation}
\scalebox{0.95}{
    \begin{tikzpicture}[baseline=-0.25em,scale=0.5]
    	\begin{pgfonlayer}{nodelayer}
		\node [style={gauge_gp}] (2) at (-7, 13.5) {};
		\node [style={gauge_gp}] (3) at (-5, 13.5) {};
		\node [style={flavour_gp}] (4) at (-7, 15.5) {};
		\node [style={flavour_gp}] (5) at (-5, 15.5) {};
		\node [style=none] (6) at (3, 14.5) {};
		\node [style=none] (7) at (-3, 14.5) {};
		\node [style=none] (8) at (0, 15.25) {3D mirror};
		\node [style={flavour_gp}] (9) at (6.5, 15.5) {};
		\node [style={flavour_gp}] (10) at (8.5, 15.5) {};
		\node [style={gauge_gp}] (11) at (6.5, 13.5) {};
		\node [style={gauge_gp}] (12) at (8.5, 13.5) {};
		\node [style={gauge_gp}] (13) at (4.5, 13.5) {};
		\node [style={gauge_gp}] (14) at (10.5, 13.5) {};
		\node [style=none] (15) at (6.5, 12.5) {2};
		\node [style=none] (16) at (8.5, 12.5) {2};
		\node [style=none] (17) at (4.5, 12.5) {1};
		\node [style=none] (18) at (10.5, 12.5) {1};
		\node [style=none] (19) at (6.5, 16.5) {1};
		\node [style=none] (20) at (8.5, 16.5) {2};
		\node [style=none] (21) at (-7, 16.5) {3};
		\node [style=none] (22) at (-5, 16.5) {2};
		\node [style=none] (23) at (-7, 12.5) {2};
		\node [style=none] (24) at (-5, 12.5) {2};
		\node [style=none] (25) at (-6, 10) {};
		\node [style=none] (26) at (-6, 8) {};
		\node [style=none] (27) at (7.5, 10) {};
		\node [style=none] (28) at (7.5, 8) {};
		\node [style={gauge_gp}] (29) at (-7, 4) {};
		\node [style={gauge_gp}] (30) at (-5, 4) {};
		\node [style={flavour_gp}] (31) at (-7, 6) {};
		\node [style={flavour_gp}] (32) at (-5, 6) {};
		\node [style=none] (33) at (-7, 7) {3};
		\node [style=none] (34) at (-5, 7) {2};
		\node [style=none] (35) at (-7, 3) {2};
		\node [style=none] (36) at (-5, 3) {1};
		\node [style={flavour_gp}] (37) at (6.5, 6) {};
		\node [style={gauge_gp}] (39) at (6.5, 4) {};
		\node [style={gauge_gp}] (40) at (8.5, 4) {};
		\node [style={gauge_gp}] (41) at (4.5, 4) {};
		\node [style={gauge_gp}] (42) at (10.5, 4) {};
		\node [style=none] (43) at (6.5, 3) {2};
		\node [style=none] (44) at (8.5, 3) {1};
		\node [style=none] (45) at (4.5, 3) {1};
		\node [style=none] (46) at (10.5, 3) {1};
		\node [style=none] (47) at (6.5, 7) {2};
		\node [style=none] (48) at (10.5, 7) {1};
		\node [style={flavour_gp}] (49) at (10.5, 6) {};
		\node [style=none] (50) at (-6, 2) {};
		\node [style=none] (51) at (-6, 0) {};
		\node [style=none] (52) at (7.5, 2) {};
		\node [style=none] (53) at (7.5, 0) {};
		\node [style={gauge_gp}] (54) at (-7, -4) {};
		\node [style={gauge_gp}] (55) at (-5, -4) {};
		\node [style={flavour_gp}] (56) at (-7, -2) {};
		\node [style={flavour_gp}] (57) at (-5, -2) {};
		\node [style=none] (58) at (-7, -1) {3};
		\node [style=none] (59) at (-5, -1) {2};
		\node [style=none] (60) at (-7, -5) {1};
		\node [style=none] (61) at (-5, -5) {1};
		\node [style={flavour_gp}] (62) at (4.5, -2) {};
		\node [style={gauge_gp}] (63) at (6.5, -4) {};
		\node [style={gauge_gp}] (64) at (8.5, -4) {};
		\node [style={gauge_gp}] (65) at (4.5, -4) {};
		\node [style={gauge_gp}] (66) at (10.5, -4) {};
		\node [style=none] (67) at (6.5, -5) {1};
		\node [style=none] (68) at (8.5, -5) {1};
		\node [style=none] (69) at (4.5, -5) {1};
		\node [style=none] (70) at (10.5, -5) {1};
		\node [style=none] (71) at (4.5, -1) {1};
		\node [style=none] (72) at (10.5, -1) {1};
		\node [style={flavour_gp}] (73) at (10.5, -2) {};
		\node [style=none] (74) at (3, 5) {};
		\node [style=none] (75) at (-3, 5) {};
		\node [style=none] (76) at (0, 5.75) {3D mirror};
		\node [style=none] (77) at (3, -3) {};
		\node [style=none] (78) at (-3, -3) {};
		\node [style=none] (79) at (0, -2.25) {3D mirror};
		\node [style={flavour_gp}] (80) at (8.5, -2) {};
		\node [style=none] (81) at (8.5, -1) {1};
		\node [style=none] (82) at (-7, -6) {};
		\node [style=none] (83) at (-8, -8) {};
		\node [style=none] (88) at (-3.5, -9) {2};
		\node [style=none] (89) at (-8.5, -9) {3};
		\node [style=none] (90) at (-2.5, -12) {1};
		\node [style=none] (91) at (-9.5, -12) {1};
		\node [style=none] (92) at (-5, -6) {};
		\node [style=none] (93) at (-4, -8) {};
		\node [style=none] (98) at (3.5, -9) {1};
		\node [style=none] (100) at (2.5, -12) {1};
		\node [style=none] (104) at (6.5, -12) {1};
		\node [style=none] (105) at (5.5, -9) {1};
		\node [style=none] (106) at (4.5, -8) {};
		\node [style=none] (107) at (5.5, -6) {};
		\node [style=none] (108) at (9.5, -6) {};
		\node [style=none] (109) at (10.5, -8) {};
		\node [style=none] (110) at (2, -10.75) {};
		\node [style=none] (111) at (-2, -10.75) {};
		\node [style=none] (112) at (0, -10) {3D mirror};
		\node [style={gauge_gp}] (113) at (-3.5, -12) {};
		\node [style={flavour_gp}] (114) at (-3.5, -10) {};
		\node [style={flavour_gp}] (115) at (-8.5, -10) {};
		\node [style={gauge_gp}] (116) at (-8.5, -12) {};
		\node [style=none] (117) at (11, -9) {2};
		\node [style=none] (118) at (12, -12) {1};
		\node [style={gauge_gp}] (119) at (11, -12) {};
		\node [style={flavour_gp}] (120) at (11, -10) {};
		\node [style={flavour_gp}] (121) at (5.5, -10) {};
		\node [style={flavour_gp}] (122) at (3.5, -10) {};
		\node [style={gauge_gp}] (123) at (3.5, -12) {};
		\node [style={gauge_gp}] (124) at (5.5, -12) {};
		\node [style=none] (125) at (-6.75, -16.25) {};
		\node [style=none] (126) at (-8, -14.25) {};
		\node [style=none] (127) at (-5.25, -16.25) {};
		\node [style=none] (128) at (-4, -14.25) {};
		\node [style=none] (129) at (4.5, -14.25) {};
		\node [style=none] (130) at (6.75, -16.25) {};
		\node [style=none] (131) at (8.25, -16.25) {};
		\node [style=none] (132) at (10.5, -14.25) {};
		\node [style=none, scale=1.25] (135) at (-6, 18) {\underline{Decay and fission}};
		\node [style=none, scale=1.25] (136) at (7.5, 18) {\underline{Quiver subtraction}};
		\node [style=none] (137) at (-6, 11.25) {$(X)$};
		\node [style=none] (138) at (7.5, 11.25) {$(Y)$};
		\node [style=none] (139) at (-8.5, -13.25) {$(*)$};
		\node [style=none] (140) at (4.5, -13.25) {$(*)$};
		\node [style=none] (141) at (-3.5, -13.25) {$(\dag)$};
		\node [style=none] (142) at (11, -13.25) {$(\dag)$};
		\node [style=none] (271) at (-6, -17) {$\varnothing$};
		\node [style=none] (272) at (7.5, -17) {$\varnothing$};
	\end{pgfonlayer}
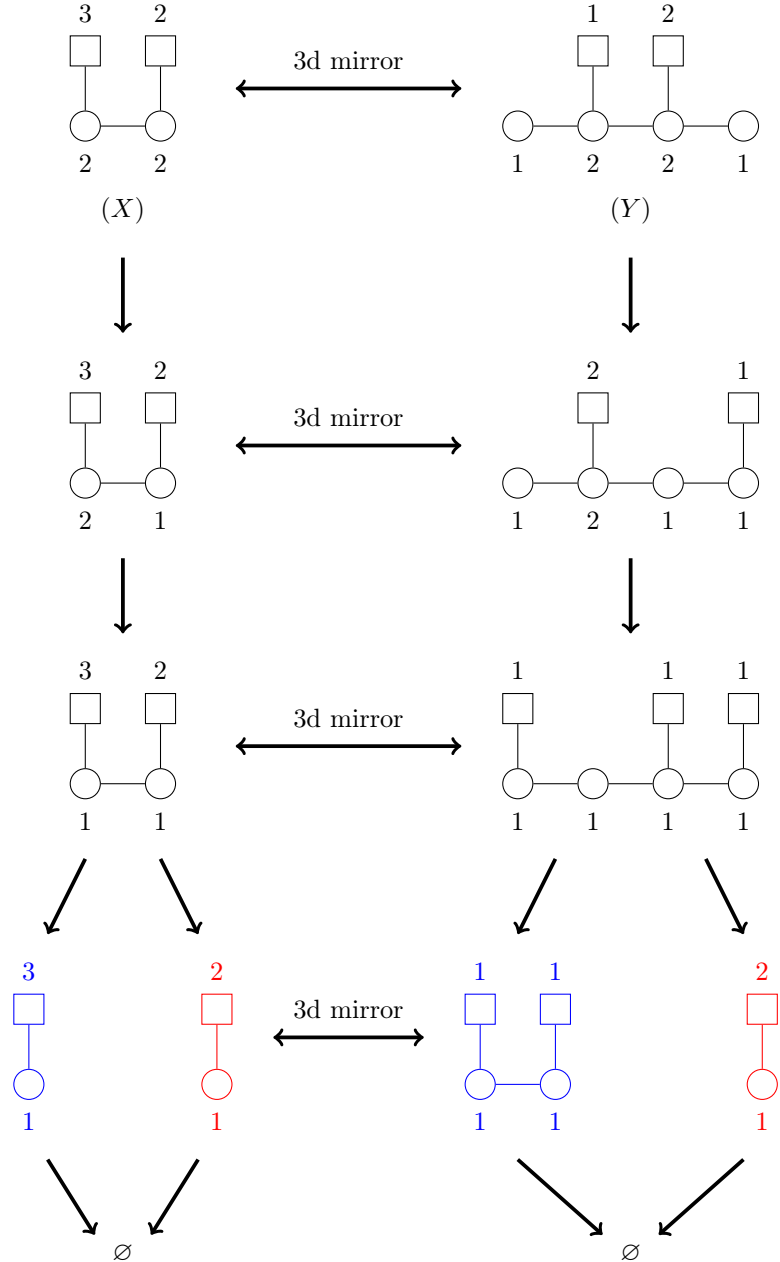
	\begin{pgfonlayer}{edgelayer}
		\draw (2) to (3);
		\draw (3) to (5);
		\draw (4) to (2);
		\draw [style=double arrow] (7.center) to (6.center);
		\draw (12) to (11);
		\draw (11) to (9);
		\draw (13) to (11);
		\draw (14) to (12);
		\draw (12) to (10);
		\draw [style=arrow] (25.center) to (26.center);
		\draw [style=arrow] (27.center) to (28.center);
		\draw (29) to (30);
		\draw (30) to (32);
		\draw (31) to (29);
		\draw (40) to (39);
		\draw (39) to (37);
		\draw (41) to (39);
		\draw (42) to (40);
		\draw (49) to (42);
		\draw [style=arrow] (50.center) to (51.center);
		\draw [style=arrow] (52.center) to (53.center);
		\draw (54) to (55);
		\draw (55) to (57);
		\draw (56) to (54);
		\draw (64) to (63);
		\draw (65) to (63);
		\draw (66) to (64);
		\draw (73) to (66);
		\draw [style=double arrow] (75.center) to (74.center);
		\draw [style=double arrow] (78.center) to (77.center);
		\draw (62) to (65);
		\draw (80) to (64);
		\draw [style=arrow] (82.center) to (83.center);
		\draw [style=arrow] (92.center) to (93.center);
		\draw [style=arrow] (107.center) to (106.center);
		\draw [style=arrow] (108.center) to (109.center);
		\draw [style=double arrow] (111.center) to (110.center);
		\draw (114) to (113);
		\draw (115) to (116);
		\draw (120) to (119);
		\draw (124) to (121);
		\draw (122) to (123);
		\draw (123) to (124);
		\draw [style=arrow] (132.center) to (131.center);
		\draw [style=arrow] (129.center) to (130.center);
		\draw [style=arrow] (126.center) to (125.center);
		\draw [style=arrow] (128.center) to (127.center);
	\end{pgfonlayer}
\end{tikzpicture}
}
\end{equation}
\end{widetext}

The symbols help identify the mirror pairs. The 3D mirror relation generated by these two algorithms was already introduced in Refs.\ \cite{Gu:2022dac,Fission}. In this paper, we extend this relation to the quiver addition and the new growth and fusion algorithm as well. 

\subsection{Growth and fusion}
The growth and fusion algorithm is the antithesis of the decay and fission algorithm. Whereas the latter algorithm is one of Higgsing, the former algorithm is one of unHiggsing. This means that given a quiver Q, we are looking at possible quivers that Higgs into quiver Q.
The rules for the growth algorithm are as follows:
\begin{itemize}
    \item \textbf{Adding a single U(1):} There are two choices; either
    \begin{enumerate}
        \item[(a)] Increase the rank of a unitary node by 1, or
        \item[(b)] add a new $\U(1)$ node that is connected to the rest of the quiver.
    \end{enumerate}
    This growth process is allowed as long as the resulting quiver has good gauge nodes (in the sense of Ref.\ \cite{Gaiotto:2008ak}). This means that for each gauge group $\U(k_i)$ in the quiver, the total number of hypers connected to it must be greater than or equal to $2k_i$. 

    Using our previous example, we can apply growth to obtain the following quiver:
    \begin{equation}
    \scalebox{0.85}{
        \begin{tikzpicture}[baseline=-0.25em,scale=0.5]
    	\begin{pgfonlayer}{nodelayer}
		\node [style={gauge_gp}] (2) at (-5.5, -1) {};
		\node [style={gauge_gp}] (3) at (-3.5, -1) {};
		\node [style={flavour_gp}] (4) at (-5.5, 1) {};
		\node [style={flavour_gp}] (5) at (-3.5, 1) {};
		\node [style=none] (6) at (2, 0) {};
		\node [style=none] (7) at (-2, 0) {};
		\node [style=none] (8) at (0, 0.75) {Growth};
		\node [style={flavour_gp}] (9) at (3.5, 1) {};
		\node [style={flavour_gp}] (10) at (5.5, 1) {};
		\node [style={gauge_gp}] (11) at (3.5, -1) {};
		\node [style={gauge_gp}] (12) at (5.5, -1) {};
		\node [style={gauge_gp}] (14) at (7.5, -1) {};
		\node [style=none] (15) at (3.5, -2) {2};
		\node [style=none] (16) at (5.5, -2) {2};
		\node [style=none] (18) at (7.5, -2) {1};
		\node [style=none] (19) at (3.5, 2) {3};
		\node [style=none] (20) at (5.5, 2) {2};
		\node [style=none] (21) at (-5.5, 2) {3};
		\node [style=none] (22) at (-3.5, 2) {2};
		\node [style=none] (23) at (-5.5, -2) {2};
		\node [style=none] (24) at (-3.5, -2) {2};
	\end{pgfonlayer}
	\begin{pgfonlayer}{edgelayer}
		\draw (2) to (3);
		\draw (3) to (5);
		\draw (4) to (2);
		\draw (12) to (11);
		\draw (11) to (9);
		\draw (14) to (12);
		\draw (12) to (10);
		\draw [style=arrow] (7.center) to (6.center);
	\end{pgfonlayer}
    \end{tikzpicture}
    }
    \end{equation}
    
    where choice (b) was utilized as a new $\U(1)$ gauge group connected to the rest of the quiver. This is, of course, just one of many possible choices. 
    \item \textbf{Adding multiple nodes:} This possibility is allowed by adding one of the diagrams in Table 1 of Ref.\ \cite{Fission}. These are mostly finite Dynkin diagrams of Lie algebra $\mathfrak{g}$.\footnotemark\ The addition of these diagrams is allowed only if the resulting gauge groups remain good.

Continue growing the previous quiver, and we obtain the following:
    \begin{equation}
        \scalebox{0.85}{
        \begin{tikzpicture}[baseline=-0.25em,scale=0.5]
    	\begin{pgfonlayer}{nodelayer}
    		\node [style={gauge_gp}] (2) at (-7.5, -1) {};
    		\node [style={gauge_gp}] (3) at (-5.5, -1) {};
    		\node [style={flavour_gp}] (4) at (-7.5, 1) {};
    		\node [style={flavour_gp}] (5) at (-5.5, 1) {};
    		\node [style=none] (6) at (2, 0) {};
    		\node [style=none] (7) at (-2, 0) {};
    		\node [style=none] (8) at (0, 0.75) {Growth};
    		\node [style={flavour_gp}] (9) at (3.5, 1) {};
    		\node [style={flavour_gp}] (10) at (5.5, 1) {};
    		\node [style={gauge_gp}] (11) at (3.5, -1) {};
    		\node [style={gauge_gp}] (12) at (5.5, -1) {};
    		\node [style={gauge_gp}] (14) at (7.5, -1) {};
    		\node [style=none] (15) at (3.5, -2) {3};
    		\node [style=none] (16) at (5.5, -2) {3};
    		\node [style=none] (18) at (7.5, -2) {1};
    		\node [style=none] (19) at (3.5, 2) {3};
    		\node [style=none] (20) at (5.5, 2) {2};
    		\node [style=none] (21) at (-7.5, 2) {3};
    		\node [style=none] (22) at (-5.5, 2) {2};
    		\node [style=none] (23) at (-7.5, -2) {2};
    		\node [style=none] (24) at (-5.5, -2) {2};
    		\node [style=none] (25) at (0, -0.75) {$a_2$};
    		\node [style=none] (26) at (-3.5, -2) {1};
    		\node [style={gauge_gp}] (27) at (-3.5, -1) {};
    	\end{pgfonlayer}
    	\begin{pgfonlayer}{edgelayer}
    		\draw (2) to (3);
    		\draw (3) to (5);
    		\draw (4) to (2);
    		\draw (12) to (11);
    		\draw (11) to (9);
    		\draw (14) to (12);
    		\draw (12) to (10);
    		\draw [style=arrow] (7.center) to (6.center);
    		\draw (27) to (3);
    	\end{pgfonlayer}
    \end{tikzpicture}
    }
    \end{equation}

    \newpage
    
where we also labeled the transverse space $a_2$, which takes the shape of the $a_2$ finite Dynkin diagram and hence the two $\U(2)$s on the left each have their ranks increased by 1 to $\U(3)$s. Similar to the (b) case, this extension can also result in more gauge nodes than before.
\end{itemize}
The rules for fusion are the following:
\begin{itemize}
    \item \textbf{Fusing two diagrams}: Given a quiver Q, one can add \emph{any} quiver $\mathrm{Q}'$ as long as all the gauge groups of $\mathrm{Q}'$ are good and the resulting quiver $\mathrm{Q}''=\mathrm{Q}+\mathrm{Q}'$ has only good gauge groups. If $\mathrm{Q}'$ is not a quiver associated with an isolated symplectic singularity (see  Ref.\ \cite{Bourget:2025wsp} for the most up-to-date list) then this step is not elementary. One can perform decay and fission to easily find all the elementary steps in between. 
\end{itemize}
These rules are made to be the antithesis of the decay and fission rules. 

Now, we combine this with quiver addition, which was already given in Sec.\ 5.1 of Ref.\ \cite{Bourget:2021siw}. While the original quiver addition algorithm was developed within the context of affine Grassmannians, which fixed the number of gauge groups, we must relax this constraint to achieve the general version of the algorithm, allowing the quiver to possess more gauge groups after performing quiver addition. Utilizing both algorithms on the 3D mirror pair \eqref{eqn:mirrorpair} gives us the following result:

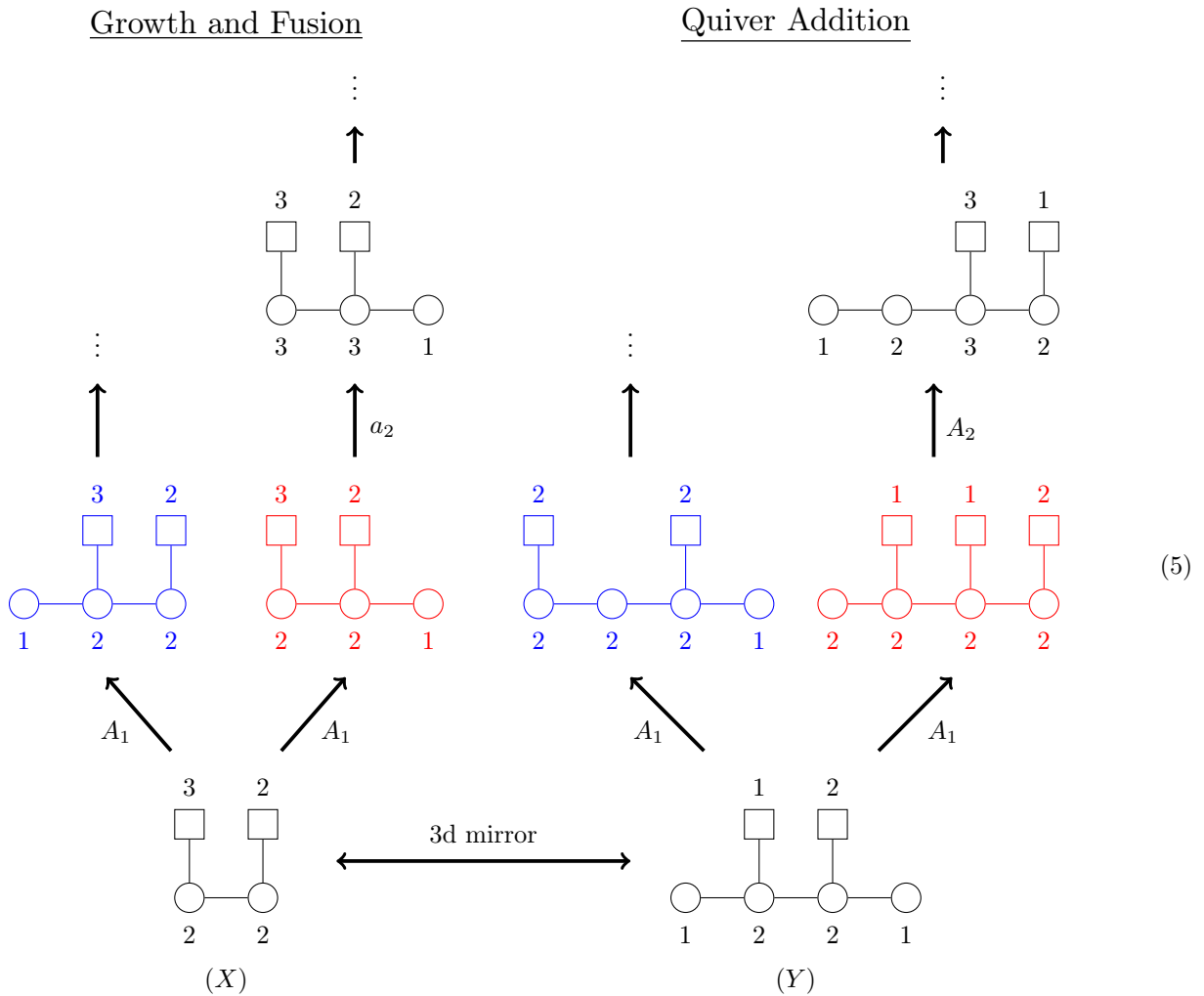
\begin{widetext}
\begin{equation}
\scalebox{0.75}{
    \begin{tikzpicture}[baseline=-0.25em,scale=0.5]
    \begin{pgfonlayer}{nodelayer}
		\node [style=none] (0) at (-7, -12.25) {$(X)$};
		\node [style=none] (1) at (8.5, -12.25) {$(Y)$};
		\node [style={gauge_gp}] (2) at (-8, -10) {};
		\node [style={gauge_gp}] (3) at (-6, -10) {};
		\node [style={flavour_gp}] (4) at (-8, -8) {};
		\node [style={flavour_gp}] (5) at (-6, -8) {};
		\node [style=none] (6) at (4, -9) {};
		\node [style=none] (7) at (-4, -9) {};
		\node [style=none] (8) at (0, -8.25) {3D mirror};
		\node [style={flavour_gp}] (9) at (7.5, -8) {};
		\node [style={flavour_gp}] (10) at (9.5, -8) {};
		\node [style={gauge_gp}] (11) at (7.5, -10) {};
		\node [style={gauge_gp}] (12) at (9.5, -10) {};
		\node [style={gauge_gp}] (13) at (5.5, -10) {};
		\node [style={gauge_gp}] (14) at (11.5, -10) {};
		\node [style=none] (15) at (7.5, -11) {2};
		\node [style=none] (16) at (9.5, -11) {2};
		\node [style=none] (17) at (5.5, -11) {1};
		\node [style=none] (18) at (11.5, -11) {1};
		\node [style=none] (19) at (7.5, -7) {1};
		\node [style=none] (20) at (9.5, -7) {2};
		\node [style=none] (21) at (-8, -7) {3};
		\node [style=none] (22) at (-6, -7) {2};
		\node [style=none] (23) at (-8, -11) {2};
		\node [style=none] (24) at (-6, -11) {2};
		\node [style=none] (89) at (-10.5, 2) {3};
		\node [style=none] (90) at (-5.5, -2) {2};
		\node [style=none] (91) at (-8.5, -2) {2};
		\node [style=none] (98) at (1.5, 2) {2};
		\node [style=none] (100) at (1.5, -2) {2};
		\node [style=none] (104) at (3.5, -2) {2};
		\node [style={gauge_gp}] (113) at (-5.5, -1) {};
		\node [style={flavour_gp}] (114) at (-5.5, 1) {};
		\node [style={flavour_gp}] (115) at (-8.5, 1) {};
		\node [style={gauge_gp}] (116) at (-8.5, -1) {};
		\node [style=none] (117) at (15.25, 2) {2};
		\node [style=none] (118) at (15.25, -2) {2};
		\node [style={gauge_gp}] (119) at (15.25, -1) {};
		\node [style={flavour_gp}] (120) at (15.25, 1) {};
		\node [style={flavour_gp}] (122) at (1.5, 1) {};
		\node [style={gauge_gp}] (123) at (1.5, -1) {};
		\node [style={gauge_gp}] (124) at (3.5, -1) {};
		\node [style=none, scale=1.25] (135) at (-7, 14.75) {\underline{Growth and fusion}};
		\node [style=none, scale=1.25] (136) at (8.5, 14.75) {\underline{Quiver addition}};
		\node [style={gauge_gp}] (137) at (-10.5, -1) {};
		\node [style={gauge_gp}] (138) at (-12.5, -1) {};
		\node [style={flavour_gp}] (139) at (-10.5, 1) {};
		\node [style=none] (140) at (-12.5, -2) {1};
		\node [style=none] (141) at (-10.5, -2) {2};
		\node [style=none] (142) at (-8.5, 2) {2};
		\node [style={gauge_gp}] (143) at (-3.5, -1) {};
		\node [style={flavour_gp}] (144) at (-3.5, 1) {};
		\node [style={gauge_gp}] (145) at (-1.5, -1) {};
		\node [style=none] (146) at (-3.5, -2) {2};
		\node [style=none] (147) at (-5.5, 2) {3};
		\node [style=none] (148) at (-3.5, 2) {2};
		\node [style=none] (149) at (-1.5, -2) {1};
		\node [style=none] (150) at (-8.5, -6) {};
		\node [style=none] (151) at (-10, -4) {};
		\node [style=none] (152) at (-5.5, -6) {};
		\node [style=none] (153) at (-4, -4) {};
		\node [style=none] (154) at (5.5, 2) {2};
		\node [style=none] (155) at (5.5, -2) {2};
		\node [style=none] (156) at (7.5, -2) {1};
		\node [style={flavour_gp}] (158) at (5.5, 1) {};
		\node [style={gauge_gp}] (159) at (5.5, -1) {};
		\node [style={gauge_gp}] (160) at (7.5, -1) {};
		\node [style=none] (161) at (13.25, 2) {1};
		\node [style=none] (162) at (13.25, -2) {2};
		\node [style={gauge_gp}] (163) at (13.25, -1) {};
		\node [style={flavour_gp}] (164) at (13.25, 1) {};
		\node [style=none] (165) at (11.25, 2) {1};
		\node [style=none] (166) at (11.25, -2) {2};
		\node [style={gauge_gp}] (167) at (11.25, -1) {};
		\node [style={flavour_gp}] (168) at (11.25, 1) {};
		\node [style=none] (170) at (9.5, -2) {2};
		\node [style={gauge_gp}] (171) at (9.5, -1) {};
		\node [style=none] (172) at (6.5, -6) {};
		\node [style=none] (173) at (4.75, -4) {};
		\node [style=none] (174) at (10.75, -6) {};
		\node [style=none] (175) at (12.5, -4) {};
		\node [style=none] (176) at (-3.5, 3) {};
		\node [style=none] (177) at (-3.5, 5) {};
		\node [style={gauge_gp}] (178) at (-5.5, 7) {};
		\node [style={gauge_gp}] (179) at (-3.5, 7) {};
		\node [style={flavour_gp}] (180) at (-5.5, 9) {};
		\node [style={flavour_gp}] (181) at (-3.5, 9) {};
		\node [style=none] (182) at (-5.5, 10) {3};
		\node [style=none] (183) at (-3.5, 10) {2};
		\node [style=none] (184) at (-5.5, 6) {3};
		\node [style=none] (185) at (-3.5, 6) {3};
		\node [style={gauge_gp}] (186) at (-1.5, 7) {};
		\node [style=none] (187) at (-1.5, 6) {1};
		\node [style={flavour_gp}] (189) at (13.25, 9) {};
		\node [style={gauge_gp}] (190) at (11.25, 7) {};
		\node [style={gauge_gp}] (191) at (13.25, 7) {};
		\node [style={gauge_gp}] (192) at (9.25, 7) {};
		\node [style={gauge_gp}] (193) at (15.25, 7) {};
		\node [style=none] (194) at (11.25, 6) {2};
		\node [style=none] (195) at (13.25, 6) {3};
		\node [style=none] (196) at (9.25, 6) {1};
		\node [style=none] (197) at (15.25, 6) {2};
		\node [style=none] (199) at (13.25, 10) {3};
		\node [style={flavour_gp}] (200) at (15.25, 9) {};
		\node [style=none] (201) at (15.25, 10) {1};
		\node [style=none] (202) at (12.25, 3) {};
		\node [style=none] (203) at (12.25, 5) {};
		\node [style=none] (204) at (-10.5, 3) {};
		\node [style=none] (205) at (-10.5, 5) {};
		\node [style=none] (206) at (-10.5, 5) {};
		\node [style=none] (207) at (-10.5, 6.25) {$\vdots$};
		\node [style=none] (208) at (-3.5, 11) {};
		\node [style=none] (209) at (-3.5, 12) {};
		\node [style=none] (210) at (-3.5, 13.25) {$\vdots$};
		\node [style=none] (211) at (12.5, 11) {};
		\node [style=none] (212) at (12.5, 12) {};
		\node [style=none] (213) at (12.5, 13.25) {$\vdots$};
		\node [style=none] (214) at (-10, -5.5) {$A_1$};
		\node [style=none] (215) at (-4, -5.5) {$A_1$};
		\node [style=none] (216) at (5, -5.5) {$A_1$};
		\node [style=none] (217) at (12.5, -5.5) {$A_1$};
		\node [style=none] (218) at (-2.75, 3.75) {$a_2$};
		\node [style=none] (219) at (13, 3.75) {$A_2$};
		\node [style=none] (220) at (4.5, 3) {};
		\node [style=none] (221) at (4.5, 5) {};
		\node [style=none] (222) at (4.5, 5) {};
		\node [style=none] (223) at (4.5, 6.25) {$\vdots$};
		\node [style=none] (224) at (-10.5, -3) {$(*)$};
		\node [style=none] (225) at (4.5, -3) {$(*)$};
		\node [style=none] (226) at (12.5, -3) {$(\dag)$};
		\node [style=none] (227) at (-3.5, -3) {$(\dag)$};
	\end{pgfonlayer}
	\begin{pgfonlayer}{edgelayer}
		\draw (2) to (3);
		\draw (3) to (5);
		\draw (4) to (2);
		\draw [style=double arrow] (7.center) to (6.center);
		\draw (12) to (11);
		\draw (11) to (9);
		\draw (13) to (11);
		\draw (14) to (12);
		\draw (12) to (10);
		\draw (114) to (113);
		\draw (115) to (116);
		\draw (120) to (119);
		\draw (122) to (123);
		\draw (123) to (124);
		\draw (144) to (143);
		\draw (113) to (143);
		\draw (143) to (145);
		\draw (138) to (137);
		\draw (137) to (116);
		\draw [style=arrow] (150.center) to (151.center);
		\draw [style=arrow] (152.center) to (153.center);
		\draw (139) to (137);
		\draw (158) to (159);
		\draw (159) to (160);
		\draw (159) to (124);
		\draw (164) to (163);
		\draw (168) to (167);
		\draw (171) to (167);
		\draw (167) to (163);
		\draw (163) to (119);
		\draw [style=arrow] (172.center) to (173.center);
		\draw [style=arrow] (174.center) to (175.center);
		\draw [style=arrow] (176.center) to (177.center);
		\draw (178) to (179);
		\draw (179) to (181);
		\draw (180) to (178);
		\draw (186) to (179);
		\draw (191) to (190);
		\draw (192) to (190);
		\draw (193) to (191);
		\draw (191) to (189);
		\draw (200) to (193);
		\draw [style=arrow] (202.center) to (203.center);
		\draw [style=arrow] (204.center) to (206.center);
		\draw [style=arrow] (208.center) to (209.center);
		\draw [style=arrow] (211.center) to (212.center);
		\draw [style=arrow] (220.center) to (222.center);
	\end{pgfonlayer}
\end{tikzpicture}
}
\end{equation}

\footnotetext[3]{Strictly speaking, in both cases (a) and (b) previously, where a single $\U(1)$ is added, if the resulting gauge nodes are balanced, it is a special case of this with $\mathfrak{g}=\mathfrak{su}(2)$.}

\end{widetext}

Where, once again, every quiver on the left has its 3D mirror on the right (the symbols help distinguish the pairs). This is a \emph{subset} of all possible mirror pairs that can be generated in two unHiggsing steps, which showcases the power of this bootstrapping method in generating new mirror pairs. A key practical question is that if we perform growth and fusion to obtain a quiver, how do we know which of the myriad quiver additions to perform on quiver $Y$ to obtain the 3D mirror? At first sight, quiver addition for any quiver may look like it has infinite possibilities. However, this is significantly constrained by eliminating additions that lead to two decoupled theories; for instance, connecting any quiver $\mathrm{Q}'$ only to a $\U(1)$ node in quiver $Y$ results in the decoupled theory quiver $Y\times \mathrm{Q}'$ after decoupling that $\U(1)$. Furthermore, if we constrain ourselves to $A_n$ and $a_k$ type transverse spaces in the growth and fusion algorithm\footnote{This is important to make sure we are only dealing with quivers made of unitary gauge groups.}, under 3D mirror symmetry, they translate to $a_n$ and $A_k$ transverse spaces in the mirror, respectively. With this, the possible results become much more finite, and in most cases we have seen, we are able to precisely identify the exact 3D mirror pair. Any remaining ambiguity can be resolved by matching decay and fission on one side with quiver subtractions on the other side to make sure daughter theories match. This way, we can remove degeneracies without performing Hilbert series computations (which is the last resort).

\section{Beyond Dynkin mirror pairs: sunshine quivers}
\label{sec:nonlinear}

This section aims to significantly extend the landscape of 3D mirror pairs beyond the well-known Dynkin quivers. We introduce a novel class of nonlinear quivers, which we term \emph{sunshine quivers}, due to their distinctive diagrammatic shape. The theories discussed here are composed of only unitary gauge and flavor nodes unless explicitly stated otherwise, and all possess Lagrangian descriptions.

We systematically explore families of these sunshine quiver mirror pairs. This section starts with introducing the glueing method as an alternative way to construct quivers and their 3D mirror. This method establishes dualities on the Abelian case, which serves as a foundation for extending our analysis to more complex, non-Abelian theories and even structures beyond sunshine quivers, such as forked rays and higher genus quivers. Each proposed mirror pair is supported by explicit examples and verified through Hilbert series computations. These quivers can be generated systematically using the algorithms introduced in Sec.\ \ref{sec:algo}. 

\subsection{Sunshine quiver structure and terminology}\label{subsec:notation}

Sunshine quivers are characterized by two distinct components: a central loop of unitary gauge nodes and linear rays that extend from this loop.

\begin{enumerate}[label=(\roman*),noitemsep]
    \item \textbf{Central Loop}: The loop is a closed cycle of unitary gauge nodes, which we call \emph{loop nodes}, connected sequentially by bifundamental hypermultiplets, also referred to as \emph{bonds}. We require that no other closed cycles exist within the quiver. The bond connecting two consecutive loop nodes is a \emph{loop edge}, and its \emph{multiplicity} is defined as the number of bonds on that edge.
    \item \textbf{Rays}: A ray is a linear chain of gauge nodes that extends from a single loop node and terminates at a flavor node. The \emph{length} of a ray is the number of nodes it contains, excluding the loop node from which it originates, but including the final flavor node. A single loop node can have multiple rays emanating from it.
\end{enumerate}
To uniquely identify a sunshine quiver, we propose a representation based on the following ordered list of parameters: $[n, \{k_{1}, ..., k_{n}\}, \{m_{1}, ..., m_{n}\}, \{r_{j, l}\}_{j\in \{1,...,n\}, l\in \mathbb{Z}^+}]$.

\begin{enumerate}[label=(\arabic*),noitemsep]
    \item $n$: The number of gauge nodes in the central loop.
    \item $k_{i}$: The rank of the $i$th loop node (i.e.\ $\U(k_i)$), with $i \in \{1, ..., n\}$.
    \item $m_{i}$: The multiplicity of the loop edge between the $i$th and $(i+1)$th loop node, with $i \in \{1, ..., n\}$. The indices are taken modulo $n$ to account for the closed loop.
    \item $r_{j, l}$: The set of rays, where:
    \begin{enumerate}[label=(\alph*),noitemsep]
        \item[$\vartriangleright$] $j\in \{1,...,n\}$ represents the position of the loop node from which the ray emerges.
        \item[$\vartriangleright$] $l\in \mathbb{Z}^+$ is the length of the ray.
    \end{enumerate}
\end{enumerate}
For a sunshine quiver to be specified using this notation, we require that the flavor node at the end of each ray is rank 1. To satisfy this rule, it is useful to note that we can ``explode'' a $\U(k)$ flavor node into $k$ $\U(1)$ flavor nodes, as follows:
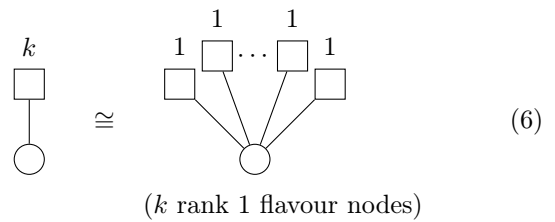
\vspace{-1.5em}

\begin{equation}
\label{eqn:expl_flav}
    \begin{tikzpicture}[baseline=-0.25em,scale=0.5]
        \begin{pgfonlayer}{nodelayer}
		\node [style=none] (0) at (0, 0) {$\cong$};
		\node [style={gauge_gp}] (1) at (-2, -1) {};
		\node [style={flavour_gp}] (2) at (-2, 1) {};
		\node [style=none] (3) at (-2, 2) {$k$};
		\node [style={gauge_gp}] (4) at (4, -1) {};
		\node [style={flavour_gp}] (5) at (2, 1) {};
		\node [style={flavour_gp}] (6) at (3, 1.75) {};
		\node [style={flavour_gp}] (7) at (5, 1.75) {};
		\node [style={flavour_gp}] (8) at (6, 1) {};
		\node [style=none] (9) at (4, 1.75) {$\cdots$};
		\node [style=none] (17) at (4, 3.75) {($k$ rank 1 flavor nodes)};
		\node [style=none] (23) at (2, 2) {1};
		\node [style=none] (24) at (3, 2.75) {1};
		\node [style=none] (25) at (5, 2.75) {1};
		\node [style=none] (26) at (6, 2) {1};
	\end{pgfonlayer}
	\begin{pgfonlayer}{edgelayer}
		\draw (2) to (1);
		\draw (5) to (4);
		\draw (8) to (4);
		\draw (7) to (4);
		\draw (4) to (6);
	\end{pgfonlayer}
\end{tikzpicture}
\end{equation}
$[5, \{1,2,2,1,1\}, \{1,1,1,2,3\}, \{r_{1,2}, r_{1,3}, r_{2,1}^3, r_{3,1}^3\}]$, as an example, defines the quiver \eqref{eqn:sunshinelabel}, where we write $r_{2,1}^3$ to represent $\{r_{2,1}, r_{2,1}, r_{2,1}\}$ for brevity. The grey numbers on each node in this diagram denote the counting index of loop nodes, and the black text denotes rank as usual.

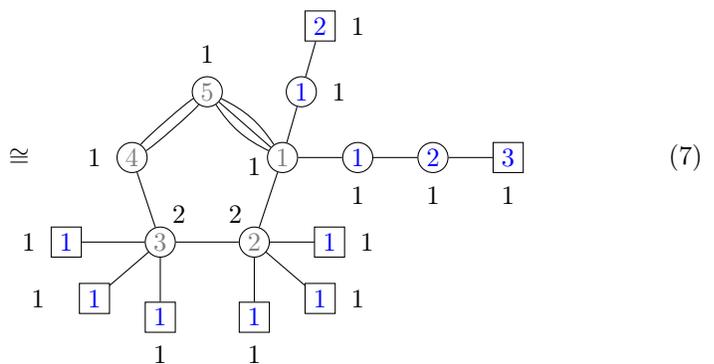
\begin{widetext}
\begin{equation}
    \label{eqn:sunshinelabel}
    \begin{tikzpicture}[baseline=-0.25em,scale=0.5]
	\begin{pgfonlayer}{nodelayer}
		\node [style={gauge_gp}] (113) at (3.75, -2.25) {};
		\node [style={gauge_gp}] (115) at (6.25, -2.25) {};
		\node [style={gauge_gp}] (116) at (3, 0) {};
		\node [style={gauge_gp}] (117) at (7, 0) {};
		\node [style={gauge_gp}] (118) at (5, 1.75) {};
		\node [style={gauge_gp}] (119) at (7.5, 1.75) {};
		\node [style={gauge_gp}] (120) at (9, 0) {};
		\node [style={gauge_gp}] (121) at (11, 0) {};
		\node [style={flavour_gp}] (122) at (13, 0) {};
		\node [style={flavour_gp}] (123) at (8, 3.5) {};
		\node [style={flavour_gp}] (124) at (6.25, -4.25) {};
		\node [style={flavour_gp}] (125) at (3.75, -4.25) {};
		\node [style={flavour_gp}] (126) at (1.25, -2.25) {};
		\node [style=none] (127) at (3.75, -5.25) {1};
		\node [style=none] (128) at (6.25, -5.25) {1};
		\node [style=none] (129) at (0.25, -2.25) {1};
		\node [style=none] (130) at (9, 3.5) {1};
		\node [style=none] (131) at (9, -1) {1};
		\node [style=none] (132) at (11, -1) {1};
		\node [style=none] (133) at (13, -1) {1};
		\node [style=none] (134) at (8.5, 1.75) {1};
		\node [style=none] (135) at (6.25, -0.25) {1};
		\node [style=none] (136) at (5.75, -1.5) {2};
		\node [style=none] (137) at (4.25, -1.5) {2};
		\node [style=none] (138) at (5, 2.75) {1};
		\node [style=none] (139) at (2, 0) {1};
		\node [style=none] (140) at (5, 1.75) {\textcolor{lightgray}{5}};
		\node [style=none] (141) at (7, 0) {\textcolor{lightgray}{1}};
		\node [style=none] (142) at (6.25, -2.25) {\textcolor{lightgray}{2}};
		\node [style=none] (143) at (3.75, -2.25) {\textcolor{lightgray}{3}};
		\node [style=none] (144) at (3, 0) {\textcolor{lightgray}{4}};
		\node [style={gauge_gp}] (153) at (-11.5, -2.25) {};
		\node [style={gauge_gp}] (154) at (-9, -2.25) {};
		\node [style={gauge_gp}] (155) at (-12.25, 0) {};
		\node [style={gauge_gp}] (156) at (-8.25, 0) {};
		\node [style={gauge_gp}] (157) at (-10.25, 1.75) {};
		\node [style={gauge_gp}] (158) at (-7.75, 1.75) {};
		\node [style={gauge_gp}] (159) at (-6.25, 0) {};
		\node [style={gauge_gp}] (160) at (-4.25, 0) {};
		\node [style={flavour_gp}] (161) at (-2.25, 0) {};
		\node [style={flavour_gp}] (162) at (-7.25, 3.5) {};
		\node [style={flavour_gp}] (163) at (-9, -4.25) {};
		\node [style={flavour_gp}] (164) at (-11.5, -4.25) {};
		\node [style=none] (167) at (-9, -5.25) {3};
		\node [style=none] (168) at (-11.5, -5.25) {3};
		\node [style=none] (169) at (-6.25, 3.5) {1};
		\node [style=none] (170) at (-6.25, -1) {1};
		\node [style=none] (171) at (-4.25, -1) {1};
		\node [style=none] (172) at (-2.25, -1) {1};
		\node [style=none] (173) at (-6.75, 1.75) {1};
		\node [style=none] (174) at (-9, -0.25) {1};
		\node [style=none] (175) at (-8, -2.5) {2};
		\node [style=none] (176) at (-12.5, -2.5) {2};
		\node [style=none] (177) at (-10.25, 2.75) {1};
		\node [style=none] (178) at (-12.75, -0.75) {1};
		\node [style=none] (179) at (0, 0) {$\cong$};
		\node [style={flavour_gp}] (180) at (8.25, -2.25) {};
		\node [style=none] (181) at (9.25, -2.25) {1};
		\node [style={flavour_gp}] (183) at (2, -3.75) {};
		\node [style=none] (184) at (0.5, -3.75) {1};
		\node [style={flavour_gp}] (186) at (8, -3.75) {};
		\node [style=none] (187) at (9, -3.75) {1};
	\end{pgfonlayer}
	\begin{pgfonlayer}{edgelayer}
		\draw (115) to (113);
		\draw (117) to (115);
		\draw (116) to (113);
		\draw (118) to (117);
		\draw [bend right=15] (118) to (117);
		\draw [bend left=15] (118) to (117);
		\draw [bend left=15, looseness=0.25] (116) to (118);
		\draw [bend left=15, looseness=0.25] (118) to (116);
		\draw (123) to (119);
		\draw (119) to (117);
		\draw (117) to (120);
		\draw (120) to (121);
		\draw (121) to (122);
		\draw (126) to (113);
		\draw (113) to (125);
		\draw (124) to (115);
		\draw (154) to (153);
		\draw (156) to (154);
		\draw (155) to (153);
		\draw (157) to (156);
		\draw [bend right=15] (157) to (156);
		\draw [bend left=15] (157) to (156);
		\draw [bend left=15, looseness=0.25] (155) to (157);
		\draw [bend left=15, looseness=0.25] (157) to (155);
		\draw (162) to (158);
		\draw (158) to (156);
		\draw (156) to (159);
		\draw (159) to (160);
		\draw (160) to (161);
		\draw (153) to (164);
		\draw (163) to (154);
		\draw (183) to (143.center);
		\draw (186) to (142.center);
		\draw (142.center) to (180);
	\end{pgfonlayer}
\end{tikzpicture}
\end{equation}
\end{widetext}

We make the simplifying assumption that the central loop consists of at least three nodes ($n \geq 3$), and that sunshine quivers have at least three loop nodes emitting rays. With this assumption in the definition of sunshine quivers, all sunshine quivers are dual to another sunshine quiver.\footnote{For 3D mirror pairs when there are fewer than three loop nodes or fewer than three rays, refer to Appendix \ref{appx:no-ray}.}

\vspace{-1.5em}
\subsection{The gluing method}\label{subsec:glueing_method}
\vspace{-1em}

This section introduces a method of building cyclical sunshine quivers by gluing together smaller quivers. This provides an alternate and, in some cases, faster method to find 3D mirror pairs, as compared with the algorithms developed in Sec.\ \ref{sec:algo}. 

The setup is as follows: consider two good linear unitary quivers $A$ and $A'$, which both have at least two flavors in total. Their mirror quivers, denoted $B$ and $B'$, respectively, can be identified by applying Hanany-Witten transitions to the D3-D5-NS5 brane system of the quiver. Then, it is possible to glue together a rank 1 flavor node from $A$ and $A'$ into a $\U(1)$ gauge node. These flavors can be exploded from a $\U(k)$ flavor node, as stated in \eqref{eqn:expl_flav}. For example, we can glue together the flavor nodes of $[1]-(1)-(1)-(1)-[1]$ and $[1]-(1)-(1)-[1]$ as follows:
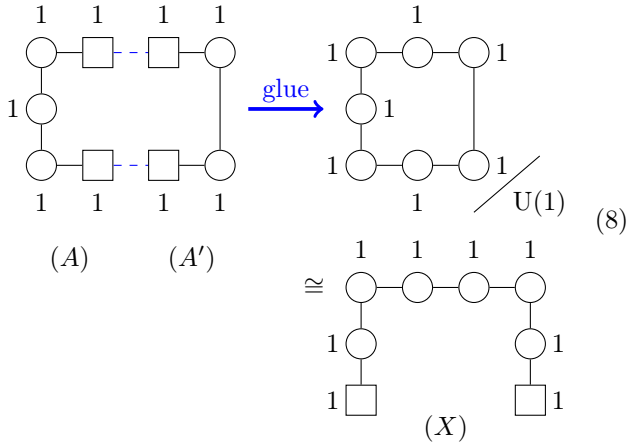
\begin{equation} \label{eqn:gluingA1}
    \scalebox{1}{
    \begin{tikzpicture}[baseline=-0.25em,scale=0.5]
	\begin{pgfonlayer}{nodelayer}
		\node [style={gauge_gp}] (0) at (-6.5, 1.5) {};
		\node [style={gauge_gp}] (1) at (-6.5, 3) {};
		\node [style={gauge_gp}] (2) at (-6.5, 4.5) {};
		\node [style={flavour_gp}] (3) at (-5, 1.5) {};
		\node [style={flavour_gp}] (4) at (-5, 4.5) {};
		\node [style={gauge_gp}] (5) at (-1.75, 1.5) {};
		\node [style={gauge_gp}] (7) at (-1.75, 4.5) {};
		\node [style={flavour_gp}] (8) at (-3.25, 1.5) {};
		\node [style={flavour_gp}] (9) at (-3.25, 4.5) {};
		\node [style=none] (10) at (-5, 4.5) {};
		\node [style=none] (11) at (-3.25, 4.5) {};
		\node [style=none] (12) at (-5, 1.5) {};
		\node [style=none] (13) at (-3.25, 1.5) {};
		\node [style=none] (14) at (-1, 3) {};
		\node [style=none] (15) at (1, 3) {};
		\node [style=none] (16) at (0, 3.5) {glue};
		\node [style={gauge_gp}] (17) at (2, 1.5) {};
		\node [style={gauge_gp}] (18) at (2, 3) {};
		\node [style={gauge_gp}] (19) at (2, 4.5) {};
		\node [style={gauge_gp}] (20) at (3.5, 1.5) {};
		\node [style={gauge_gp}] (21) at (3.5, 4.5) {};
		\node [style={gauge_gp}] (22) at (5, 1.5) {};
		\node [style={gauge_gp}] (23) at (5, 4.5) {};
		\node [style=none] (27) at (6.75, 1.75) {};
		\node [style=none] (28) at (5, 0.25) {};
		\node [style=none] (29) at (6.75, 0.5) {U(1)};
		\node [style=none] (30) at (-5.75, -1) {$(A)$};
		\node [style=none] (31) at (-2.5, -1) {$(A')$};
		\node [style=none] (32) at (3.5, -6.25) {$(X)$};
		\node [style=none] (33) at (3.5, -0.5) {\rotatebox{270}{$\cong$}};
		\node [style={gauge_gp}] (34) at (1.25, -4) {};
		\node [style={gauge_gp}] (35) at (1.25, -2.5) {};
		\node [style={gauge_gp}] (36) at (2.75, -2.5) {};
		\node [style={gauge_gp}] (37) at (4.25, -2.5) {};
		\node [style={gauge_gp}] (38) at (5.75, -2.5) {};
		\node [style={gauge_gp}] (39) at (5.75, -4) {};
		\node [style={flavour_gp}] (40) at (1.25, -5.5) {};
		\node [style={flavour_gp}] (41) at (5.75, -5.5) {};
		\node [style=none] (42) at (-6.5, 5.5) {1};
		\node [style=none] (43) at (-5, 5.5) {1};
		\node [style=none] (44) at (-1.75, 5.5) {1};
		\node [style=none] (45) at (-3.25, 5.5) {1};
		\node [style=none] (46) at (-7.25, 3) {1};
		\node [style=none] (47) at (-6.5, 0.5) {1};
		\node [style=none] (48) at (-5, 0.5) {1};
		\node [style=none] (49) at (-3.25, 0.5) {1};
		\node [style=none] (50) at (-1.75, 0.5) {1};
		\node [style=none] (51) at (1.25, 4.5) {1};
		\node [style=none] (52) at (3.5, 5.5) {1};
		\node [style=none] (53) at (2.75, 3) {1};
		\node [style=none] (54) at (5.75, 4.5) {1};
		\node [style=none] (55) at (1.25, 1.5) {1};
		\node [style=none] (56) at (3.5, 0.5) {1};
		\node [style=none] (57) at (5.75, 1.5) {1};
		\node [style=none] (58) at (0.5, -5.5) {1};
		\node [style=none] (59) at (6.5, -5.5) {1};
		\node [style=none] (60) at (0.5, -4) {1};
		\node [style=none] (61) at (1.25, -1.5) {1};
		\node [style=none] (62) at (2.75, -1.5) {1};
		\node [style=none] (63) at (4.25, -1.5) {1};
		\node [style=none] (64) at (5.75, -1.5) {1};
		\node [style=none] (65) at (6.5, -4) {1};
	\end{pgfonlayer}
	\begin{pgfonlayer}{edgelayer}
		\draw (3) to (0);
		\draw (0) to (1);
		\draw (2) to (4);
		\draw (2) to (1);
		\draw (8) to (5);
		\draw (7) to (9);
		\draw (5) to (7);
		\draw (17) to (18);
		\draw (19) to (18);
		\draw (17) to (20);
		\draw (19) to (21);
		\draw (22) to (23);
		\draw (21) to (23);
		\draw (20) to (22);
		\draw (27.center) to (28.center);
		\draw (34) to (35);
		\draw (35) to (36);
		\draw (37) to (38);
		\draw (38) to (39);
		\draw (37) to (36);
		\draw (40) to (34);
		\draw (41) to (39);
		\draw [style=arrow] (14.center) to (15.center);
		\draw [style=dashed] (10.center) to (11.center);
		\draw [style=dashed] (13.center) to (12.center);
	\end{pgfonlayer}
\end{tikzpicture}
}
\end{equation}

$X$ is the sunshine quiver, $[7,\{1^7\},\{1^7\}]$, with no rays. Given $A$ and $A'$ in the above gluing, we know what $B$ and $B'$ are---$[4]-(1)$ and $[3]-(1)$ respectively---but how do we glue these together? It is worth noting that gluing together the flavor nodes decreased the dimension of the Higgs branch symmetry, and so, in the mirror theory, the dimension of the Coulomb branch symmetry must decrease. Therefore, we must glue together two $\U(1)$ gauge nodes in $B$ and $B'$. Flavors joined to each $\U(1)$ gauge node at the point of gluing are all attached to the new $\U(1)$ in the resulting quiver. This gives
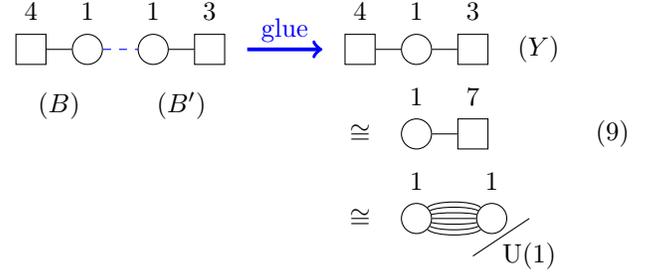
\begin{equation} \label{eqn:gluingA2}
    \begin{tikzpicture}[baseline=-0.25em,scale=0.5]
	\begin{pgfonlayer}{nodelayer}
		\node [style={gauge_gp}] (2) at (-5.25, 2.25) {};
		\node [style={flavour_gp}] (4) at (-6.75, 2.25) {};
		\node [style={gauge_gp}] (6) at (-3.5, 2.25) {};
		\node [style={flavour_gp}] (8) at (-2, 2.25) {};
		\node [style=none] (9) at (-5.25, 2.25) {};
		\node [style=none] (10) at (-3.5, 2.25) {};
		\node [style=none] (13) at (-1, 2.25) {};
		\node [style=none] (14) at (1, 2.25) {};
		\node [style=none] (23) at (-6.75, 3.25) {4};
		\node [style=none] (24) at (-2, 3.25) {3};
		\node [style={gauge_gp}] (25) at (3.5, 0) {};
		\node [style={flavour_gp}] (26) at (5, 0) {};
		\node [style=none] (28) at (5, 1) {7};
		\node [style=none] (31) at (-6, 0.75) {$(B)$};
		\node [style=none] (32) at (-2.75, 0.75) {$(B')$};
		\node [style=none] (33) at (6.75, 2.25) {$(Y)$};
		\node [style=none] (34) at (0, 2.75) {glue};
		\node [style=none] (35) at (2, -2.25) {$\cong$};
		\node [style={gauge_gp}] (36) at (3.5, -2.25) {};
		\node [style={gauge_gp}] (39) at (5.5, -2.25) {};
		\node [style=none] (41) at (6.5, -2.25) {};
		\node [style=none] (42) at (5, -3.25) {};
		\node [style=none] (43) at (6.5, -3.25) {U(1)};
		\node [style=none] (44) at (-5.25, 3.25) {1};
		\node [style=none] (45) at (-3.5, 3.25) {1};
		\node [style=none] (46) at (3.5, 1) {1};
		\node [style=none] (47) at (5.5, -1.25) {1};
		\node [style=none] (48) at (3.5, -1.25) {1};
		\node [style={flavour_gp}] (50) at (2, 2.25) {};
		\node [style={gauge_gp}] (51) at (3.5, 2.25) {};
		\node [style={flavour_gp}] (52) at (5, 2.25) {};
		\node [style=none] (55) at (2, 3.25) {4};
		\node [style=none] (56) at (5, 3.25) {3};
		\node [style=none] (58) at (3.5, 3.25) {1};
		\node [style=none] (59) at (2, 0) {$\cong$};
	\end{pgfonlayer}
	\begin{pgfonlayer}{edgelayer}
		\draw (2) to (4);
		\draw (6) to (8);
		\draw [style=dashed] (9.center) to (10.center);
		\draw [style=arrow] (13.center) to (14.center);
		\draw (25) to (26);
		\draw [bend right=45, looseness=0.50] (36) to (39);
		\draw [bend left=45, looseness=0.50] (36) to (39);
		\draw (41.center) to (42.center);
		\draw [bend left, looseness=0.50] (36) to (39);
		\draw (36) to (39);
		\draw [bend right, looseness=0.50] (36) to (39);
		\draw [bend left=15, looseness=0.50] (36) to (39);
		\draw [bend right=15, looseness=0.50] (36) to (39);
		\draw (51) to (52);
		\draw (51) to (50);
	\end{pgfonlayer}
\end{tikzpicture}
\end{equation}

The quivers resulting from the gluings \eqref{eqn:gluingA1} and \eqref{eqn:gluingA2}, $X$ and $Y$ respectively, are a 3D mirror pair. This simple example shows that the gluing method is a potential way to combine smaller 3D mirror quivers into a larger 3D mirror. In this case, there was only one choice of gluing due to the simplicity of the quivers, but for quivers with more flavor and gauge nodes, there can be multiple choices. Looking at the Coulomb and Higgs branch symmetries of the quivers may enlighten us as to which gluing is the correct choice, which can be further verified and confirmed with explicit Hilbert series. We will develop this idea further throughout this section for Abelian and non-Abelian sunshine quivers.

\vspace{-.5em}
\subsection{Abelian sunshine mirror pairs}\label{subsec:abelian_mirror_pairs}
\vspace{-.5em}
In Abelian quivers, all gauge nodes are $\U(1)$ and most flavor nodes are rank 1. Hence, for brevity and clarity in quiver diagrams, we omit all gauge node labels and only retain flavor node labels for rank $>1$.

We propose a correspondence between the properties of an Abelian sunshine quiver and its mirror. The number of consecutive balanced $\U(1)$ gauge nodes (a ``balanced chain'' of $\U(1)$s) is a mirror to bond multiplicity or number of flavors\footnote{Recall that gauging a $\U(1)$ combines all flavor nodes, and number of flavors translates to bond multiplicity of the edge between the original gauge node endpoint and the new $\U(1)$ node.} in its dual quiver.

The mathematical basis behind this duality is straightforward: a chain of $k-1$ balanced $\U(1)$ gauge nodes (i.e.\ $k$ edges) contributes $\SU(k)$ to the Coulomb branch $G_\mathcal{C}$, whereas $k$ flavors (by definition) or a bond of multiplicity $k$ contributes $\SU(k)$ to the Higgs branch $G_\mathcal{H}$. This stems from the relation between the transverse space $A_n$ and $a_k$ that appears in the growth and fusion algorithm.

More specifically, a balanced $\U(1)$ chain on a ray is dual to loop bond multiplicity, and a balanced $\U(1)$ chain in the loop (separating two overbalanced loop nodes) corresponds to the number of flavors.

We can use these dualities to build arbitrarily large Abelian sunshine quivers. Starting with the simplest such theory --- consider a loop of $\U(1)$ gauge nodes connected by single bonds, which all have a rank 1 flavor node attached (rays of length 1). These are all self-dual. In the notation we introduced, this is $[n,\{1^n\},\{1^n\},\{r_{1,1},r_{2,1},r_{3,1},...,r_{n,1}\}]$ for a size-$n$ loop. For example, such quivers corresponding to loops of size 3 and 4 are as follows (note $\HS_\mathcal{C}(t)=\HS_\mathcal{H}(t)$ indicates self-duality):
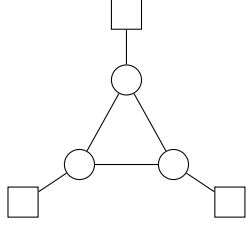
\vspace{-1em}
\begin{equation} \label{self_dual_3}
    \begin{tikzpicture}[baseline=-0.25em,scale=0.5]
	\begin{pgfonlayer}{nodelayer}
		\node [style={gauge_gp}] (111) at (0, 1.25) {};
		\node [style={flavour_gp}] (112) at (0, 3) {};
		\node [style={gauge_gp}] (113) at (-1.25, -1) {};
		\node [style={gauge_gp}] (115) at (1.25, -1) {};
		\node [style={flavour_gp}] (116) at (2.75, -2) {};
		\node [style={flavour_gp}] (139) at (-2.75, -2) {};
		\node [style=none] (140) at (0, -3.5) {$[3,\{1,1,1\},\{1,1,1\},\{r_{1,1},r_{2,1},r_{3,1}\}]$};
	\end{pgfonlayer}
	\begin{pgfonlayer}{edgelayer}
		\draw (112) to (111);
		\draw (116) to (115);
		\draw (113) to (111);
		\draw (115) to (113);
		\draw (139) to (113);
		\draw (111) to (115);
	\end{pgfonlayer}
\end{tikzpicture}
\end{equation}
\vspace{-1em}
\begin{eqnarray*}
\HS_\mathcal{C}(t)=&&\HS_\mathcal{H}(t)\nonumber\\
=&&1 + 3t^2 + 8t^3 + 12t^4 + 24t^5 + 48t^6 + 72t^7 +\\
&&117t^8 + 184t^9 + 267t^{10} + \mathcal{O}(t^{11})
\end{eqnarray*}


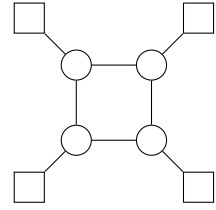
\begin{widetext}
\begin{equation} \label{eqn:sunshine4}
    \begin{tikzpicture}[baseline=-0.5em,scale=0.5]
	\begin{pgfonlayer}{nodelayer}
		\node [style={gauge_gp}] (111) at (-1, 1) {};
		\node [style={flavour_gp}] (112) at (-2.25, 2.25) {};
		\node [style={gauge_gp}] (113) at (-1, -1) {};
		\node [style={gauge_gp}] (114) at (1, 1) {};
		\node [style={gauge_gp}] (115) at (1, -1) {};
		\node [style={flavour_gp}] (116) at (2.25, -2.25) {};
		\node [style={flavour_gp}] (138) at (2.25, 2.25) {};
		\node [style={flavour_gp}] (139) at (-2.25, -2.25) {};
		\node [style=none] (140) at (0, -3.5) {
$[4,\{1,1,1,1\},\{1,1,1,1\},\{r_{1,1},r_{2,1},r_{3,1},r_{4,1}\}]$};
	\end{pgfonlayer}
	\begin{pgfonlayer}{edgelayer}
		\draw (112) to (111);
		\draw (116) to (115);
		\draw (113) to (111);
		\draw (111) to (114);
		\draw (115) to (113);
		\draw (115) to (114);
		\draw (138) to (114);
		\draw (139) to (113);
	\end{pgfonlayer}
\end{tikzpicture}
\end{equation}
\begin{equation*}
\HS_\mathcal{C}(t)=\HS_\mathcal{H}(t)
=1 + 4t^2 + 8t^3 + 20t^4 + 40t^5 + 84t^6 + 152t^7 
285t^8 + 472t^9 + 769t^{10} + \mathcal{O}(t^{11})
\end{equation*}

\subsubsection{Adding loop nodes and flavours}

One modification to the simplest sunshine quivers to obtain new 3D mirror pairs is to add flavours to one quiver and loop nodes to its mirror. For instance, starting from the simple four-loop sunshine quiver in \eqref{eqn:sunshine4}, $[4,\{1,1,1,1\},\{1,1,1,1\},\{r_{1,1},r_{2,1},r_{3,1},r_{4,1}\}]$, we can obtain the following mirror pair:

\begin{equation} \label{eqn:flav+loopnode}
\scalebox{0.9}{
    \begin{tikzpicture}[baseline=-0.25em,scale=0.5]
	\begin{pgfonlayer}{nodelayer}
		\node [style={gauge_gp}] (111) at (-6.5, 1.5) {};
		\node [style={flavour_gp}] (112) at (-7.75, 2.75) {};
		\node [style={gauge_gp}] (113) at (-6.5, -1) {};
		\node [style={gauge_gp}] (114) at (-4, 1.5) {};
		\node [style={gauge_gp}] (115) at (-4, -1) {};
		\node [style={flavour_gp}] (116) at (-2.75, -2.25) {};
		\node [style={flavour_gp}] (138) at (-2.75, 2.75) {};
		\node [style={flavour_gp}] (139) at (-7.75, -2.25) {};
		\node [style=none] (140) at (1.5, 0) {};
		\node [style=none] (141) at (-1.5, 0) {};
		\node [style=none] (142) at (0, 0.75) {3D mirror};
		\node [style={gauge_gp}] (144) at (6, 1.5) {};
		\node [style={gauge_gp}] (146) at (4.5, 0) {};
		\node [style={gauge_gp}] (147) at (7.5, 0) {};
		\node [style={gauge_gp}] (148) at (6, -1.5) {};
		\node [style={flavour_gp}] (149) at (6, -3) {};
		\node [style={flavour_gp}] (150) at (9, 0) {};
		\node [style={flavour_gp}] (151) at (3, 0) {};
		\node [style=none] (153) at (6.75, 3) {2};
		\node [style={gauge_gp}] (159) at (14.5, 1.5) {};
		\node [style={gauge_gp}] (160) at (13, 0) {};
		\node [style={gauge_gp}] (161) at (16, 0) {};
		\node [style={gauge_gp}] (162) at (14.5, -1.5) {};
		\node [style={flavour_gp}] (163) at (14.5, -3) {};
		\node [style={flavour_gp}] (164) at (17.5, 0) {};
		\node [style={flavour_gp}] (165) at (11.5, 0) {};
		\node [style=none] (169) at (10.25, 0) {$\cong$};
		\node [style=none] (171) at (-5.25, -6.75) {($X$)};
		\node [style=none] (172) at (10.25, -6.75) {($Y$)};
		\node [style=none] (173) at (-5.25, -4.25) {$[5,\{1,1,1,1,1\},\{1,1,1,1,1\},$};
		\node [style=none] (174) at (10.25, -4.25) {$[4,\{1,1,1,1\},\{1,1,1,1\},$};
		\node [style=none] (175) at (-5.25, -5.5) {$\{r_{1,1},r_{2,1},r_{3,1},r_{4,1}\}]$};
		\node [style=none] (176) at (10.25, -5.5) {$\{r_{1,1}^{2},r_{2,1},r_{3,1},r_{4,1}\}]$};
		\node [style=none] (180) at (2.5, -8.25) {$\HS_\mathcal{C}^X(t) = 1 + 7t^2 + 12t^3 + 41t^4 + 88t^5 + 215t^6 + 428t^7 + 902t^8 + 1656t^9 + 3114t^{10} + \mathcal{O}(t^{11}) = \HS_\mathcal{H}^Y(t)$};
		\node [style=none] (181) at (2.5, -9.5) {$\HS_\mathcal{H}^X(t) = 1 + 4t^2 + 6t^3 + 16t^4 + 32t^5 + 64t^6 + 112t^7 + 209t^8 + 342t^9 + 570t^{10} + \mathcal{O}(t^{11}) = \HS_\mathcal{C}^Y(t)$};
		\node [style={gauge_gp}] (182) at (-5.25, 2.25) {};
		\node [style={flavour_gp}] (183) at (13.25, 2.75) {};
		\node [style={flavour_gp}] (184) at (15.75, 2.75) {};
		\node [style={flavour_gp}] (185) at (6, 3) {};
	\end{pgfonlayer}
	\begin{pgfonlayer}{edgelayer}
		\draw (112) to (111);
		\draw (116) to (115);
		\draw (113) to (111);
		\draw (115) to (113);
		\draw (115) to (114);
		\draw (138) to (114);
		\draw (139) to (113);
		\draw [style=double arrow] (141.center) to (140.center);
		\draw (149) to (148);
		\draw (146) to (144);
		\draw (148) to (146);
		\draw (148) to (147);
		\draw (150) to (147);
		\draw (151) to (146);
		\draw (144) to (147);
		\draw (163) to (162);
		\draw (160) to (159);
		\draw (162) to (160);
		\draw (162) to (161);
		\draw (164) to (161);
		\draw (165) to (160);
		\draw (159) to (161);
		\draw (182) to (114);
		\draw (182) to (111);
		\draw (183) to (159);
		\draw (184) to (159);
		\draw (185) to (144);
	\end{pgfonlayer}
\end{tikzpicture}
}
\end{equation}
\vspace{-.5em}

We can do this for arbitrarily many nodes $k\in\mathbb{Z}^+$:
\vspace{-.5em}
\begin{equation}\label{eqn:add_flav}
    \begin{tikzpicture}[baseline=-0.25em,scale=0.5]
	\begin{pgfonlayer}{nodelayer}
		\node [style={gauge_gp}] (111) at (-7, 1) {};
		\node [style={flavour_gp}] (112) at (-8.25, 2.25) {};
		\node [style={gauge_gp}] (113) at (-7, -1.5) {};
		\node [style={gauge_gp}] (114) at (-3.5, 1) {};
		\node [style={gauge_gp}] (115) at (-3.5, -1.5) {};
		\node [style={flavour_gp}] (116) at (-2.25, -2.75) {};
		\node [style={flavour_gp}] (138) at (-2.25, 2.25) {};
		\node [style={flavour_gp}] (139) at (-8.25, -2.75) {};
		\node [style=none] (140) at (2, 0) {};
		\node [style=none] (141) at (-2, 0) {};
		\node [style=none] (142) at (0, 0.75) {3D mirror};
		\node [style={gauge_gp}] (144) at (6.25, 1.5) {};
		\node [style={gauge_gp}] (146) at (4.75, 0) {};
		\node [style={gauge_gp}] (147) at (7.75, 0) {};
		\node [style={gauge_gp}] (148) at (6.25, -1.5) {};
		\node [style={flavour_gp}] (149) at (6.25, -3) {};
		\node [style={flavour_gp}] (150) at (9.25, 0) {};
		\node [style={flavour_gp}] (151) at (3.25, 0) {};
		\node [style=none] (153) at (6.25, 4) {$k$};
		\node [style={gauge_gp}] (156) at (-6.75, 2.25) {};
		\node [style={flavour_gp}] (158) at (6.25, 3) {};
		\node [style={gauge_gp}] (159) at (-3.75, 2.25) {};
		\node [style=none] (160) at (-6, 3) {};
		\node [style=none] (161) at (-4.5, 3) {};
		\node [style=none] (162) at (-5.25, 3) {$\cdots$};
		\node [style=none] (163) at (-5.25, 4.5) {$k-1$ nodes};
		\node [style={gauge_gp}] (165) at (14.75, 1.5) {};
		\node [style={gauge_gp}] (166) at (13.25, 0) {};
		\node [style={gauge_gp}] (167) at (16.25, 0) {};
		\node [style={gauge_gp}] (168) at (14.75, -1.5) {};
		\node [style={flavour_gp}] (169) at (14.75, -3) {};
		\node [style={flavour_gp}] (170) at (17.75, 0) {};
		\node [style={flavour_gp}] (171) at (11.75, 0) {};
		\node [style=none] (172) at (14.75, 4.75) {$k$ U(1) flavor nodes};
		\node [style={flavour_gp}] (174) at (16.5, 3) {};
		\node [style={flavour_gp}] (175) at (13, 3) {};
		\node [style=none] (176) at (12, 3.25) {};
		\node [style=none] (177) at (12.5, 3.75) {};
		\node [style=none] (178) at (14.5, 3.75) {};
		\node [style=none] (179) at (15, 3.75) {};
		\node [style=none] (180) at (14.75, 4.25) {};
		\node [style=none] (181) at (17.5, 3.25) {};
		\node [style=none] (182) at (17, 3.75) {};
		\node [style=none] (183) at (14.75, 3) {$\cdots$};
		\node [style=none] (184) at (10.5, 0) {$\cong$};
		\node [style=none] (185) at (-5.25, -4.25) {$[4+k,\{1,1,1,1,...\},\{1,1,1,1,...\},$};
		\node [style=none] (186) at (10.5, -4.25) {$[4,\{1,1,1,1\},\{1,1,1,1\},$};
		\node [style=none] (187) at (-5.25, -5.5) {$\{r_{1,1},r_{2,1},r_{3,1},r_{4,1}\}]$};
		\node [style=none] (188) at (10.5, -5.5) {$\{r_{1,1}^{k},r_{2,1},r_{3,1},r_{4,1}\}]$};
		\node [style=none] (189) at (-7, 3) {};
		\node [style=none] (190) at (-6.5, 3.5) {};
		\node [style=none] (191) at (-5.5, 3.5) {};
		\node [style=none] (192) at (-5, 3.5) {};
		\node [style=none] (193) at (-5.25, 4) {};
		\node [style=none] (194) at (-3.25, 3) {};
		\node [style=none] (195) at (-3.75, 3.5) {};
	\end{pgfonlayer}
	\begin{pgfonlayer}{edgelayer}
		\draw (112) to (111);
		\draw (116) to (115);
		\draw (113) to (111);
		\draw (115) to (113);
		\draw (115) to (114);
		\draw (138) to (114);
		\draw (139) to (113);
		\draw [style=double arrow] (141.center) to (140.center);
		\draw (149) to (148);
		\draw (146) to (144);
		\draw (148) to (146);
		\draw (148) to (147);
		\draw (150) to (147);
		\draw (151) to (146);
		\draw (144) to (147);
		\draw (156) to (111);
		\draw (158) to (144);
		\draw (159) to (114);
		\draw (161.center) to (159);
		\draw (160.center) to (156);
		\draw (169) to (168);
		\draw (166) to (165);
		\draw (168) to (166);
		\draw (168) to (167);
		\draw (170) to (167);
		\draw (171) to (166);
		\draw (165) to (167);
		\draw (175) to (165);
		\draw (165) to (174);
		\draw (180.center) to (178.center);
		\draw (180.center) to (179.center);
		\draw [bend right=45, looseness=0.75] (181.center) to (182.center);
		\draw (182.center) to (179.center);
		\draw (178.center) to (177.center);
		\draw [bend left=315, looseness=0.75] (177.center) to (176.center);
		\draw (193.center) to (191.center);
		\draw (193.center) to (192.center);
		\draw [bend right=45, looseness=0.75] (194.center) to (195.center);
		\draw (195.center) to (192.center);
		\draw (191.center) to (190.center);
		\draw [bend left=315, looseness=0.75] (190.center) to (189.center);
	\end{pgfonlayer}
\end{tikzpicture}
\end{equation}

This pair can be produced by gluing together the following two linear quiver 3D mirror pairs:
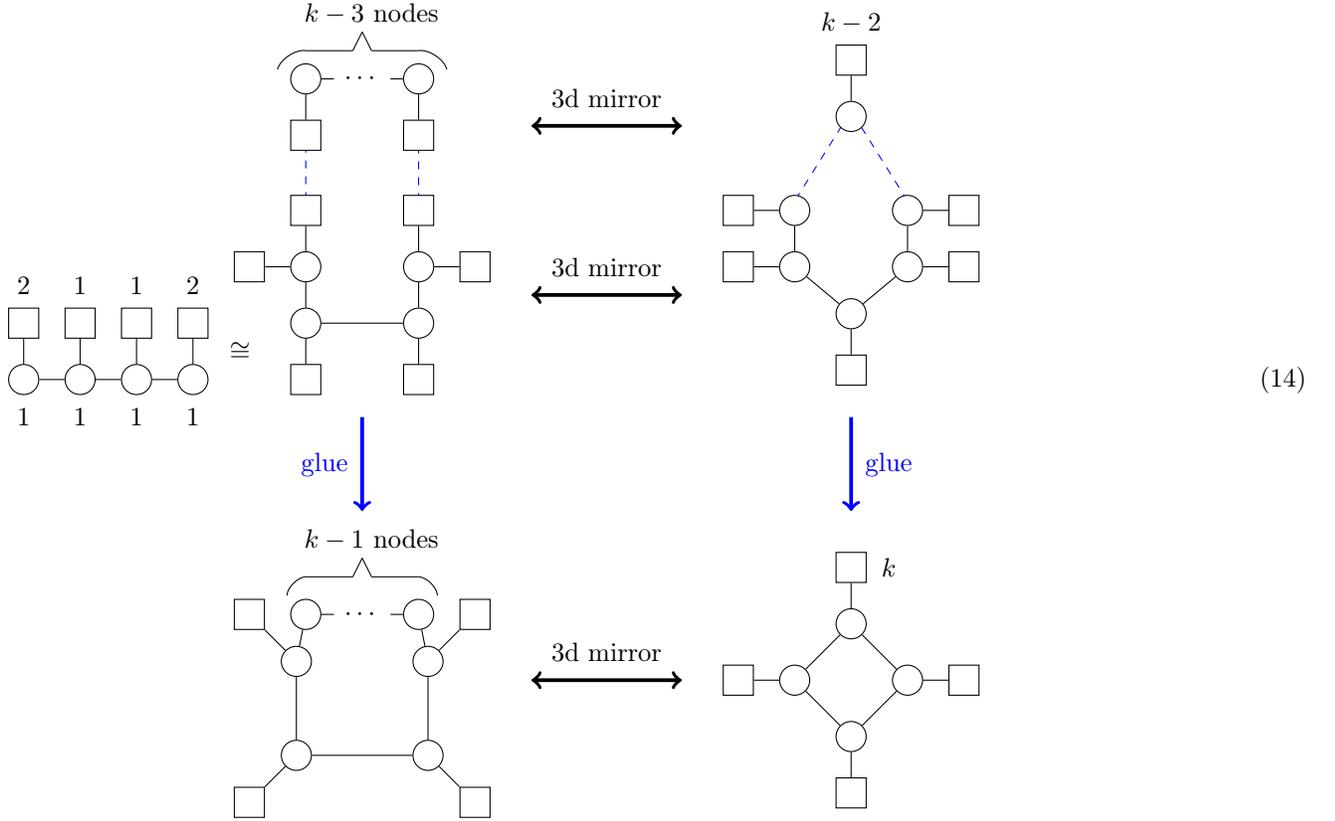
\begin{equation}
\scalebox{0.95}{
\begin{tikzpicture}[baseline=-0.25em,scale=0.5]
	\begin{pgfonlayer}{nodelayer}
		\node [style={flavour_gp}] (2) at (-4, 6.75) {};
		\node [style={flavour_gp}] (3) at (-6.5, 6.75) {};
		\node [style=none] (4) at (2, 7.25) {};
		\node [style=none] (5) at (-2, 7.25) {};
		\node [style=none] (6) at (0, 8) {3D mirror};
		\node [style=none] (11) at (2, 2.25) {};
		\node [style=none] (12) at (-2, 2.25) {};
		\node [style=none] (13) at (0, 3) {3D mirror};
		\node [style={gauge_gp}] (15) at (6, 6.75) {};
		\node [style={flavour_gp}] (17) at (6, 8.25) {};
		\node [style=none] (24) at (-5.25, -0.75) {};
		\node [style=none] (25) at (-5.25, -3.25) {};
		\node [style=none] (49) at (6, 9.25) {$k-2$};
		\node [style=none] (69) at (-6.25, -2) {glue};
		\node [style=none] (70) at (-6.5, 6.5) {};
		\node [style=none] (71) at (-6.5, 5) {};
		\node [style=none] (72) at (-4, 6.5) {};
		\node [style=none] (73) at (-4, 5) {};
		\node [style=none] (74) at (5.75, 6.5) {};
		\node [style=none] (75) at (4.5, 4.75) {};
		\node [style=none] (76) at (6.25, 6.5) {};
		\node [style=none] (77) at (7.5, 4.75) {};
		\node [style=none] (78) at (6, -1) {};
		\node [style=none] (79) at (6, -3.5) {};
		\node [style=none] (80) at (7, -2.25) {glue};
		\node [style={gauge_gp}] (82) at (-7.5, 13) {};
		\node [style={gauge_gp}] (83) at (-6, 13) {};
		\node [style={gauge_gp}] (84) at (-4.5, 13) {};
		\node [style={gauge_gp}] (85) at (-3, 13) {};
		\node [style={flavour_gp}] (86) at (-3, 14.5) {};
		\node [style={flavour_gp}] (87) at (-4.5, 14.5) {};
		\node [style={flavour_gp}] (88) at (-6, 14.5) {};
		\node [style={flavour_gp}] (89) at (-7.5, 14.5) {};
		\node [style=none] (90) at (-7.5, 15.5) {2};
		\node [style=none] (91) at (-6, 15.5) {1};
		\node [style=none] (92) at (-4.5, 15.5) {1};
		\node [style=none] (93) at (-3, 15.5) {2};
		\node [style=none] (94) at (-3, 12) {1};
		\node [style=none] (95) at (-7.5, 12) {1};
		\node [style=none] (96) at (-6, 12) {1};
		\node [style=none] (97) at (-4.5, 12) {1};
		\node [style={gauge_gp}] (98) at (-6.5, 1.75) {};
		\node [style={gauge_gp}] (99) at (-6.5, 3.25) {};
		\node [style={gauge_gp}] (100) at (-4, 3.25) {};
		\node [style={gauge_gp}] (101) at (-4, 1.75) {};
		\node [style={flavour_gp}] (102) at (-4, 0.25) {};
		\node [style={flavour_gp}] (103) at (-4, 4.75) {};
		\node [style={flavour_gp}] (104) at (-6.5, 4.75) {};
		\node [style={flavour_gp}] (105) at (-6.5, 0.25) {};
		\node [style={flavour_gp}] (106) at (-8, 3.25) {};
		\node [style={flavour_gp}] (107) at (-2.5, 3.25) {};
		\node [style=none] (125) at (-5.25, 8.25) {$\cdots$};
		\node [style={gauge_gp}] (131) at (4.5, 3) {};
		\node [style={gauge_gp}] (132) at (4.5, 4.5) {};
		\node [style={gauge_gp}] (133) at (7.5, 4.5) {};
		\node [style={gauge_gp}] (134) at (7.5, 3) {};
		\node [style={flavour_gp}] (135) at (9, 3) {};
		\node [style={flavour_gp}] (136) at (9, 4.5) {};
		\node [style={flavour_gp}] (137) at (3, 4.5) {};
		\node [style={flavour_gp}] (138) at (3, 3) {};
		\node [style={gauge_gp}] (151) at (6, 1.5) {};
		\node [style={flavour_gp}] (152) at (6, 0) {};
		\node [style={gauge_gp}] (155) at (-6.75, -7.25) {};
		\node [style={flavour_gp}] (156) at (-8, -6) {};
		\node [style={gauge_gp}] (157) at (-6.75, -9.75) {};
		\node [style={gauge_gp}] (158) at (-3.75, -7.25) {};
		\node [style={gauge_gp}] (159) at (-3.75, -9.75) {};
		\node [style={flavour_gp}] (160) at (-2.5, -11) {};
		\node [style={flavour_gp}] (161) at (-2.5, -6) {};
		\node [style={flavour_gp}] (162) at (-8, -11) {};
		\node [style=none] (163) at (2, -8) {};
		\node [style=none] (164) at (-2, -8) {};
		\node [style=none] (165) at (0, -7.25) {3D mirror};
		\node [style={gauge_gp}] (166) at (6, -6.5) {};
		\node [style={gauge_gp}] (167) at (4.5, -8) {};
		\node [style={gauge_gp}] (168) at (7.5, -8) {};
		\node [style={gauge_gp}] (169) at (6, -9.5) {};
		\node [style={flavour_gp}] (170) at (6, -11) {};
		\node [style={flavour_gp}] (171) at (9, -8) {};
		\node [style={flavour_gp}] (172) at (3, -8) {};
		\node [style=none] (173) at (7, -5) {$k$};
		\node [style={gauge_gp}] (174) at (-6.5, -6) {};
		\node [style={flavour_gp}] (175) at (6, -5) {};
		\node [style={gauge_gp}] (176) at (-4, -6) {};
		\node [style=none] (177) at (-5.75, -6) {};
		\node [style=none] (178) at (-4.75, -6) {};
		\node [style=none] (179) at (-5.25, -6) {$\cdots$};
		\node [style=none] (187) at (-5.75, 8.25) {};
		\node [style=none] (188) at (-4.75, 8.25) {};
		\node [style=none] (189) at (-7.25, 8.5) {};
		\node [style=none] (190) at (-6.5, 9) {};
		\node [style=none] (191) at (-5.5, 9) {};
		\node [style=none] (192) at (-5, 9) {};
		\node [style=none] (193) at (-5.25, 9.5) {};
		\node [style=none] (194) at (-3.25, 8.5) {};
		\node [style=none] (195) at (-4, 9) {};
		\node [style=none] (196) at (-5, 10) {$k-3$ nodes};
		\node [style={gauge_gp}] (197) at (-6.5, 8.25) {};
		\node [style={gauge_gp}] (198) at (-4, 8.25) {};
		\node [style=none] (199) at (-7, -5.5) {};
		\node [style=none] (200) at (-6.5, -5) {};
		\node [style=none] (201) at (-5.5, -5) {};
		\node [style=none] (202) at (-5, -5) {};
		\node [style=none] (203) at (-5.25, -4.5) {};
		\node [style=none] (204) at (-3.5, -5.5) {};
		\node [style=none] (205) at (-4, -5) {};
		\node [style=none] (206) at (-5, -4) {$k-1$ nodes};
		\node [style=none] (209) at (-5.25, 11) {\rotatebox{270}{$\cong$}};
	\end{pgfonlayer}
	\begin{pgfonlayer}{edgelayer}
		\draw [style=double arrow] (5.center) to (4.center);
		\draw [style=double arrow] (12.center) to (11.center);
		\draw (17) to (15);
		\draw [style=dashed] (74.center) to (75.center);
		\draw [style=dashed] (77.center) to (76.center);
		\draw [style=dashed] (73.center) to (72.center);
		\draw [style=dashed] (70.center) to (71.center);
		\draw [style=arrow] (78.center) to (79.center);
		\draw [style=arrow] (24.center) to (25.center);
		\draw (88) to (83);
		\draw (83) to (84);
		\draw (84) to (87);
		\draw (86) to (85);
		\draw (85) to (84);
		\draw (82) to (83);
		\draw (82) to (89);
		\draw (104) to (99);
		\draw (100) to (103);
		\draw (102) to (101);
		\draw (101) to (100);
		\draw (98) to (99);
		\draw (98) to (105);
		\draw (106) to (99);
		\draw (107) to (100);
		\draw (137) to (132);
		\draw (133) to (136);
		\draw (135) to (134);
		\draw (134) to (133);
		\draw (131) to (132);
		\draw (131) to (138);
		\draw (98) to (101);
		\draw (151) to (131);
		\draw (151) to (134);
		\draw (152) to (151);
		\draw (156) to (155);
		\draw (160) to (159);
		\draw (157) to (155);
		\draw (159) to (157);
		\draw (159) to (158);
		\draw (161) to (158);
		\draw (162) to (157);
		\draw [style=double arrow] (164.center) to (163.center);
		\draw (170) to (169);
		\draw (167) to (166);
		\draw (169) to (167);
		\draw (169) to (168);
		\draw (171) to (168);
		\draw (172) to (167);
		\draw (166) to (168);
		\draw (174) to (155);
		\draw (175) to (166);
		\draw (176) to (158);
		\draw (178.center) to (176);
		\draw (177.center) to (174);
		\draw (195.center) to (192.center);
		\draw (192.center) to (193.center);
		\draw (193.center) to (191.center);
		\draw (191.center) to (190.center);
		\draw [bend right=45, looseness=0.75] (190.center) to (189.center);
		\draw [bend left=45, looseness=0.75] (195.center) to (194.center);
		\draw (197) to (187.center);
		\draw (197) to (3);
		\draw (198) to (2);
		\draw (198) to (188.center);
		\draw (205.center) to (202.center);
		\draw (202.center) to (203.center);
		\draw (203.center) to (201.center);
		\draw (201.center) to (200.center);
		\draw [bend right=45, looseness=0.75] (200.center) to (199.center);
		\draw [bend left=45, looseness=0.75] (205.center) to (204.center);
	\end{pgfonlayer}
\end{tikzpicture}
}
\end{equation}

\newpage
\subsubsection{Adding Ray Nodes and Bond Multiplicity}

Another modification we can consider towards generating new 3D mirror pairs is introducing higher multiplicity ($>1$) to loop bonds and extending ray length (adding ray nodes) to its mirror. Again, starting from the basic four-loop quiver \eqref{eqn:sunshine4}, we can obtain the following pair:
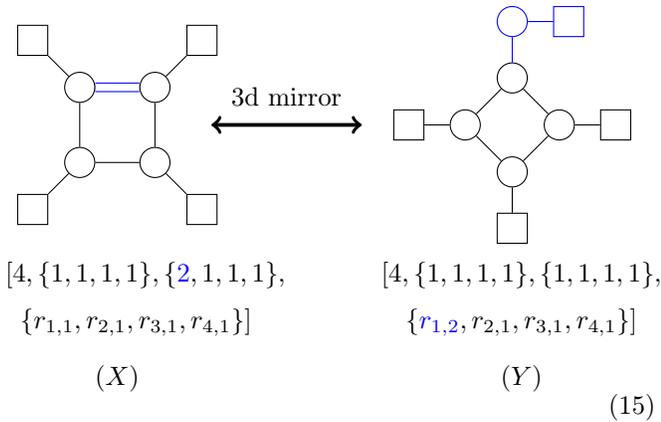
\vfill

\begin{equation}
\begin{tikzpicture}[baseline=-0.25em,scale=0.5]
	\begin{pgfonlayer}{nodelayer}
		\node [style={gauge_gp}] (111) at (-5.25, 1) {};
		\node [style={flavour_gp}] (112) at (-6.5, 2.25) {};
		\node [style={gauge_gp}] (113) at (-5.25, -1) {};
		\node [style={gauge_gp}] (114) at (-3.25, 1) {};
		\node [style={gauge_gp}] (115) at (-3.25, -1) {};
		\node [style={flavour_gp}] (116) at (-2, -2.25) {};
		\node [style={flavour_gp}] (138) at (-2, 2.25) {};
		\node [style={flavour_gp}] (139) at (-6.5, -2.25) {};
		\node [style=none] (140) at (2.25, 0) {};
		\node [style=none] (141) at (-1.75, 0) {};
		\node [style=none] (142) at (0.25, 0.75) {3D mirror};
		\node [style={gauge_gp}] (144) at (6.25, 1.25) {};
		\node [style={gauge_gp}] (146) at (5, 0) {};
		\node [style={gauge_gp}] (147) at (7.5, 0) {};
		\node [style={gauge_gp}] (148) at (6.25, -1.25) {};
		\node [style={flavour_gp}] (149) at (6.25, -2.75) {};
		\node [style={flavour_gp}] (150) at (9, 0) {};
		\node [style={flavour_gp}] (151) at (3.5, 0) {};
		\node [style=none] (154) at (-4.25, -6.75) {($X$)};
		\node [style=none] (155) at (6.5, -6.75) {($Y$)};
		\node [style={flavour_gp}] (158) at (7.75, 2.75) {};
		\node [style={gauge_gp}] (159) at (6.25, 2.75) {};
		\node [style=none] (162) at (-3.5, -4) {$[4,\{1,1,1,1\},\{2,1,1,1\},$};
		\node [style=none] (163) at (6.5, -4) {$[4,\{1,1,1,1\},\{1,1,1,1\},$};
		\node [style=none] (164) at (-3.75, -5.25) {$\{r_{1,1},r_{2,1},r_{3,1},r_{4,1}\}]$};
		\node [style=none] (165) at (6.5, -5.25) {$\{r_{1,2},r_{2,1},r_{3,1},r_{4,1}\}]$};
	\end{pgfonlayer}
	\begin{pgfonlayer}{edgelayer}
		\draw (112) to (111);
		\draw (116) to (115);
		\draw (113) to (111);
		\draw (115) to (113);
		\draw (115) to (114);
		\draw (138) to (114);
		\draw (139) to (113);
		\draw [style=double arrow] (141.center) to (140.center);
		\draw (149) to (148);
		\draw (146) to (144);
		\draw (148) to (146);
		\draw (148) to (147);
		\draw (150) to (147);
		\draw (151) to (146);
		\draw (144) to (147);
		\draw [style=doubleline] (111) to (114);
		\draw (158) to (159);
		\draw (159) to (144);
	\end{pgfonlayer}
\end{tikzpicture}
\end{equation}
\vfill

\begin{eqnarray*}
\HS_\mathcal{C}^X(t) =&& \HS_\mathcal{H}^Y(t)\\
=&& 1 + 4t^2 + 4t^3 + 20t^4 + 24t^5 + 70t^6 +
96t^7 + 221t^8 + 300t^9 + 598t^{10} + \mathcal{O}(t^{11})\\
\HS_\mathcal{H}^X(t) =&& \HS_\mathcal{C}^Y(t)\\
=&& 1 + 7t^2 + 10t^3 + 43t^4 + 80t^5 + 213t^6 +
406t^7 + 884t^8 + 1584t^9 + 3042t^{10} + \mathcal{O}(t^{11})
\end{eqnarray*}
\vfill

Ray length/bond multiplicity can be arbitrarily large $k\in\mathbb{Z}^+$: 
\vfill

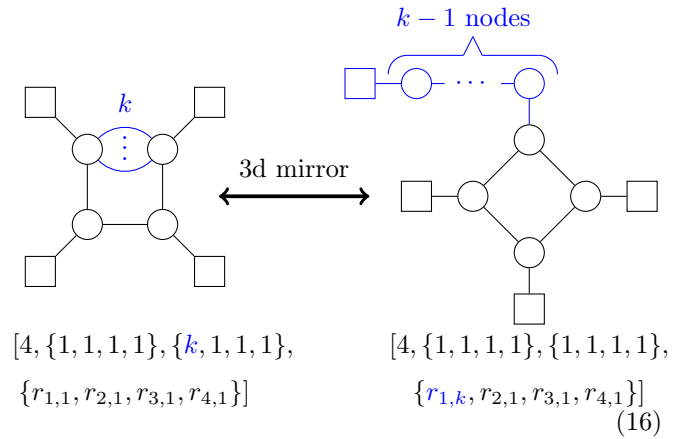
\begin{equation} \label{eqn:add_ray}
    \begin{tikzpicture}[baseline=-0.25em,scale=0.5]
	\begin{pgfonlayer}{nodelayer}
		\node [style={gauge_gp}] (111) at (-5.5, 1.25) {};
		\node [style={flavour_gp}] (112) at (-6.75, 2.5) {};
		\node [style={gauge_gp}] (113) at (-5.5, -0.75) {};
		\node [style={gauge_gp}] (114) at (-3.5, 1.25) {};
		\node [style={gauge_gp}] (115) at (-3.5, -0.75) {};
		\node [style={flavour_gp}] (116) at (-2.25, -2) {};
		\node [style={flavour_gp}] (138) at (-2.25, 2.5) {};
		\node [style={flavour_gp}] (139) at (-6.75, -2) {};
		\node [style=none] (140) at (2, 0) {};
		\node [style=none] (141) at (-2, 0) {};
		\node [style=none] (142) at (0, 0.75) {3D mirror};
		\node [style={gauge_gp}] (144) at (6.25, 1.5) {};
		\node [style={gauge_gp}] (146) at (4.75, 0) {};
		\node [style={gauge_gp}] (147) at (7.75, 0) {};
		\node [style={gauge_gp}] (148) at (6.25, -1.5) {};
		\node [style={flavour_gp}] (149) at (6.25, -3) {};
		\node [style={flavour_gp}] (150) at (9.25, 0) {};
		\node [style={flavour_gp}] (151) at (3.25, 0) {};
		\node [style={flavour_gp}] (158) at (1.75, 3) {};
		\node [style={gauge_gp}] (159) at (3.25, 3) {};
		\node [style=none] (160) at (-4.5, 1.5) {$\vdots$};
		\node [style=none] (161) at (-4.5, 2.5) {$k$};
		\node [style={gauge_gp}] (162) at (6.25, 3) {};
		\node [style=none] (163) at (4, 3) {};
		\node [style=none] (164) at (5.5, 3) {};
		\node [style=none] (165) at (4.75, 3) {$\cdots$};
		\node [style=none] (166) at (4.5, 4.75) {$k-1$ nodes};
		\node [style=none] (167) at (-3.75, -4) {$[4,\{1,1,1,1\},\{k,1,1,1\},$};
		\node [style=none] (168) at (6.25, -4) {$[4,\{1,1,1,1\},\{1,1,1,1\},$};
		\node [style=none] (169) at (-4.25, -5.25) {$\{r_{1,1},r_{2,1},r_{3,1},r_{4,1}\}]$};
		\node [style=none] (170) at (6.25, -5.25) {$\{r_{1,k},r_{2,1},r_{3,1},r_{4,1}\}]$};
		\node [style=none] (171) at (7, 3.25) {};
		\node [style=none] (172) at (6.5, 3.75) {};
		\node [style=none] (173) at (5, 3.75) {};
		\node [style=none] (174) at (4.5, 3.75) {};
		\node [style=none] (175) at (4.75, 4.25) {};
		\node [style=none] (176) at (2.5, 3.25) {};
		\node [style=none] (177) at (3, 3.75) {};
	\end{pgfonlayer}
	\begin{pgfonlayer}{edgelayer}
		\draw (112) to (111);
		\draw (116) to (115);
		\draw (113) to (111);
		\draw (115) to (113);
		\draw (115) to (114);
		\draw (138) to (114);
		\draw (139) to (113);
		\draw [style=double arrow] (141.center) to (140.center);
		\draw (149) to (148);
		\draw (146) to (144);
		\draw (148) to (146);
		\draw (148) to (147);
		\draw (150) to (147);
		\draw (151) to (146);
		\draw (144) to (147);
		\draw (158) to (159);
		\draw [bend left=45] (111) to (114);
		\draw [bend right=45] (111) to (114);
		\draw (162) to (144);
		\draw (164.center) to (162);
		\draw (159) to (163.center);
		\draw (175.center) to (173.center);
		\draw (175.center) to (174.center);
		\draw [bend left=45] (176.center) to (177.center);
		\draw (177.center) to (174.center);
		\draw (173.center) to (172.center);
		\draw [bend left=45] (172.center) to (171.center);
	\end{pgfonlayer}
\end{tikzpicture}
\end{equation}
\vfill

We can obtain this result by gluing together two smaller linear quiver 3D mirror pairs, or by gluing linear quivers to the simple four-loop sunshine quiver, as follows:

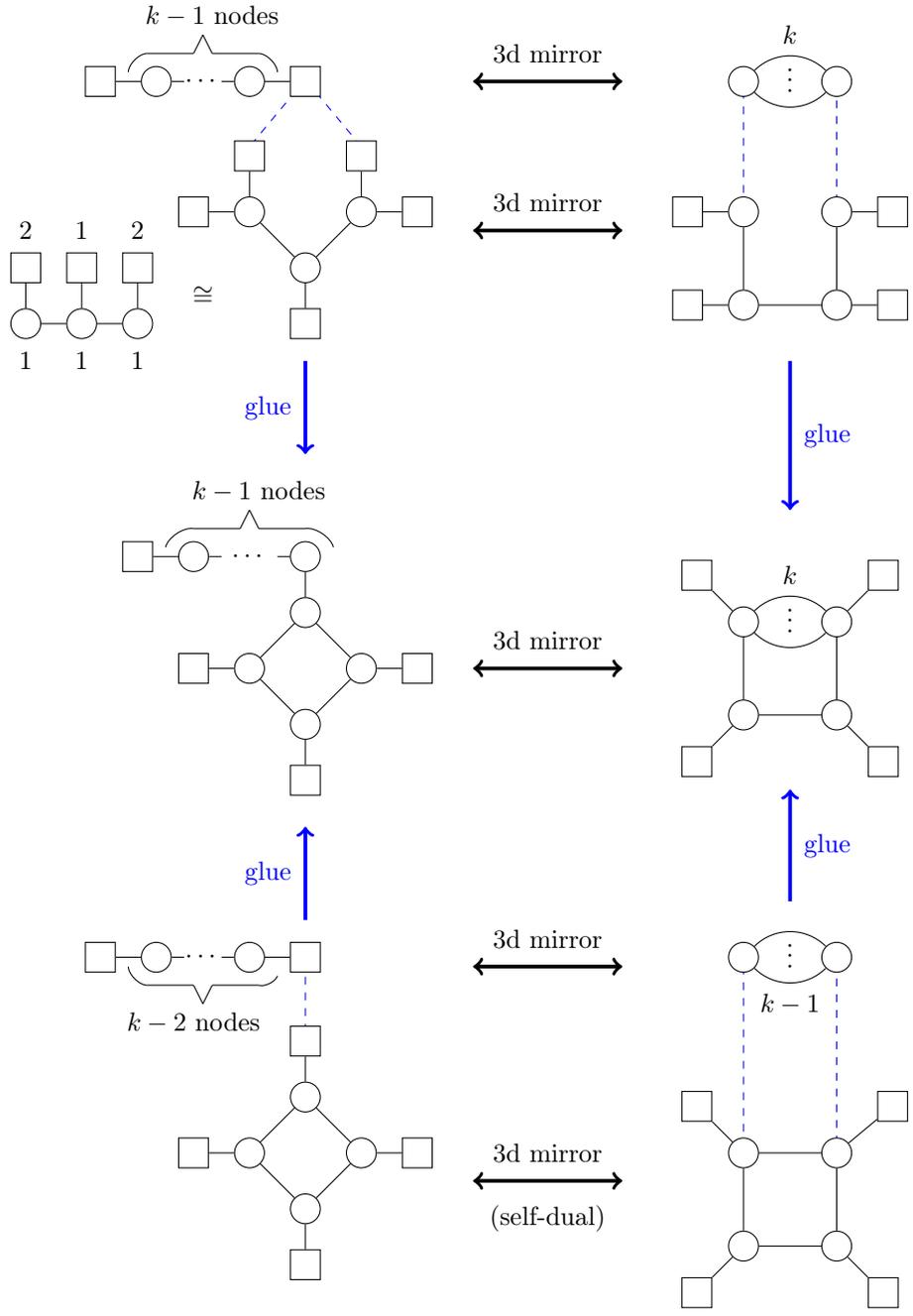
\begin{equation}
\scalebox{0.85}{
    \begin{tikzpicture}[baseline=-0.25em,scale=0.5]
    \begin{pgfonlayer}{nodelayer}
		\node [style=none] (4) at (2, 15.75) {};
		\node [style=none] (5) at (-2, 15.75) {};
		\node [style=none] (6) at (0, 16.5) {3D mirror};
		\node [style=none] (13) at (0, 12.5) {3D mirror};
		\node [style={gauge_gp}] (15) at (5.25, 15.75) {};
		\node [style=none] (24) at (-6.5, 8.25) {};
		\node [style=none] (25) at (-6.5, 5.75) {};
		\node [style=none] (49) at (6.5, 17) {$k$};
		\node [style=none] (69) at (-7.5, 7) {glue};
		\node [style=none] (70) at (-6.75, 15.5) {};
		\node [style=none] (71) at (-8, 14) {};
		\node [style=none] (72) at (-6.25, 15.5) {};
		\node [style=none] (73) at (-5, 14) {};
		\node [style=none] (74) at (5.25, 15.5) {};
		\node [style=none] (75) at (5.25, 12.5) {};
		\node [style=none] (76) at (7.75, 15.5) {};
		\node [style=none] (77) at (7.75, 12.5) {};
		\node [style=none] (78) at (6.5, 8.25) {};
		\node [style=none] (79) at (6.5, 4.25) {};
		\node [style=none] (80) at (7.5, 6.25) {glue};
		\node [style=none] (81) at (-9.25, 10) {$\cong$};
		\node [style={gauge_gp}] (82) at (-14, 9.25) {};
		\node [style={gauge_gp}] (84) at (-12.5, 9.25) {};
		\node [style={gauge_gp}] (85) at (-11, 9.25) {};
		\node [style={flavour_gp}] (86) at (-11, 10.75) {};
		\node [style={flavour_gp}] (87) at (-12.5, 10.75) {};
		\node [style={flavour_gp}] (89) at (-14, 10.75) {};
		\node [style=none] (90) at (-14, 11.75) {2};
		\node [style=none] (92) at (-12.5, 11.75) {1};
		\node [style=none] (93) at (-11, 11.75) {2};
		\node [style=none] (94) at (-11, 8.25) {1};
		\node [style=none] (95) at (-14, 8.25) {1};
		\node [style=none] (97) at (-12.5, 8.25) {1};
		\node [style={gauge_gp}] (98) at (-6.5, 10.75) {};
		\node [style={gauge_gp}] (99) at (-8, 12.25) {};
		\node [style={gauge_gp}] (100) at (-5, 12.25) {};
		\node [style={flavour_gp}] (103) at (-5, 13.75) {};
		\node [style={flavour_gp}] (104) at (-8, 13.75) {};
		\node [style={flavour_gp}] (105) at (-6.5, 9.25) {};
		\node [style={flavour_gp}] (106) at (-9.5, 12.25) {};
		\node [style={flavour_gp}] (107) at (-3.5, 12.25) {};
		\node [style={gauge_gp}] (131) at (5.25, 9.75) {};
		\node [style={gauge_gp}] (132) at (5.25, 12.25) {};
		\node [style={gauge_gp}] (133) at (7.75, 12.25) {};
		\node [style={gauge_gp}] (134) at (7.75, 9.75) {};
		\node [style={flavour_gp}] (135) at (9.25, 9.75) {};
		\node [style={flavour_gp}] (136) at (9.25, 12.25) {};
		\node [style={flavour_gp}] (137) at (3.75, 12.25) {};
		\node [style={flavour_gp}] (138) at (3.75, 9.75) {};
		\node [style={gauge_gp}] (187) at (7.75, 15.75) {};
		\node [style=none] (188) at (6.5, 16) {$\vdots$};
		\node [style=none] (190) at (14, 9.25) {};
		\node [style={gauge_gp}] (193) at (5.25, 1.25) {};
		\node [style={flavour_gp}] (194) at (4, 2.5) {};
		\node [style={gauge_gp}] (195) at (5.25, -1.25) {};
		\node [style={gauge_gp}] (196) at (7.75, 1.25) {};
		\node [style={gauge_gp}] (197) at (7.75, -1.25) {};
		\node [style={flavour_gp}] (198) at (9, -2.5) {};
		\node [style={flavour_gp}] (199) at (9, 2.5) {};
		\node [style={flavour_gp}] (200) at (4, -2.5) {};
		\node [style=none] (201) at (2, 0) {};
		\node [style=none] (202) at (-2, 0) {};
		\node [style=none] (203) at (0, 0.75) {3D mirror};
		\node [style={gauge_gp}] (204) at (-6.5, 1.5) {};
		\node [style={gauge_gp}] (205) at (-8, 0) {};
		\node [style={gauge_gp}] (206) at (-5, 0) {};
		\node [style={gauge_gp}] (207) at (-6.5, -1.5) {};
		\node [style={flavour_gp}] (208) at (-6.5, -3) {};
		\node [style={flavour_gp}] (209) at (-3.5, 0) {};
		\node [style={flavour_gp}] (210) at (-9.5, 0) {};
		\node [style=none] (213) at (6.5, 1.5) {$\vdots$};
		\node [style=none] (214) at (6.5, 2.5) {$k$};
		\node [style=none] (216) at (-8.75, 3) {};
		\node [style=none] (217) at (-7.25, 3) {};
		\node [style=none] (218) at (-8, 3) {$\cdots$};
		\node [style={flavour_gp}] (223) at (-11, 3) {};
		\node [style=none] (236) at (-10.25, 3.25) {};
		\node [style=none] (237) at (-9.5, 3.75) {};
		\node [style=none] (238) at (-8.25, 3.75) {};
		\node [style=none] (239) at (-7.75, 3.75) {};
		\node [style=none] (240) at (-8, 4.25) {};
		\node [style=none] (241) at (-5.75, 3.25) {};
		\node [style=none] (242) at (-6.5, 3.75) {};
		\node [style=none] (243) at (-7.75, 4.75) {$k-1$ nodes};
		\node [style={gauge_gp}] (244) at (-9.5, 3) {};
		\node [style={gauge_gp}] (245) at (-6.5, 3) {};
		\node [style=none] (246) at (2, 11.75) {};
		\node [style=none] (247) at (-2, 11.75) {};
		\node [style={flavour_gp}] (248) at (-6.5, -7.75) {};
		\node [style={flavour_gp}] (249) at (-12, -7.75) {};
		\node [style=none] (250) at (2, -8) {};
		\node [style=none] (251) at (-2, -8) {};
		\node [style=none] (252) at (0, -7.25) {3D mirror};
		\node [style=none] (253) at (0, -13) {3D mirror};
		\node [style={gauge_gp}] (254) at (5.25, -7.75) {};
		\node [style=none] (255) at (6.5, -9) {$k-1$};
		\node [style=none] (256) at (-6.5, -8) {};
		\node [style=none] (257) at (-6.5, -9.75) {};
		\node [style=none] (258) at (5.25, -8) {};
		\node [style=none] (259) at (5.25, -12.75) {};
		\node [style=none] (260) at (7.75, -8) {};
		\node [style=none] (261) at (7.75, -12.75) {};
		\node [style={gauge_gp}] (262) at (-6.5, -14.5) {};
		\node [style={gauge_gp}] (263) at (-8, -13) {};
		\node [style={gauge_gp}] (264) at (-5, -13) {};
		\node [style={flavour_gp}] (265) at (-6.5, -16) {};
		\node [style={flavour_gp}] (266) at (-9.5, -13) {};
		\node [style={flavour_gp}] (267) at (-3.5, -13) {};
		\node [style=none] (268) at (-9.75, -7.75) {};
		\node [style=none] (269) at (-8.75, -7.75) {};
		\node [style=none] (270) at (-9.25, -7.75) {$\cdots$};
		\node [style={gauge_gp}] (271) at (5.25, -15.5) {};
		\node [style={gauge_gp}] (272) at (5.25, -13) {};
		\node [style={gauge_gp}] (273) at (7.75, -13) {};
		\node [style={gauge_gp}] (274) at (7.75, -15.5) {};
		\node [style={flavour_gp}] (275) at (9, -16.75) {};
		\node [style={flavour_gp}] (276) at (9.25, -11.75) {};
		\node [style={flavour_gp}] (277) at (4, -11.75) {};
		\node [style={flavour_gp}] (278) at (4, -16.75) {};
		\node [style={gauge_gp}] (279) at (7.75, -7.75) {};
		\node [style=none] (280) at (6.5, -7.5) {$\vdots$};
		\node [style={gauge_gp}] (281) at (-8, -7.75) {};
		\node [style={gauge_gp}] (282) at (-10.5, -7.75) {};
		\node [style=none] (283) at (-11.25, -8) {};
		\node [style=none] (284) at (-10.75, -8.5) {};
		\node [style=none] (285) at (-9.5, -8.5) {};
		\node [style=none] (286) at (-9, -8.5) {};
		\node [style=none] (287) at (-9.25, -9) {};
		\node [style=none] (288) at (-7.25, -8.25) {};
		\node [style=none] (289) at (-7.75, -8.5) {};
		\node [style=none] (290) at (-9.5, -9.5) {$k-2$ nodes};
		\node [style=none] (291) at (2, -13.75) {};
		\node [style=none] (292) at (-2, -13.75) {};
		\node [style={gauge_gp}] (293) at (-6.5, -11.5) {};
		\node [style={flavour_gp}] (294) at (-6.5, -10) {};
		\node [style=none] (295) at (-6.5, -6.75) {};
		\node [style=none] (296) at (-6.5, -4.25) {};
		\node [style=none] (297) at (-7.5, -5.5) {glue};
		\node [style=none] (298) at (6.5, -6.25) {};
		\node [style=none] (299) at (6.5, -3.25) {};
		\node [style=none] (300) at (7.5, -4.75) {glue};
		\node [style={flavour_gp}] (301) at (-6.5, 15.75) {};
		\node [style={flavour_gp}] (302) at (-12, 15.75) {};
		\node [style=none] (303) at (-9.75, 15.75) {};
		\node [style=none] (304) at (-8.75, 15.75) {};
		\node [style=none] (305) at (-9.25, 15.75) {$\cdots$};
		\node [style={gauge_gp}] (306) at (-8, 15.75) {};
		\node [style={gauge_gp}] (307) at (-10.5, 15.75) {};
		\node [style=none] (308) at (-11.25, 16) {};
		\node [style=none] (309) at (-10.75, 16.5) {};
		\node [style=none] (310) at (-9.5, 16.5) {};
		\node [style=none] (311) at (-9, 16.5) {};
		\node [style=none] (312) at (-9.25, 17) {};
		\node [style=none] (313) at (-7.25, 16) {};
		\node [style=none] (314) at (-7.75, 16.5) {};
		\node [style=none] (315) at (-9, 17.5) {$k-1$ nodes};
		\node [style=none] (316) at (0, -14.75) {(self-dual)};
	\end{pgfonlayer}
	\begin{pgfonlayer}{edgelayer}
		\draw [style=double arrow] (5.center) to (4.center);
		\draw [style=dashed] (74.center) to (75.center);
		\draw [style=dashed] (77.center) to (76.center);
		\draw [style=dashed] (73.center) to (72.center);
		\draw [style=dashed] (70.center) to (71.center);
		\draw [style=arrow] (78.center) to (79.center);
		\draw [style=arrow] (24.center) to (25.center);
		\draw (84) to (87);
		\draw (86) to (85);
		\draw (85) to (84);
		\draw (82) to (89);
		\draw (104) to (99);
		\draw (100) to (103);
		\draw (98) to (99);
		\draw (98) to (105);
		\draw (106) to (99);
		\draw (107) to (100);
		\draw (137) to (132);
		\draw (133) to (136);
		\draw (135) to (134);
		\draw (134) to (133);
		\draw (131) to (132);
		\draw (131) to (138);
		\draw [bend left=45] (187) to (15);
		\draw [bend left=45] (15) to (187);
		\draw (100) to (98);
		\draw (82) to (84);
		\draw (131) to (134);
		\draw (194) to (193);
		\draw (198) to (197);
		\draw (195) to (193);
		\draw (197) to (195);
		\draw (197) to (196);
		\draw (199) to (196);
		\draw (200) to (195);
		\draw [style=double arrow] (202.center) to (201.center);
		\draw (208) to (207);
		\draw (205) to (204);
		\draw (207) to (205);
		\draw (207) to (206);
		\draw (209) to (206);
		\draw (210) to (205);
		\draw (204) to (206);
		\draw [bend left=45] (193) to (196);
		\draw [bend right=45] (193) to (196);
		\draw (242.center) to (239.center);
		\draw (239.center) to (240.center);
		\draw (240.center) to (238.center);
		\draw (238.center) to (237.center);
		\draw [bend right=45, looseness=0.75] (237.center) to (236.center);
		\draw [bend left=45, looseness=0.75] (242.center) to (241.center);
		\draw (245) to (217.center);
		\draw (245) to (204);
		\draw (244) to (216.center);
		\draw (244) to (223);
		\draw [style=double arrow] (247.center) to (246.center);
		\draw [style=double arrow] (251.center) to (250.center);
		\draw [style=dashed] (258.center) to (259.center);
		\draw [style=dashed] (261.center) to (260.center);
		\draw [style=dashed] (257.center) to (256.center);
		\draw (262) to (263);
		\draw (262) to (265);
		\draw (266) to (263);
		\draw (267) to (264);
		\draw (277) to (272);
		\draw (273) to (276);
		\draw (275) to (274);
		\draw (274) to (273);
		\draw (271) to (272);
		\draw (271) to (278);
		\draw [bend left=45] (279) to (254);
		\draw [bend left=45] (254) to (279);
		\draw (264) to (262);
		\draw (271) to (274);
		\draw (282) to (268.center);
		\draw (282) to (249);
		\draw (281) to (248);
		\draw (281) to (269.center);
		\draw (289.center) to (286.center);
		\draw (286.center) to (287.center);
		\draw (287.center) to (285.center);
		\draw (285.center) to (284.center);
		\draw [bend left=45, looseness=0.75] (284.center) to (283.center);
		\draw [bend right=45, looseness=0.75] (289.center) to (288.center);
		\draw [style=double arrow] (292.center) to (291.center);
		\draw (293) to (263);
		\draw (264) to (293);
		\draw (294) to (293);
		\draw (272) to (273);
		\draw [style=arrow] (298.center) to (299.center);
		\draw [style=arrow] (295.center) to (296.center);
		\draw (307) to (303.center);
		\draw (307) to (302);
		\draw (306) to (301);
		\draw (306) to (304.center);
		\draw (314.center) to (311.center);
		\draw (311.center) to (312.center);
		\draw (312.center) to (310.center);
		\draw (310.center) to (309.center);
		\draw [bend right=45, looseness=0.75] (309.center) to (308.center);
		\draw [bend left=45, looseness=0.75] (314.center) to (313.center);
	\end{pgfonlayer}
\end{tikzpicture}
}
\end{equation}

\vspace{-1em}
\subsubsection{Multiple Abelian modifications}
\vspace{-.5em}

\noindent
The two aforementioned Abelian modifications can be summarized by the following conjecture as seen in Fig.\ \ref{fig:conj}. 

Finally, note that these two modifications can be implemented sequentially. 
Introducing loop nodes with multiple higher multiplicity bonds---i.e.\ introducing overbalanced $\U(1)$ chains, composed of $\U(1)$ gauge nodes connected by higher multiplicity bonds---generalizes Eq.\ \eqref{eqn:expl_flav}, which corresponds to multiple rays (possibly of different length) attached to the same loop node. For instance:

\begin{equation}
\scalebox{0.85}{
    \begin{tikzpicture}[baseline=-0.25em,scale=0.5]
    \begin{pgfonlayer}{nodelayer}
		\node [style={gauge_gp}] (111) at (-9, 1.25) {};
		\node [style={flavour_gp}] (112) at (-10.25, 2.5) {};
		\node [style={gauge_gp}] (113) at (-9, -1.25) {};
		\node [style={gauge_gp}] (114) at (-5, 1.25) {};
		\node [style={gauge_gp}] (115) at (-5, -1.25) {};
		\node [style={flavour_gp}] (116) at (-3.75, -2.5) {};
		\node [style={flavour_gp}] (138) at (-3.75, 2.5) {};
		\node [style={flavour_gp}] (139) at (-10.25, -2.5) {};
		\node [style=none] (140) at (3, 0) {};
		\node [style=none] (141) at (-3, 0) {};
		\node [style=none] (142) at (0, 0.75) {3D mirror};
		\node [style={gauge_gp}] (144) at (7.75, 1.5) {};
		\node [style={gauge_gp}] (146) at (6.25, 0) {};
		\node [style={gauge_gp}] (147) at (9.25, 0) {};
		\node [style={gauge_gp}] (148) at (7.75, -1.5) {};
		\node [style={flavour_gp}] (149) at (7.75, -3) {};
		\node [style={flavour_gp}] (150) at (10.75, 0) {};
		\node [style={flavour_gp}] (151) at (4.75, 0) {};
		\node [style={flavour_gp}] (158) at (3, 5.25) {};
		\node [style={gauge_gp}] (159) at (4, 4.25) {};
		\node [style={gauge_gp}] (162) at (6, 2.25) {};
		\node [style=none] (163) at (4.75, 3.5) {};
		\node [style=none] (164) at (5.25, 3) {};
		\node [style=none] (165) at (5, 3.5) {$\vdots$};
		\node [style=none] (167) at (2, 2) {$\ell-1$ nodes};
		\node [style={gauge_gp}] (168) at (-8, 2.75) {};
		\node [style={gauge_gp}] (169) at (-6, 2.75) {};
		\node [style=none] (170) at (-8.5, 2) {$\cdots$};
		\node [style=none] (171) at (-5.5, 2) {$\cdots$};
		\node [style=none] (172) at (-7, 3) {$\vdots$};
		\node [style=none] (176) at (11, 3.5) {};
		\node [style=none] (177) at (10.5, 3) {};
		\node [style=none] (178) at (10.75, 3.5) {$\vdots$};
		\node [style={gauge_gp}] (180) at (9.75, 2.25) {};
		\node [style={gauge_gp}] (181) at (11.75, 4.25) {};
		\node [style={flavour_gp}] (182) at (12.75, 5.25) {};
		\node [style={gauge_gp}] (183) at (7.75, 2.75) {};
		\node [style={gauge_gp}] (184) at (7.75, 5.25) {};
		\node [style={flavour_gp}] (185) at (7.75, 6.5) {};
		\node [style=none] (186) at (7.75, 4.5) {};
		\node [style=none] (187) at (7.75, 3.5) {};
		\node [style=none] (188) at (7.75, 4.25) {$\vdots$};
		\node [style=none] (189) at (5.5, 5.5) {$m-1$};
		\node [style=none] (190) at (-9.25, 2.75) {$\ell$};
		\node [style=none] (191) at (-4.75, 2.75) {$n$};
		\node [style=none] (192) at (-7, 3.75) {$m$};
		\node [style=none] (195) at (-7, -4.25) {$[6,\{1,1,1,1,1,1\},\{1,1,1,1,\ell,m,n\},$};
		\node [style=none] (196) at (7.75, -4.25) {$[4,\{1,1,1,1\},\{1,1,1,1\},$};
		\node [style=none] (197) at (-7, -5.5) {$\{r_{1,1},r_{2,1},r_{3,1},r_{4,1}\}]$};
		\node [style=none] (198) at (7.75, -5.5) {$\{r_{1,\ell},r_{1,m},r_{1,n},r_{2,1},r_{3,1},r_{4,1}\}]$};
		\node [style=none] (205) at (3.25, 4.5) {};
		\node [style=none] (206) at (3.25, 3.75) {};
		\node [style=none] (207) at (4, 3) {};
		\node [style=none] (208) at (4.25, 2.75) {};
		\node [style=none] (209) at (3.75, 2.5) {};
		\node [style=none] (210) at (6.25, 1.5) {};
		\node [style=none] (211) at (5.5, 1.5) {};
		\node [style=none] (217) at (9.5, 1.5) {};
		\node [style=none] (218) at (10.25, 1.5) {};
		\node [style=none] (219) at (11.5, 2.75) {};
		\node [style=none] (220) at (11.75, 3) {};
		\node [style=none] (221) at (12, 2.5) {};
		\node [style=none] (222) at (12.5, 4.5) {};
		\node [style=none] (223) at (12.5, 3.75) {};
		\node [style=none] (224) at (14, 2.25) {$n-1$ nodes};
		\node [style=none] (225) at (7.5, 5.75) {};
		\node [style=none] (226) at (7, 5) {};
		\node [style=none] (227) at (7, 4.25) {};
		\node [style=none] (228) at (6.5, 4.5) {};
		\node [style=none] (229) at (7, 3.75) {};
		\node [style=none] (230) at (7, 2.75) {};
		\node [style=none] (231) at (7.5, 2.25) {};
		\node [style=none] (232) at (5.5, 4.75) {nodes};
	\end{pgfonlayer}
	\begin{pgfonlayer}{edgelayer}
		\draw (112) to (111);
		\draw (116) to (115);
		\draw (113) to (111);
		\draw (115) to (113);
		\draw (115) to (114);
		\draw (138) to (114);
		\draw (139) to (113);
		\draw [style=double arrow] (141.center) to (140.center);
		\draw (149) to (148);
		\draw (146) to (144);
		\draw (148) to (146);
		\draw (148) to (147);
		\draw (150) to (147);
		\draw (151) to (146);
		\draw (144) to (147);
		\draw (158) to (159);
		\draw (162) to (144);
		\draw (164.center) to (162);
		\draw (159) to (163.center);
		\draw [bend left=45] (111) to (168);
		\draw [bend right=45] (111) to (168);
		\draw [bend right=45] (169) to (114);
		\draw [bend right=45] (114) to (169);
		\draw [bend left=45] (168) to (169);
		\draw [bend left=45] (169) to (168);
		\draw (177.center) to (180);
		\draw (180) to (144);
		\draw (181) to (176.center);
		\draw (182) to (181);
		\draw (185) to (184);
		\draw (183) to (144);
		\draw (187.center) to (183);
		\draw (186.center) to (184);
		\draw (209.center) to (207.center);
		\draw (209.center) to (208.center);
		\draw [bend left=45] (210.center) to (211.center);
		\draw (211.center) to (208.center);
		\draw (207.center) to (206.center);
		\draw [bend left=45] (206.center) to (205.center);
		\draw [bend left=45] (218.center) to (217.center);
		\draw [bend right=45] (223.center) to (222.center);
		\draw (223.center) to (220.center);
		\draw (218.center) to (219.center);
		\draw (221.center) to (219.center);
		\draw (221.center) to (220.center);
		\draw [bend right=45] (225.center) to (226.center);
		\draw (227.center) to (226.center);
		\draw (227.center) to (228.center);
		\draw (228.center) to (229.center);
		\draw (230.center) to (229.center);
		\draw [bend right, looseness=1.25] (230.center) to (231.center);
	\end{pgfonlayer}
\end{tikzpicture}
}
\end{equation}

We can iterate these two abelian modifications arbitrarily many times to obtain infinitely many abelian sunshine quiver pairs. Self-dual quivers can be found by applying these two operations to the same quiver. For example, $[6,\{1,1,1,1,1,1\},\{1,1,1,\ell,m,n\},\{r_{1,\ell},r_{1,m},r_{1,n},r_{2,1},r_{3,1},r_{4,1}\}]$ is self-dual.

We end this section with an illustration of how one can apply modifications sequentially to obtain a complicated Abelian sunshine quiver pair. 

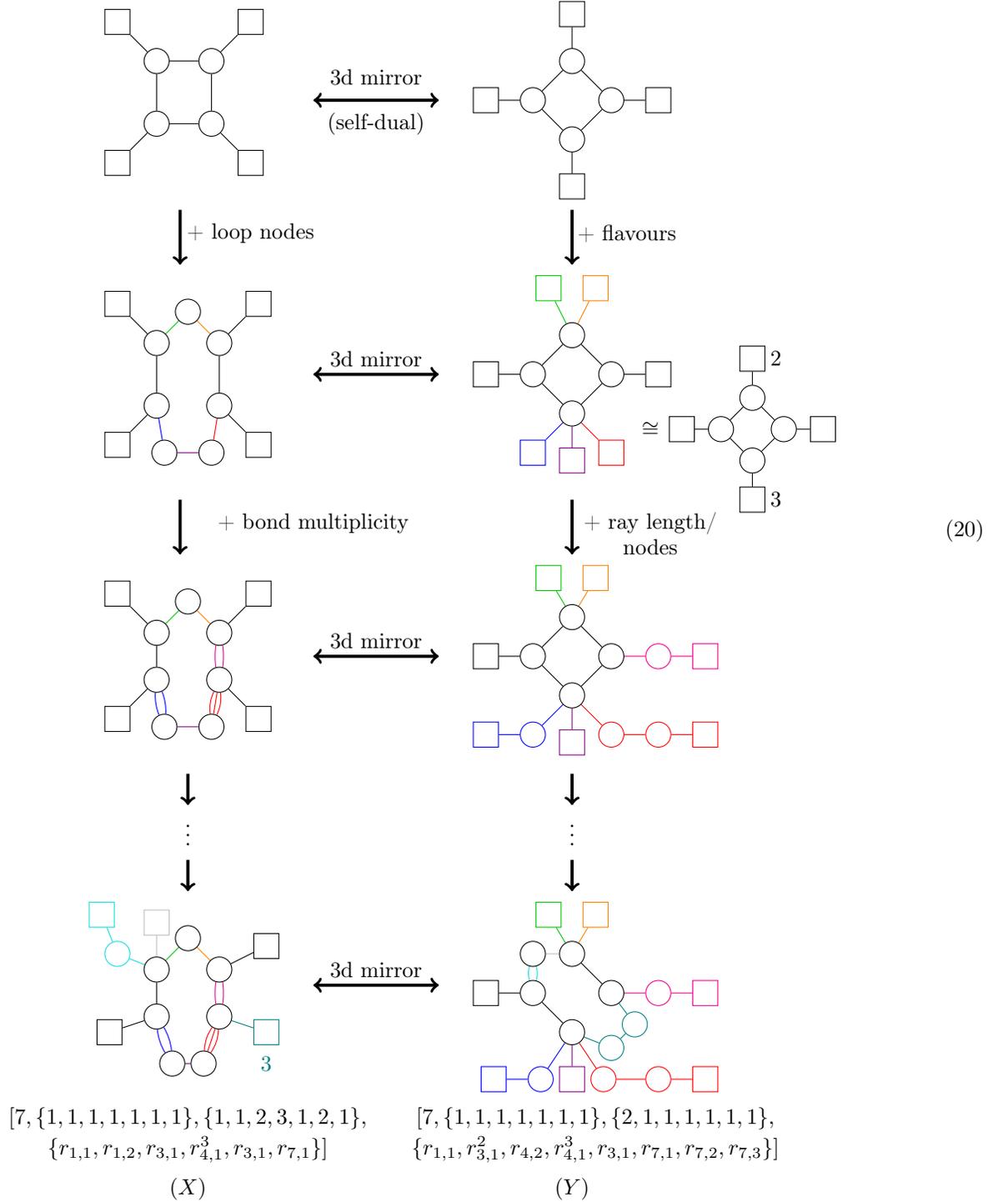
\begin{equation}
    \begin{tikzpicture}[baseline=-0.25em,scale=0.5]
	\begin{pgfonlayer}{nodelayer}
		\node [style={gauge_gp}] (111) at (-7, 15) {};
		\node [style={flavour_gp}] (112) at (-8.25, 16.25) {};
		\node [style={gauge_gp}] (113) at (-7, 13) {};
		\node [style={gauge_gp}] (114) at (-5.25, 15) {};
		\node [style={gauge_gp}] (115) at (-5.25, 13) {};
		\node [style={flavour_gp}] (116) at (-4, 11.75) {};
		\node [style={flavour_gp}] (138) at (-4, 16.25) {};
		\node [style={flavour_gp}] (139) at (-8.25, 11.75) {};
		\node [style=none] (140) at (2, 13.75) {};
		\node [style=none] (141) at (-2, 13.75) {};
		\node [style=none] (142) at (0, 14.5) {3D mirror};
		\node [style={gauge_gp}] (144) at (6.25, 15) {};
		\node [style={gauge_gp}] (146) at (5, 13.75) {};
		\node [style={gauge_gp}] (147) at (7.5, 13.75) {};
		\node [style={gauge_gp}] (148) at (6.25, 12.5) {};
		\node [style={flavour_gp}] (149) at (6.25, 11) {};
		\node [style={flavour_gp}] (150) at (9, 13.75) {};
		\node [style={flavour_gp}] (151) at (3.5, 13.75) {};
		\node [style=none] (160) at (0, 13) {(self-dual)};
		\node [style={flavour_gp}] (161) at (6.25, 16.5) {};
		\node [style={gauge_gp}] (162) at (-7, 6) {};
		\node [style={flavour_gp}] (163) at (-8.25, 7.25) {};
		\node [style={gauge_gp}] (164) at (-7, 4) {};
		\node [style={gauge_gp}] (165) at (-5, 6) {};
		\node [style={gauge_gp}] (166) at (-5, 4) {};
		\node [style={flavour_gp}] (167) at (-3.75, 2.75) {};
		\node [style={flavour_gp}] (168) at (-3.75, 7.25) {};
		\node [style={flavour_gp}] (169) at (-8.25, 2.75) {};
		\node [style=none] (170) at (-6.25, 10.25) {};
		\node [style=none] (171) at (-6.25, 8.5) {};
		\node [style={gauge_gp}] (172) at (-6.75, 2.5) {};
		\node [style={gauge_gp}] (173) at (-5.25, 2.5) {};
		\node [style={gauge_gp}] (174) at (-6, 7) {};
		\node [style={gauge_gp}] (175) at (6.25, 6.25) {};
		\node [style={gauge_gp}] (176) at (5, 5) {};
		\node [style={gauge_gp}] (177) at (7.5, 5) {};
		\node [style={gauge_gp}] (178) at (6.25, 3.75) {};
		\node [style={flavour_gp}] (180) at (9, 5) {};
		\node [style={flavour_gp}] (181) at (3.5, 5) {};
		\node [style=none] (183) at (6.25, 10.25) {};
		\node [style=none] (184) at (6.25, 8.5) {};
		\node [style=none] (185) at (-4, 9.5) {+ loop nodes};
		\node [style={gauge_gp}] (186) at (12, 4.25) {};
		\node [style={gauge_gp}] (187) at (11, 3.25) {};
		\node [style={gauge_gp}] (188) at (13, 3.25) {};
		\node [style={gauge_gp}] (189) at (12, 2.25) {};
		\node [style={flavour_gp}] (190) at (12, 1) {};
		\node [style={flavour_gp}] (191) at (14.25, 3.25) {};
		\node [style={flavour_gp}] (192) at (9.75, 3.25) {};
		\node [style={flavour_gp}] (193) at (12, 5.5) {};
		\node [style=none] (194) at (12.75, 5.5) {2};
		\node [style=none] (195) at (12.75, 1) {3};
		\node [style=none] (196) at (8.75, 3.25) {$\cong$};
		\node [style={flavour_gp}] (197) at (5.5, 7.75) {};
		\node [style={flavour_gp}] (198) at (7, 7.75) {};
		\node [style={flavour_gp}] (199) at (7.5, 2.5) {};
		\node [style={flavour_gp}] (200) at (6.25, 2.25) {};
		\node [style={flavour_gp}] (201) at (5, 2.5) {};
		\node [style=none] (202) at (8, 9.5) {+ flavours};
		\node [style={gauge_gp}] (203) at (-7, -3.25) {};
		\node [style={flavour_gp}] (204) at (-8.25, -2) {};
		\node [style={gauge_gp}] (205) at (-7, -4.75) {};
		\node [style={gauge_gp}] (206) at (-5, -3.25) {};
		\node [style={gauge_gp}] (207) at (-5, -4.75) {};
		\node [style={flavour_gp}] (208) at (-3.75, -6) {};
		\node [style={flavour_gp}] (209) at (-3.75, -2) {};
		\node [style={flavour_gp}] (210) at (-8.25, -6) {};
		\node [style=none] (211) at (-6.25, 1) {};
		\node [style=none] (212) at (-6.25, -0.75) {};
		\node [style={gauge_gp}] (213) at (-6.75, -6.25) {};
		\node [style={gauge_gp}] (214) at (-5.25, -6.25) {};
		\node [style={gauge_gp}] (215) at (-6, -2.25) {};
		\node [style={gauge_gp}] (216) at (6.25, -2.75) {};
		\node [style={gauge_gp}] (217) at (5, -4) {};
		\node [style={gauge_gp}] (218) at (7.5, -4) {};
		\node [style={gauge_gp}] (219) at (6.25, -5.25) {};
		\node [style={flavour_gp}] (221) at (3.5, -4) {};
		\node [style=none] (222) at (6.25, 1) {};
		\node [style=none] (223) at (6.25, -0.75) {};
		\node [style=none] (224) at (-2, 0.25) {+ bond multiplicity};
		\node [style={flavour_gp}] (236) at (5.5, -1.5) {};
		\node [style={flavour_gp}] (237) at (7, -1.5) {};
		\node [style={flavour_gp}] (239) at (6.25, -6.75) {};
		\node [style=none] (241) at (8.75, 0.25) {+ ray length/};
		\node [style={gauge_gp}] (244) at (5, -6.5) {};
		\node [style={flavour_gp}] (245) at (3.5, -6.5) {};
		\node [style={gauge_gp}] (246) at (7.5, -6.5) {};
		\node [style={gauge_gp}] (247) at (9, -6.5) {};
		\node [style={flavour_gp}] (248) at (10.5, -6.5) {};
		\node [style={flavour_gp}] (249) at (10.5, -4) {};
		\node [style={gauge_gp}] (250) at (9, -4) {};
		\node [style=none] (251) at (2, 5) {};
		\node [style=none] (252) at (-2, 5) {};
		\node [style=none] (253) at (0, 5.5) {3D mirror};
		\node [style=none] (254) at (2, -4) {};
		\node [style=none] (255) at (-2, -4) {};
		\node [style=none] (256) at (0, -3.5) {3D mirror};
		\node [style=none] (257) at (6.25, -7.75) {};
		\node [style=none] (258) at (6.25, -8.75) {};
		\node [style=none] (259) at (-6, -7.75) {};
		\node [style=none] (260) at (-6, -8.75) {};
		\node [style=none] (310) at (6.25, -10.5) {};
		\node [style=none] (311) at (6.25, -11.5) {};
		\node [style=none] (312) at (-6, -10.5) {};
		\node [style=none] (313) at (-6, -11.5) {};
		\node [style=none] (314) at (-6, -9.5) {$\vdots$};
		\node [style=none] (315) at (6.25, -9.5) {$\vdots$};
		\node [style={gauge_gp}] (316) at (6.25, -13.5) {};
		\node [style={gauge_gp}] (317) at (5, -14.75) {};
		\node [style={gauge_gp}] (318) at (7.5, -14.75) {};
		\node [style={gauge_gp}] (319) at (6.25, -16) {};
		\node [style={flavour_gp}] (320) at (3.5, -14.75) {};
		\node [style={flavour_gp}] (321) at (5.5, -12.25) {};
		\node [style={flavour_gp}] (322) at (7, -12.25) {};
		\node [style={flavour_gp}] (323) at (6.25, -17.5) {};
		\node [style={gauge_gp}] (324) at (5.25, -17.5) {};
		\node [style={flavour_gp}] (325) at (3.75, -17.5) {};
		\node [style={gauge_gp}] (326) at (7.25, -17.5) {};
		\node [style={gauge_gp}] (327) at (9, -17.5) {};
		\node [style={flavour_gp}] (328) at (10.5, -17.5) {};
		\node [style={flavour_gp}] (329) at (10.5, -14.75) {};
		\node [style={gauge_gp}] (330) at (9, -14.75) {};
		\node [style={gauge_gp}] (331) at (5, -13.5) {};
		\node [style={gauge_gp}] (332) at (-7, -14) {};
		\node [style={gauge_gp}] (333) at (-7, -15.5) {};
		\node [style={gauge_gp}] (334) at (-5, -14) {};
		\node [style={gauge_gp}] (335) at (-5, -15.5) {};
		\node [style={flavour_gp}] (336) at (-8.5, -16) {};
		\node [style={gauge_gp}] (337) at (-6.5, -17) {};
		\node [style={gauge_gp}] (338) at (-5.5, -17) {};
		\node [style={gauge_gp}] (339) at (-6, -13) {};
		\node [style={flavour_gp}] (340) at (-7, -12.5) {};
		\node [style={gauge_gp}] (341) at (-8.25, -13.5) {};
		\node [style={flavour_gp}] (342) at (-8.75, -12.25) {};
		\node [style=none] (343) at (2, -14.5) {};
		\node [style=none] (344) at (-2, -14.5) {};
		\node [style=none] (345) at (0, -14) {3D mirror};
		\node [style={flavour_gp}] (346) at (-3.5, -16) {};
		\node [style=none] (347) at (-3.5, -17) {3};
		\node [style={gauge_gp}] (348) at (7.5, -16.5) {};
		\node [style={gauge_gp}] (349) at (8.25, -15.75) {};
		\node [style={flavour_gp}] (350) at (-3.5, -13.25) {};
		\node [style=none] (351) at (-6, -18.75) {$[7,\{1,1,1,1,1,1,1\},\{1,1,2,3,1,2,1\},$};
		\node [style=none] (352) at (-6, -19.75) {$\{r_{1,1},r_{1,2},r_{3,1},r_{4,1}^3,r_{3,1},r_{7,1}\}]$};
		\node [style=none] (353) at (-6, -21) {($X$)};
		\node [style=none] (354) at (7, -18.75) {$[7,\{1,1,1,1,1,1,1\},\{2,1,1,1,1,1,1\},$};
		\node [style=none] (355) at (7, -19.75) {$\{r_{1,1},r_{3,1}^2,r_{4,2},r_{4,1}^3,r_{3,1},r_{7,1},r_{7,2},r_{7,3}\}]$};
		\node [style=none] (356) at (6.25, -21) {($Y$)};
		\node [style=none] (357) at (8.75, -0.5) {nodes};
		\node [style=none] (361) at (-11, 3.25) {};
	\end{pgfonlayer}
	\begin{pgfonlayer}{edgelayer}
		\draw (112) to (111);
		\draw (116) to (115);
		\draw (113) to (111);
		\draw (115) to (113);
		\draw (115) to (114);
		\draw (138) to (114);
		\draw (139) to (113);
		\draw [style=double arrow] (141.center) to (140.center);
		\draw (149) to (148);
		\draw (146) to (144);
		\draw (148) to (146);
		\draw (148) to (147);
		\draw (150) to (147);
		\draw (151) to (146);
		\draw (144) to (147);
		\draw (161) to (144);
		\draw (114) to (111);
		\draw (163) to (162);
		\draw (167) to (166);
		\draw (164) to (162);
		\draw (166) to (165);
		\draw (168) to (165);
		\draw (169) to (164);
		\draw [style=arrow] (170.center) to (171.center);
		\draw (164) to (172);
		\draw (173) to (166);
		\draw (172) to (173);
		\draw (162) to (174);
		\draw (174) to (165);
		\draw (176) to (175);
		\draw (178) to (176);
		\draw (178) to (177);
		\draw (180) to (177);
		\draw (181) to (176);
		\draw (175) to (177);
		\draw [style=arrow] (183.center) to (184.center);
		\draw (190) to (189);
		\draw (187) to (186);
		\draw (189) to (187);
		\draw (189) to (188);
		\draw (191) to (188);
		\draw (192) to (187);
		\draw (186) to (188);
		\draw (193) to (186);
		\draw (197) to (175);
		\draw (198) to (175);
		\draw (199) to (178);
		\draw (200) to (178);
		\draw (201) to (178);
		\draw (204) to (203);
		\draw (208) to (207);
		\draw (205) to (203);
		\draw (209) to (206);
		\draw (210) to (205);
		\draw [style=arrow] (211.center) to (212.center);
		\draw (214) to (207);
		\draw (213) to (214);
		\draw (203) to (215);
		\draw (215) to (206);
		\draw (217) to (216);
		\draw (219) to (217);
		\draw (219) to (218);
		\draw (221) to (217);
		\draw (216) to (218);
		\draw [style=arrow] (222.center) to (223.center);
		\draw (236) to (216);
		\draw (237) to (216);
		\draw (239) to (219);
		\draw [bend left=15] (214) to (207);
		\draw [bend left=15] (207) to (214);
		\draw (244) to (219);
		\draw (245) to (244);
		\draw (248) to (247);
		\draw (247) to (246);
		\draw (246) to (219);
		\draw [bend right=15, looseness=0.75] (205) to (213);
		\draw [bend right=15, looseness=0.75] (213) to (205);
		\draw (218) to (250);
		\draw (250) to (249);
		\draw [style=double arrow] (252.center) to (251.center);
		\draw [style=double arrow] (255.center) to (254.center);
		\draw [style=arrow] (257.center) to (258.center);
		\draw [style=arrow] (259.center) to (260.center);
		\draw [bend left=15, looseness=0.50] (206) to (207);
		\draw [bend right=15, looseness=0.50] (206) to (207);
		\draw [style=arrow] (310.center) to (311.center);
		\draw [style=arrow] (312.center) to (313.center);
		\draw (319) to (317);
		\draw (320) to (317);
		\draw (321) to (316);
		\draw (322) to (316);
		\draw (323) to (319);
		\draw (324) to (319);
		\draw (325) to (324);
		\draw (328) to (327);
		\draw (327) to (326);
		\draw (326) to (319);
		\draw (318) to (330);
		\draw (330) to (329);
		\draw (333) to (332);
		\draw (336) to (333);
		\draw (338) to (335);
		\draw (337) to (338);
		\draw (332) to (339);
		\draw (339) to (334);
		\draw [bend left=15] (338) to (335);
		\draw [bend left=15] (335) to (338);
		\draw [bend right=15, looseness=0.75] (333) to (337);
		\draw [bend right=15, looseness=0.75] (337) to (333);
		\draw [bend left=15, looseness=0.50] (334) to (335);
		\draw [bend right=15, looseness=0.50] (334) to (335);
		\draw (340) to (332);
		\draw (342) to (341);
		\draw (341) to (332);
		\draw [style=double arrow] (344.center) to (343.center);
		\draw (318) to (349);
		\draw (349) to (348);
		\draw (348) to (319);
		\draw (346) to (335);
		\draw [bend right=15] (331) to (317);
		\draw [bend right=15] (317) to (331);
		\draw (331) to (316);
		\draw (350) to (334);
		\draw (316) to (318);
	\end{pgfonlayer}
\end{tikzpicture}
\end{equation}
$\HS_\mathcal{C}^X(t) = 1 + 12 t^2 + 4 t^3 + 92 t^4 + 60 t^5 + 536 t^6 + 508 t^7 + 2598 t^8 + 3070 t^9 + 10978 t^{10} + \mathcal{O}(t^{11}) = \HS_\mathcal{H}^Y(t)$\\[0.25em]
$\HS_\mathcal{H}^X(t) = 1 + 27 t^2 + 16 t^3 + 362 t^4 + 436 t^5 + 3428 t^6 + 5844 t^7 + 26521 t^8 + 53208 t^9 + 177029  t^{10} + \mathcal{O}(t^{11}) = \HS_\mathcal{C}^Y(t)$
\\\\
See Appendix \ref{appx:check} for more examples of sunshine quivers, and a consistency check with Ref.\ \cite{Dey:2020hfe}.

\begin{figure*}
    \label{fig:conj}
    \begin{tikzpicture}[baseline=-0.25em,scale=0.5]
	\begin{pgfonlayer}{nodelayer}
		\node [style={gauge_gp}] (4) at (-10.25, -4.75) {};
		\node [style={gauge_gp}] (5) at (-10.25, 4.5) {};
		\node [style={gauge_gp}] (6) at (-5.75, 2.25) {};
		\node [style={gauge_gp}] (7) at (-5.75, -2.75) {};
		\node [style=none] (8) at (-8.25, 4.25) {};
		\node [style=none] (9) at (-7.25, 3.75) {};
		\node [style={gauge_gp}] (10) at (-9, 4.5) {};
		\node [style={gauge_gp}] (11) at (-6.5, 3.25) {};
		\node [style={gauge_gp}] (12) at (-9, -4.75) {};
		\node [style={gauge_gp}] (13) at (-6.5, -3.75) {};
		\node [style={gauge_gp}] (14) at (-5, -1.5) {};
		\node [style={gauge_gp}] (15) at (-5, 1) {};
		\node [style=none] (16) at (-5, -0.75) {};
		\node [style=none] (17) at (-5, 0.25) {};
		\node [style=none] (18) at (-8.25, -4.5) {};
		\node [style=none] (19) at (-7.25, -4) {};
		\node [style=none] (23) at (-5, 0) {\textcolor{rgb,255: red,0; green,200; blue,0}{$\vdots$}};
		\node [style=none] (24) at (-7.75, -4.25) {\textcolor{rgb,255: red,128; green,0; blue,128}{$\cdots$}};
		\node [style=none] (25) at (-7.75, 4) {\textcolor{magenta}{$\cdots$}};
		\node [style={flavour_gp}] (28) at (-10.25, 9.75) {};
		\node [style=none] (29) at (-3, -4.5) {};
		\node [style={flavour_gp}] (30) at (-0.75, -6) {};
		\node [style=none] (31) at (-3.5, -4.25) {\textcolor{red}{$\cdots$}};
		\node [style={gauge_gp}] (32) at (11, 3) {};
		\node [style={gauge_gp}] (33) at (11, -3) {};
		\node [style={gauge_gp}] (34) at (13, 0) {};
		\node [style=none] (35) at (12, -1.5) {\textcolor{red}{$\cdots$}};
		\node [style={pink_flavour}] (36) at (12, 4.25) {};
		\node [style={purple_flavour}] (37) at (12.75, -4.5) {};
		\node [style={green_flavour}] (38) at (15, 0) {};
		\node [style=none] (40) at (-4, -4) {};
		\node [style={gauge_gp}] (41) at (-2.25, -5) {};
		\node [style={gauge_gp}] (42) at (-4.75, -3.5) {};
		\node [style={gauge_gp}] (45) at (7.5, -4) {};
		\node [style={gauge_gp}] (46) at (11, 3) {};
		\node [style={gauge_gp}] (47) at (13, 0) {};
		\node [style={gauge_gp}] (48) at (11, -3) {};
		\node [style={gauge_gp}] (49) at (-10.25, -6.25) {};
		\node [style=none] (50) at (-10.25, -7.25) {};
		\node [style={flavour_gp}] (51) at (-10.25, -11) {};
		\node [style={gauge_gp}] (52) at (-10.25, -9.5) {};
		\node [style=none] (53) at (-10.25, -8.5) {};
		\node [style=none] (54) at (-10.25, -7.75) {\textcolor{cyan}{$\vdots$}};
		\node [style={gauge_gp}] (55) at (-10.25, 5.75) {};
		\node [style={gauge_gp}] (56) at (-10.25, 8.25) {};
		\node [style=none] (57) at (-10.25, 7.5) {};
		\node [style=none] (58) at (-10.25, 6.5) {};
		\node [style=none] (59) at (-10.25, 7.25) {\textcolor{rgb,255: red,168; green,168; blue,0}{$\vdots$}};
		\node [style={gauge_gp}] (112) at (7.5, 4) {};
		\node [style={grey_flavour}] (113) at (6.25, -5.5) {};
		\node [style={grey_flavour}] (114) at (6.25, 5.5) {};
		\node [style=none] (132) at (-2.5, 0) {};
		\node [style=none] (133) at (2.25, 0) {};
		\node [style=none] (134) at (0, 0.75) {3D mirror};
		\node [style=none] (136) at (9, -3.5) {\textcolor{cyan}{$\vdots$}};
		\node [style=none] (137) at (9, 3.75) {\textcolor{rgb,255: red,168; green,168; blue,0}{$\vdots$}};
		\node [style=none] (138) at (12, 1.5) {\textcolor{blue}{$\cdots$}};
		\node [style=none] (144) at (-14.5, -3) {};
		\node [style={grey_flavour}] (145) at (-17, -5.5) {};
		\node [style={grey_flavour}] (146) at (-17, 5.5) {};
		\node [style={grey_flavour}] (147) at (-18.5, 0) {};
		\node [style=none] (148) at (-14.25, 3) {};
		\node [style=none] (149) at (-15.25, 0) {};
		\node [style={grey_flavour}] (150) at (3, 2.25) {};
		\node [style={grey_flavour}] (151) at (3, -2.25) {};
		\node [style=none] (152) at (4.75, 1.75) {};
		\node [style=none] (153) at (4.5, -1.5) {};
		\node [style={gauge_gp}] (161) at (-5.75, 2.25) {};
		\node [style=none] (162) at (-3, 4) {};
		\node [style={flavour_gp}] (163) at (-0.75, 5.5) {};
		\node [style=none] (164) at (-3.5, 3.75) {\textcolor{blue}{$\cdots$}};
		\node [style=none] (165) at (-4, 3.5) {};
		\node [style={gauge_gp}] (166) at (-2.25, 4.5) {};
		\node [style={gauge_gp}] (167) at (-4.75, 3) {};
	\end{pgfonlayer}
	\begin{pgfonlayer}{edgelayer}
		\draw [style=pink] (5) to (10);
		\draw [style=pink] (10) to (8.center);
		\draw [style=pink] (9.center) to (11);
		\draw [style=pink] (11) to (6);
		\draw [style=purple] (4) to (12);
		\draw [style=purple] (12) to (18.center);
		\draw [style=purple] (19.center) to (13);
		\draw [style=purple] (13) to (7);
		\draw [style=green] (7) to (14);
		\draw [style=green] (14) to (16.center);
		\draw [style=green] (17.center) to (15);
		\draw [style=green] (15) to (6);
		\draw [style=red, bend left] (33) to (34);
		\draw [style=red, bend right] (33) to (34);
		\draw [style=blue, bend right] (32) to (34);
		\draw [style=blue, bend right] (34) to (32);
		\draw [style=red] (29.center) to (41);
		\draw [style=red] (41) to (30);
		\draw [style=red] (42) to (40.center);
		\draw [style=cyan] (50.center) to (49);
		\draw [style=cyan] (52) to (51);
		\draw [style=cyan] (53.center) to (52);
		\draw [style=yellow] (57.center) to (56);
		\draw [style=yellow] (56) to (28);
		\draw [style=yellow] (58.center) to (55);
		\draw [style=cyan, bend right] (45) to (48);
		\draw [style=yellow, bend right] (112) to (46);
		\draw [style=cyan, bend left] (45) to (48);
		\draw [style=yellow, bend right] (46) to (112);
		\draw (36) to (46);
		\draw (38) to (47);
		\draw (37) to (48);
		\draw [style=grey] (45) to (113);
		\draw [style=grey] (114) to (112);
		\draw [style=dashed, bend right=75, looseness=1.25] (112) to (45);
		\draw [style=double arrow] (133.center) to (132.center);
		\draw [style=dashed] (145) to (144.center);
		\draw [style=dashed] (148.center) to (146);
		\draw [style=dashed] (149.center) to (147);
		\draw [style=grey] (150) to (152.center);
		\draw [style=grey] (151) to (153.center);
		\draw [style=cyan] (4) to (49);
		\draw [style=red] (7) to (42);
		\draw [style=yellow] (55) to (5);
		\draw [style=dashed, bend right=90, looseness=1.75] (5) to (4);
		\draw [style=blue] (167) to (165.center);
		\draw [style=blue] (167) to (161);
		\draw [style=blue] (166) to (162.center);
		\draw [style=blue] (163) to (166);
	\end{pgfonlayer}
\end{tikzpicture}
\caption{Conjecture summarizing Abelian modifications \eqref{eqn:add_flav} and \eqref{eqn:add_ray}. Corresponding values are indicated by color---ray length and number of loop edges on the left quiver correspond to loop bond multiplicity and number of flavors on the right quiver, respectively.}
\end{figure*}

\subsubsection{Forked Rays and Higher Genus}

A natural extension to the previous 3D mirror pairs found is to ``extend'' the rays. More precisely, this could be adding more flavors to the end of a ray of length greater than 1, or more generally, introducing forked tails off of already existing rays. This extends beyond the sunshine quiver definition we have given, and so these quivers do not fit with the previous notation. Again, the gluing method will enlighten us to how this can be done while preserving mirror symmetry. Suppose we start with the mirror pair $X=[5,\{1,1,1,1,1\},\{1,2,1,1,1\},\{r_{1,1},r_{3,1},r_{4,1},r_{5,1}\}]$ and $Y=[4,\{1^4\},\{1^4\},\{r_{1,1},r_{1,2},r_{2,1},r_{3,1},r_{4,1}\}]$. If we add a flavor to quiver $Y$ on the $r_{1,2}$ ray, there is now an $\SU(2)$ Higgs symmetry and hence quiver $X$ must pick up an $\SU(2)$ Coulomb symmetry, in the form of a single balanced node. Additionally, the $\SU(2)$ Coulomb symmetry in $Y$ is broken to a $\U(1)$. The double bond in $X$ must reduce to a single bond. This gives the following 3D mirror pair:
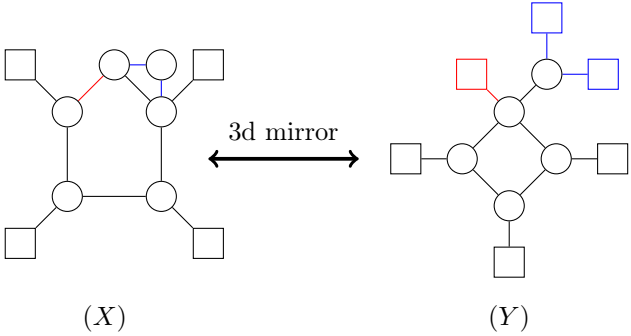
\begin{equation} \label{eqn:manyloops1}
    \begin{tikzpicture}[baseline=-0.25em,scale=0.5]
	\begin{pgfonlayer}{nodelayer}
		\node [style={gauge_gp}] (111) at (-5.75, 1.25) {};
		\node [style={flavour_gp}] (112) at (-7, 2.5) {};
		\node [style={gauge_gp}] (113) at (-5.75, -1) {};
		\node [style={gauge_gp}] (114) at (-3.25, 1.25) {};
		\node [style={gauge_gp}] (115) at (-3.25, -1) {};
		\node [style={flavour_gp}] (116) at (-2, -2.25) {};
		\node [style={flavour_gp}] (138) at (-2, 2.5) {};
		\node [style={flavour_gp}] (139) at (-7, -2.25) {};
		\node [style=none] (140) at (2, 0) {};
		\node [style=none] (141) at (-2, 0) {};
		\node [style=none] (142) at (0, 0.75) {3D mirror};
		\node [style={gauge_gp}] (144) at (6, 1.25) {};
		\node [style={gauge_gp}] (146) at (4.75, 0) {};
		\node [style={gauge_gp}] (147) at (7.25, 0) {};
		\node [style={gauge_gp}] (148) at (6, -1.25) {};
		\node [style={flavour_gp}] (149) at (6, -2.75) {};
		\node [style={flavour_gp}] (150) at (8.75, 0) {};
		\node [style={flavour_gp}] (151) at (3.25, 0) {};
		\node [style={gauge_gp}] (362) at (-4.5, 2.5) {};
		\node [style={flavour_gp}] (364) at (5, 2.25) {};
		\node [style={flavour_gp}] (366) at (8.5, 2.25) {};
		\node [style={gauge_gp}] (368) at (-3.25, 2.5) {};
		\node [style={gauge_gp}] (369) at (7, 2.25) {};
		\node [style={flavour_gp}] (377) at (7, 3.75) {};
		\node [style=none] (378) at (-4.75, -4.25) {($X$)};
		\node [style=none] (379) at (6, -4.25) {($Y$)};
		\node [style=none] (380) at (0, -5.5) {$G_\mathcal{C}^X \supset \SU(2)\times \U(1)^4 \subset G_\mathcal{H}^Y$};
		\node [style=none] (381) at (0, -6.75) {$G_\mathcal{H}^X \supset \U(1)^4 \subset G_\mathcal{C}^Y$};
	\end{pgfonlayer}
	\begin{pgfonlayer}{edgelayer}
		\draw (112) to (111);
		\draw (116) to (115);
		\draw (113) to (111);
		\draw (115) to (113);
		\draw (115) to (114);
		\draw (138) to (114);
		\draw (139) to (113);
		\draw [style=double arrow] (141.center) to (140.center);
		\draw (149) to (148);
		\draw (146) to (144);
		\draw (148) to (146);
		\draw (148) to (147);
		\draw (150) to (147);
		\draw (151) to (146);
		\draw (144) to (147);
		\draw (362) to (114);
		\draw (111) to (362);
		\draw (364) to (144);
		\draw (368) to (114);
		\draw (368) to (362);
		\draw (369) to (366);
		\draw (369) to (144);
		\draw (377) to (369);
	\end{pgfonlayer}
\end{tikzpicture}
\end{equation}

\begin{eqnarray*}
\HS_\mathcal{H}^X(t) =&& \HS_\mathcal{C}^Y(t)\\
=&& 1 + 5t^2 + 8t^3 + 21t^4 + 50t^5 + 105t^6 + 206t^7 + 414t^8 + 750t^9 + 1348t^{10} + \mathcal{O}(t^{11})
\end{eqnarray*}

Any extensions made to a ray must affect the corresponding part of the mirror, which can be seen via gluing as before. For example:

\begin{equation}
    \begin{tikzpicture}[baseline=-0.25em,scale=0.5]
	\begin{pgfonlayer}{nodelayer}
		\node [style=none] (93) at (2, 2.25) {};
		\node [style=none] (94) at (-2, 2.25) {};
		\node [style=none] (95) at (0, 3) {3D mirror};
		\node [style={gauge_gp}] (112) at (-7.25, 4) {};
		\node [style={gauge_gp}] (113) at (-7.25, 1.5) {};
		\node [style={gauge_gp}] (115) at (-5.25, 5.25) {};
		\node [style={gauge_gp}] (116) at (-3.25, 4) {};
		\node [style={gauge_gp}] (117) at (-3.25, 1.5) {};
		\node [style={flavour_gp}] (118) at (-8.5, 5.25) {};
		\node [style={flavour_gp}] (119) at (-8.5, 0.25) {};
		\node [style={flavour_gp}] (120) at (-2, 0.25) {};
		\node [style={flavour_gp}] (121) at (-1.75, 5.25) {};
		\node [style=none] (125) at (-5.25, 5.5) {};
		\node [style=none] (127) at (-3.25, 4.25) {};
		\node [style={gauge_gp}] (130) at (3.75, 4) {};
		\node [style={gauge_gp}] (133) at (3.75, 1.5) {};
		\node [style={gauge_gp}] (134) at (6.25, 4) {};
		\node [style={gauge_gp}] (137) at (6.25, 1.5) {};
		\node [style={flavour_gp}] (138) at (7.75, 4) {};
		\node [style={flavour_gp}] (139) at (7.25, 0.25) {};
		\node [style={flavour_gp}] (140) at (2.75, 0.25) {};
		\node [style={flavour_gp}] (141) at (2.25, 4) {};
		\node [style={flavour_gp}] (142) at (6.25, 5.5) {};
		\node [style=none] (147) at (6.25, 5.75) {};
		\node [style=none] (165) at (-5.25, 0.5) {};
		\node [style=none] (166) at (-5.25, -2) {};
		\node [style=none] (167) at (-4, -0.75) {glue};
		\node [style=none] (168) at (5, 0.5) {};
		\node [style=none] (169) at (5, -2) {};
		\node [style=none] (170) at (6.25, -0.75) {glue};
		\node [style={gauge_gp}] (171) at (-6.5, -4.5) {};
		\node [style={gauge_gp}] (172) at (-6.5, -7) {};
		\node [style={gauge_gp}] (173) at (-5.25, -3.25) {};
		\node [style={gauge_gp}] (174) at (-4, -4.5) {};
		\node [style={gauge_gp}] (175) at (-4, -7) {};
		\node [style={flavour_gp}] (176) at (-8, -4.5) {};
		\node [style={flavour_gp}] (177) at (-7.75, -8.25) {};
		\node [style={flavour_gp}] (178) at (-2.75, -8.25) {};
		\node [style={flavour_gp}] (179) at (-2.5, -4.5) {};
		\node [style={gauge_gp}] (180) at (-4, -3.25) {};
		\node [style=none] (181) at (2, -6) {};
		\node [style=none] (182) at (-2, -6) {};
		\node [style=none] (183) at (0, -5.25) {3D mirror};
		\node [style={gauge_gp}] (184) at (3.75, -4.5) {};
		\node [style={gauge_gp}] (185) at (3.75, -7) {};
		\node [style={gauge_gp}] (186) at (6.25, -4.5) {};
		\node [style={gauge_gp}] (187) at (6.25, -7) {};
		\node [style={flavour_gp}] (188) at (7.25, -8.25) {};
		\node [style={flavour_gp}] (189) at (2.75, -8.25) {};
		\node [style={flavour_gp}] (190) at (2.25, -4.5) {};
		\node [style={flavour_gp}] (199) at (7.75, -4.5) {};
		\node [style={gauge_gp}] (200) at (6.25, -2.75) {};
		\node [style={gauge_gp}] (201) at (4.5, -2.75) {};
		\node [style={flavour_gp}] (202) at (3, -2) {};
		\node [style={flavour_gp}] (203) at (3, -3.5) {};
		\node [style=none] (205) at (6.25, 7.5) {};
		\node [style=none] (206) at (2, 7.75) {};
		\node [style=none] (207) at (-2, 7.75) {};
		\node [style=none] (208) at (0, 8.5) {3D mirror};
		\node [style={gauge_gp}] (209) at (-3.5, 8.75) {};
		\node [style={gauge_gp}] (210) at (-5.25, 7.75) {};
		\node [style={gauge_gp}] (211) at (-3.25, 6.75) {};
		\node [style={gauge_gp}] (212) at (4.5, 7.75) {};
		\node [style={flavour_gp}] (213) at (6.25, 7.75) {};
		\node [style={flavour_gp}] (214) at (3, 6.5) {};
		\node [style={flavour_gp}] (215) at (3, 9) {};
		\node [style=none] (216) at (-5.25, 7.5) {};
		\node [style=none] (217) at (-3.25, 6.5) {};
		\node [style=none] (218) at (0, -9.75) {$G_\mathcal{C}^X \supset \U(1)^4\times\SU(2) \subset G_\mathcal{H}^Y$};
		\node [style=none] (219) at (0, -11) {$G_\mathcal{H}^X \supset \U(1)^4\times\SU(2) \subset G_\mathcal{C}^Y$};
	\end{pgfonlayer}
	\begin{pgfonlayer}{edgelayer}
		\draw [style=double arrow] (94.center) to (93.center);
		\draw (118) to (112);
		\draw (112) to (115);
		\draw (115) to (116);
		\draw (112) to (113);
		\draw (113) to (117);
		\draw (117) to (116);
		\draw (113) to (119);
		\draw (117) to (120);
		\draw (116) to (121);
		\draw (133) to (130);
		\draw (130) to (134);
		\draw (134) to (137);
		\draw (137) to (133);
		\draw (141) to (130);
		\draw (134) to (138);
		\draw (137) to (139);
		\draw (133) to (140);
		\draw (142) to (134);
		\draw [style=arrow] (165.center) to (166.center);
		\draw [style=arrow] (168.center) to (169.center);
		\draw (176) to (171);
		\draw (171) to (172);
		\draw (172) to (175);
		\draw (175) to (174);
		\draw (172) to (177);
		\draw (175) to (178);
		\draw (174) to (179);
		\draw [style=double arrow] (182.center) to (181.center);
		\draw (185) to (184);
		\draw (184) to (186);
		\draw (186) to (187);
		\draw (187) to (185);
		\draw (190) to (184);
		\draw (187) to (188);
		\draw (185) to (189);
		\draw (171) to (173);
		\draw [bend right=15] (173) to (174);
		\draw [bend left=15] (173) to (174);
		\draw (199) to (186);
		\draw (186) to (200);
		\draw (173) to (180);
		\draw (180) to (174);
		\draw (201) to (202);
		\draw (201) to (203);
		\draw (200) to (201);
		\draw [style=dashed] (205.center) to (147.center);
		\draw [style=double arrow] (207.center) to (206.center);
		\draw (211) to (209);
		\draw (209) to (210);
		\draw (212) to (213);
		\draw (215) to (212);
		\draw (212) to (214);
		\draw (210) to (211);
		\draw [style=dashed] (216.center) to (125.center);
		\draw [style=dashed] (217.center) to (127.center);
	\end{pgfonlayer}
\end{tikzpicture}
\end{equation}

\begin{eqnarray*}
\HS_\mathcal{H}^X(t) = && \HS_\mathcal{C}^Y(t)\\
=&& 1 + 8t^2 + 10t^3 + 43t^4 + 92t^5 + 234t^6 +
494t^7 + 1096t^8 + 2150t^9 + 
 4282t^{10} + \mathcal{O}(t^{11})
\end{eqnarray*}

Similar extensions are shown below:
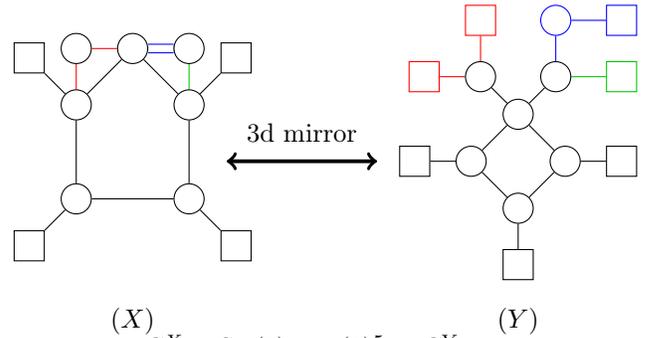
\begin{equation}
    \begin{tikzpicture}[baseline=-0.25em,scale=0.5]
	\begin{pgfonlayer}{nodelayer}
		\node [style={gauge_gp}] (111) at (-6, 1.5) {};
		\node [style={flavour_gp}] (112) at (-7.25, 2.75) {};
		\node [style={gauge_gp}] (113) at (-6, -1) {};
		\node [style={gauge_gp}] (114) at (-3, 1.5) {};
		\node [style={gauge_gp}] (115) at (-3, -1) {};
		\node [style={flavour_gp}] (116) at (-1.75, -2.25) {};
		\node [style={flavour_gp}] (138) at (-1.75, 2.75) {};
		\node [style={flavour_gp}] (139) at (-7.25, -2.25) {};
		\node [style=none] (140) at (2, 0) {};
		\node [style=none] (141) at (-2, 0) {};
		\node [style=none] (142) at (0, 0.75) {3D mirror};
		\node [style={gauge_gp}] (144) at (5.75, 1.25) {};
		\node [style={gauge_gp}] (146) at (4.5, 0) {};
		\node [style={gauge_gp}] (147) at (7, 0) {};
		\node [style={gauge_gp}] (148) at (5.75, -1.25) {};
		\node [style={flavour_gp}] (149) at (5.75, -2.75) {};
		\node [style={flavour_gp}] (150) at (8.5, 0) {};
		\node [style={flavour_gp}] (151) at (3, 0) {};
		\node [style={gauge_gp}] (362) at (-4.5, 3) {};
		\node [style={flavour_gp}] (364) at (3.25, 2.25) {};
		\node [style={flavour_gp}] (366) at (8.5, 3.75) {};
		\node [style={gauge_gp}] (369) at (-3, 3) {};
		\node [style={gauge_gp}] (370) at (-6, 3) {};
		\node [style={gauge_gp}] (371) at (4.75, 2.25) {};
		\node [style={gauge_gp}] (372) at (6.75, 2.25) {};
		\node [style={gauge_gp}] (373) at (6.75, 3.75) {};
		\node [style={flavour_gp}] (374) at (8.5, 2.25) {};
		\node [style={flavour_gp}] (379) at (4.75, 3.75) {};
		\node [style=none] (380) at (-4.5, -4.25) {($X$)};
		\node [style=none] (381) at (5.75, -4.25) {($Y$)};
		\node [style=none] (382) at (0, -5) {$G_\mathcal{C}^X \supset \SU(2)\times \U(1)^5 \subset G_\mathcal{H}^Y$};
		\node [style=none] (383) at (0, -6.25) {$G_\mathcal{H}^X \supset \SU(2)\times \U(1)^4 \subset G_\mathcal{C}^Y$};
	\end{pgfonlayer}
	\begin{pgfonlayer}{edgelayer}
		\draw (112) to (111);
		\draw (116) to (115);
		\draw (113) to (111);
		\draw (115) to (113);
		\draw (115) to (114);
		\draw (138) to (114);
		\draw (139) to (113);
		\draw [style=double arrow] (141.center) to (140.center);
		\draw (149) to (148);
		\draw (146) to (144);
		\draw (148) to (146);
		\draw (148) to (147);
		\draw (150) to (147);
		\draw (151) to (146);
		\draw (144) to (147);
		\draw (362) to (114);
		\draw (362) to (111);
		\draw (362) to (370);
		\draw (370) to (111);
		\draw [style=doubleline] (362) to (369);
		\draw (369) to (114);
		\draw (371) to (364);
		\draw (371) to (144);
		\draw (372) to (144);
		\draw (373) to (372);
		\draw (373) to (366);
		\draw (374) to (372);
		\draw (379) to (371);
	\end{pgfonlayer}
\end{tikzpicture}
\end{equation}

\begin{eqnarray*}
\HS_\mathcal{H}^X(t) =&& \HS_\mathcal{C}^Y(t)\\
=&& 1 + 9t^2 + 12t^3 + 50t^4 + 118t^5 + 304t^6 +  698t^7 + 1620t^8 + 3424t^9 + 7192t^{10} + \mathcal{O}(t^{11})
\end{eqnarray*}
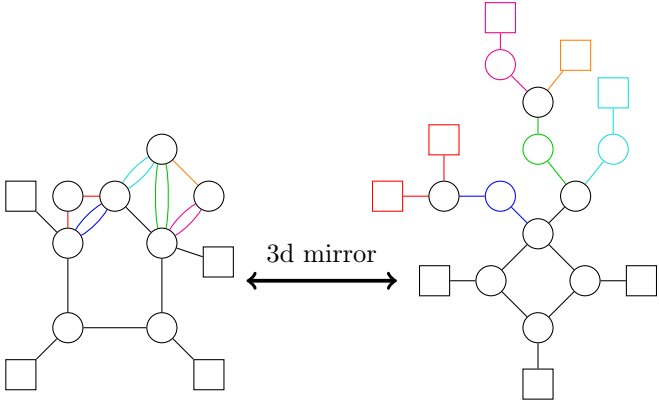
\begin{equation}
\begin{tikzpicture}[baseline=-0.25em,scale=0.5]
    \begin{pgfonlayer}{nodelayer}
		\node [style=none] (140) at (2, 0) {};
		\node [style=none] (141) at (-2, 0) {};
		\node [style=none] (142) at (0, 0.75) {3D mirror};
		\node [style={gauge_gp}] (144) at (5.75, 1.25) {};
		\node [style={gauge_gp}] (146) at (4.5, 0) {};
		\node [style={gauge_gp}] (147) at (7, 0) {};
		\node [style={gauge_gp}] (148) at (5.75, -1.25) {};
		\node [style={flavour_gp}] (149) at (5.75, -2.75) {};
		\node [style={flavour_gp}] (150) at (8.5, 0) {};
		\node [style={flavour_gp}] (151) at (3, 0) {};
		\node [style={gauge_gp}] (368) at (-6.75, 1) {};
		\node [style={flavour_gp}] (369) at (-8, 2.25) {};
		\node [style={gauge_gp}] (370) at (-6.75, -1.25) {};
		\node [style={gauge_gp}] (371) at (-4.25, 1) {};
		\node [style={gauge_gp}] (372) at (-4.25, -1.25) {};
		\node [style={flavour_gp}] (373) at (-3, -2.5) {};
		\node [style={flavour_gp}] (374) at (-2.75, 0.5) {};
		\node [style={flavour_gp}] (375) at (-8, -2.5) {};
		\node [style={gauge_gp}] (376) at (-5.5, 2.25) {};
		\node [style={gauge_gp}] (377) at (-4.25, 3.5) {};
		\node [style={gauge_gp}] (378) at (-6.75, 2.25) {};
		\node [style={gauge_gp}] (379) at (-3, 2.25) {};
		\node [style={gauge_gp}] (380) at (3.25, 2.25) {};
		\node [style={gauge_gp}] (381) at (4.75, 2.25) {};
		\node [style={flavour_gp}] (382) at (1.75, 2.25) {};
		\node [style={gauge_gp}] (384) at (6.75, 2.25) {};
		\node [style={gauge_gp}] (387) at (7.75, 3.5) {};
		\node [style={flavour_gp}] (388) at (7.75, 5) {};
		\node [style={gauge_gp}] (389) at (5.75, 3.5) {};
		\node [style={flavour_gp}] (392) at (4.75, 7) {};
		\node [style={gauge_gp}] (393) at (4.75, 5.75) {};
		\node [style={gauge_gp}] (394) at (5.75, 4.75) {};
		\node [style={flavour_gp}] (402) at (3.25, 3.75) {};
		\node [style=none] (403) at (-5.5, -4.25) {($X$)};
		\node [style=none] (404) at (5.75, -4.25) {($Y$)};
		\node [style={flavour_gp}] (407) at (6.75, 6) {};
		\node [style=none] (408) at (0, -5) {$G_\mathcal{C}^X \supset \SU(2)\times \U(1)^6 \subset G_\mathcal{H}^Y$};
		\node [style=none] (409) at (0, -6.25) {$G_\mathcal{H}^X \supset \SU(2)^4\times \U(1)^4 \subset G_\mathcal{C}^Y$};
	\end{pgfonlayer}
	\begin{pgfonlayer}{edgelayer}
		\draw [style=double arrow] (141.center) to (140.center);
		\draw (149) to (148);
		\draw (146) to (144);
		\draw (148) to (146);
		\draw (148) to (147);
		\draw (150) to (147);
		\draw (151) to (146);
		\draw (144) to (147);
		\draw (369) to (368);
		\draw (373) to (372);
		\draw (370) to (368);
		\draw (372) to (370);
		\draw (372) to (371);
		\draw (374) to (371);
		\draw (375) to (370);
		\draw (376) to (371);
	\draw [bend right=15, looseness=0.75] (376) to (368);
	\draw [bend right=15, looseness=0.75] (368) to (376);
	\draw (378) to (376);
	\draw (378) to (368);
	\draw (377) to (379);
	\draw [bend left=15, looseness=0.75] (379) to (371);
	\draw [bend right=15, looseness=0.75] (379) to (371);
	\draw [bend right=15, looseness=0.50] (377) to (371);
	\draw [bend right=15, looseness=0.50] (371) to (377);
	\draw (144) to (381);
	\draw (381) to (380);
	\draw (382) to (380);
	\draw (384) to (144);
	\draw [bend right=15, looseness=0.75] (377) to (376);
	\draw [bend right=15, looseness=0.75] (376) to (377);
	\draw (387) to (384);
	\draw (388) to (387);
	\draw (389) to (384);
	\draw (393) to (392);
	\draw (394) to (393);
	\draw (389) to (394);
	\draw (402) to (380);
	\draw (394) to (407);
	\end{pgfonlayer}
\end{tikzpicture}
\end{equation}

\begin{eqnarray*}
\HS_\mathcal{H}^X(t) =&& \HS_\mathcal{C}^Y(t)\\
=&& 1 + 19t^2 + 26t^3 + 202t^4 + 490t^5 + 
1892t^6 + 5176t^7 + 15731t^8 + 41620t^9 +
111607t^{10} + \mathcal{O}(t^{11})
\end{eqnarray*}

Another change could be adding a flavor to quiver $X$ of Eq.\ \eqref{eqn:manyloops1}, on the top right gauge node of the blue section. This breaks the $\SU(2)$ Coulomb symmetry, and so in quiver $Y$, the $\SU(2)$ Higgs symmetry is broken. In fact, this becomes a self-dual quiver\footnote{This almost looks like the quiver $[6,\{1^6\},\{1^6\},\{r_{1,1},r_{2,1},r_{3,1},r_{4,1},r_{5,1}\}]$, which has the mirror quiver $[5,\{1^5\},\{1^5\},\{r_{1,1}^2,r_{2,1},r_{3,1},r_{4,1},r_{5,1}\}]$, but now with an extra hypermultiplet/edge between nodes 2 and 6. Breaking the $\SU(2)$ symmetry in respective branches produced a self-dual quiver, which is not as geometrically clear as the previous self-dual quivers found in Section \ref{subsec:abelian_mirror_pairs}, but confirmed by computation of Hilbert series.}, as follows:
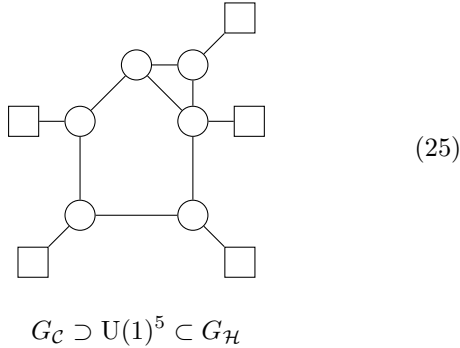
\begin{equation}
    \begin{tikzpicture}[baseline=-0.25em,scale=0.5]
	\begin{pgfonlayer}{nodelayer}
		\node [style={gauge_gp}] (0) at (-1.5, 0.75) {};
		\node [style={gauge_gp}] (1) at (-1.5, -1.75) {};
		\node [style={gauge_gp}] (2) at (0, 2.25) {};
		\node [style={gauge_gp}] (3) at (1.5, 0.75) {};
		\node [style={gauge_gp}] (4) at (1.5, -1.75) {};
		\node [style={flavour_gp}] (5) at (-3, 0.75) {};
		\node [style={flavour_gp}] (6) at (-2.75, -3) {};
		\node [style={flavour_gp}] (7) at (2.75, -3) {};
		\node [style={flavour_gp}] (8) at (3, 0.75) {};
		\node [style={gauge_gp}] (9) at (1.5, 2.25) {};
		\node [style={flavour_gp}] (10) at (2.75, 3.5) {};
	\end{pgfonlayer}
	\begin{pgfonlayer}{edgelayer}
		\draw (5) to (0);
		\draw (0) to (1);
		\draw (1) to (4);
		\draw (4) to (3);
		\draw (1) to (6);
		\draw (4) to (7);
		\draw (3) to (8);
		\draw (0) to (2);
		\draw (2) to (9);
		\draw (9) to (3);
		\draw (3) to (2);
		\draw (9) to (10);
	\end{pgfonlayer}
\end{tikzpicture}
\end{equation}

$$
G_\mathcal{C} \supset \U(1)^5 \subset G_\mathcal{H}
$$

\begin{eqnarray*}
\HS_\mathcal{C}(t) =&& \HS_\mathcal{H}(t)\\
=&& 1 + 6t^2 + 10t^3 + 33t^4 + 78t^5 + 190t^6 +  
408t^7 + 892t^8 + 1774t^9 + 3504t^{10} + \mathcal{O}(t^{11})
\end{eqnarray*}

It is plausible that there are many other self-dual quivers that can be obtained with this approach of adding hypermultiplets between nodes in the central loop, to break certain symmetry groups.

\subsection{Abelian and non-Abelian interactions}\label{subsec:ab&non-ab}
We can also use gluing to produce non-Abelian sunshine quiver mirror pairs. For example:
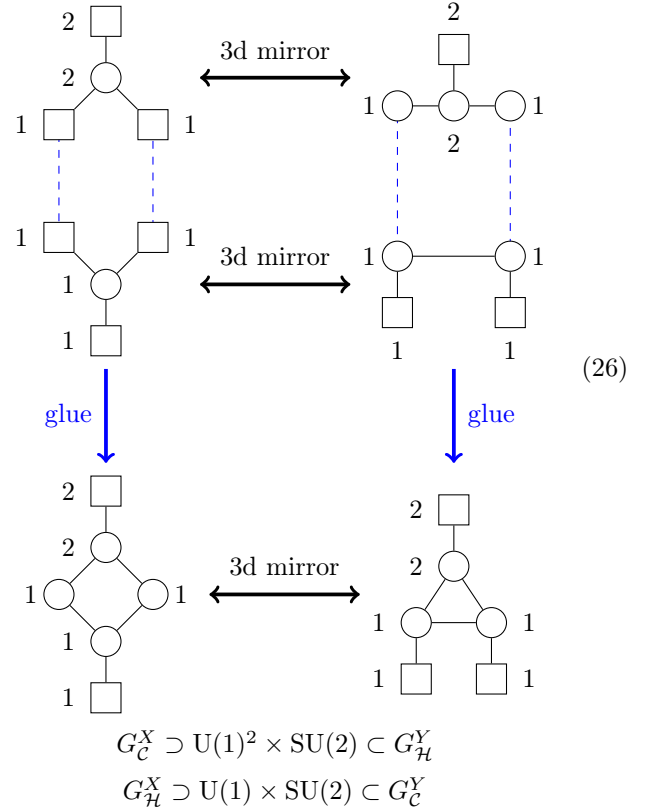
\begin{equation}
    \begin{tikzpicture}[baseline=-0.25em,scale=0.5]
	\begin{pgfonlayer}{nodelayer}
		\node [style={gauge_gp}] (0) at (-4.5, 7.75) {};
		\node [style={flavour_gp}] (1) at (-4.5, 9.25) {};
		\node [style={flavour_gp}] (2) at (-3.25, 6.5) {};
		\node [style={flavour_gp}] (3) at (-5.75, 6.5) {};
		\node [style=none] (4) at (2, 7.75) {};
		\node [style=none] (5) at (-2, 7.75) {};
		\node [style=none] (6) at (0, 8.5) {3D mirror};
		\node [style={gauge_gp}] (7) at (-4.5, 2.25) {};
		\node [style={flavour_gp}] (8) at (-4.5, 0.75) {};
		\node [style={flavour_gp}] (9) at (-3.25, 3.5) {};
		\node [style={flavour_gp}] (10) at (-5.75, 3.5) {};
		\node [style=none] (11) at (2, 2.25) {};
		\node [style=none] (12) at (-2, 2.25) {};
		\node [style=none] (13) at (0, 3) {3D mirror};
		\node [style={gauge_gp}] (14) at (3.25, 7) {};
		\node [style={gauge_gp}] (15) at (4.75, 7) {};
		\node [style={gauge_gp}] (16) at (6.25, 7) {};
		\node [style={flavour_gp}] (17) at (4.75, 8.5) {};
		\node [style={gauge_gp}] (18) at (3.25, 3) {};
		\node [style={gauge_gp}] (19) at (6.25, 3) {};
		\node [style={flavour_gp}] (20) at (3.25, 1.5) {};
		\node [style={flavour_gp}] (21) at (6.25, 1.5) {};
		\node [style=none] (24) at (-4.5, 0) {};
		\node [style=none] (25) at (-4.5, -2.5) {};
		\node [style={gauge_gp}] (26) at (-4.5, -4.75) {};
		\node [style={flavour_gp}] (27) at (-4.5, -3.25) {};
		\node [style={gauge_gp}] (28) at (-5.75, -6) {};
		\node [style={gauge_gp}] (29) at (-3.25, -6) {};
		\node [style={gauge_gp}] (30) at (-4.5, -7.25) {};
		\node [style={flavour_gp}] (31) at (-4.5, -8.75) {};
		\node [style=none] (32) at (2.25, -6) {};
		\node [style=none] (33) at (-1.75, -6) {};
		\node [style=none] (34) at (0.25, -5.25) {3D mirror};
		\node [style={gauge_gp}] (35) at (4.75, -5.25) {};
		\node [style={flavour_gp}] (36) at (4.75, -3.75) {};
		\node [style={gauge_gp}] (37) at (3.75, -6.75) {};
		\node [style={gauge_gp}] (38) at (5.75, -6.75) {};
		\node [style={flavour_gp}] (39) at (3.75, -8.25) {};
		\node [style={flavour_gp}] (40) at (5.75, -8.25) {};
		\node [style=none] (41) at (-5.5, 7.75) {2};
		\node [style=none] (42) at (-5.5, 9.25) {2};
		\node [style=none] (43) at (-6.75, 6.5) {1};
		\node [style=none] (44) at (-2.25, 6.5) {1};
		\node [style=none] (45) at (-6.75, 3.5) {1};
		\node [style=none] (46) at (-2.25, 3.5) {1};
		\node [style=none] (47) at (-5.5, 2.25) {1};
		\node [style=none] (48) at (-5.5, 0.75) {1};
		\node [style=none] (49) at (4.75, 9.5) {2};
		\node [style=none] (50) at (4.75, 6) {2};
		\node [style=none] (51) at (2.5, 7) {1};
		\node [style=none] (52) at (7, 7) {1};
		\node [style=none] (53) at (2.5, 3) {1};
		\node [style=none] (54) at (7, 3) {1};
		\node [style=none] (55) at (3.25, 0.5) {1};
		\node [style=none] (56) at (6.25, 0.5) {1};
		\node [style=none] (57) at (-5.5, -4.75) {2};
		\node [style=none] (58) at (-5.5, -3.25) {2};
		\node [style=none] (59) at (-6.5, -6) {1};
		\node [style=none] (60) at (-2.5, -6) {1};
		\node [style=none] (61) at (-5.5, -7.25) {1};
		\node [style=none] (62) at (-5.5, -8.75) {1};
		\node [style=none] (63) at (3.75, -5.25) {2};
		\node [style=none] (64) at (3.75, -3.75) {2};
		\node [style=none] (65) at (2.75, -6.75) {1};
		\node [style=none] (66) at (6.75, -6.75) {1};
		\node [style=none] (67) at (2.75, -8.25) {1};
		\node [style=none] (68) at (6.75, -8.25) {1};
		\node [style=none] (69) at (-5.5, -1.25) {glue};
		\node [style=none] (70) at (-5.75, 6.25) {};
		\node [style=none] (71) at (-5.75, 3.75) {};
		\node [style=none] (72) at (-3.25, 6.25) {};
		\node [style=none] (73) at (-3.25, 3.75) {};
		\node [style=none] (74) at (3.25, 6.75) {};
		\node [style=none] (75) at (3.25, 3.25) {};
		\node [style=none] (76) at (6.25, 6.75) {};
		\node [style=none] (77) at (6.25, 3.25) {};
		\node [style=none] (78) at (4.75, 0) {};
		\node [style=none] (79) at (4.75, -2.5) {};
		\node [style=none] (80) at (5.75, -1.25) {glue};
		\node [style=none] (81) at (0, -10) {$G_\mathcal{C}^X \supset \U(1)^2\times\SU(2) \subset G_\mathcal{H}^Y$};
		\node [style=none] (82) at (0, -11.25) {$G_\mathcal{H}^X \supset \U(1)\times\SU(2) \subset G_\mathcal{C}^Y$};
	\end{pgfonlayer}
	\begin{pgfonlayer}{edgelayer}
		\draw (1) to (0);
		\draw (3) to (0);
		\draw (0) to (2);
		\draw [style=double arrow] (5.center) to (4.center);
		\draw (8) to (7);
		\draw (10) to (7);
		\draw (7) to (9);
		\draw [style=double arrow] (12.center) to (11.center);
		\draw (15) to (14);
		\draw (15) to (16);
		\draw (17) to (15);
		\draw (18) to (20);
		\draw (18) to (19);
		\draw (19) to (21);
		\draw (27) to (26);
		\draw (31) to (30);
		\draw (28) to (26);
		\draw (26) to (29);
		\draw (30) to (28);
		\draw (30) to (29);
		\draw [style=double arrow] (33.center) to (32.center);
		\draw (36) to (35);
		\draw (37) to (39);
		\draw (37) to (38);
		\draw (38) to (40);
		\draw (37) to (35);
		\draw (35) to (38);
		\draw [style=dashed] (74.center) to (75.center);
		\draw [style=dashed] (77.center) to (76.center);
		\draw [style=dashed] (73.center) to (72.center);
		\draw [style=dashed] (70.center) to (71.center);
		\draw [style=arrow] (78.center) to (79.center);
		\draw [style=arrow] (24.center) to (25.center);
	\end{pgfonlayer}
\end{tikzpicture}
\end{equation}

\begin{eqnarray*}
\HS_\mathcal{C}^X(t) =&& \HS_\mathcal{H}^Y(t)\\
=&& 1 + 6t^2 + 12t^3 + 34t^4 + 78t^5 + 181t^6 + 364t^7 + 740t^8 + 1382t^9 + 2524t^{10} + \mathcal{O}(t^{11})
\end{eqnarray*}


Abelian and non-Abelian extensions of sunshine quivers can interact and be combined. This demonstrates the power of our systematic approach, allowing us to generate highly complex sunshine quivers.
Consider the sunshine quiver $[3,\{2,1,1\},\{1,1,1\},\{r_{1,1}^3,r_{2,4},r_{3,1}\}]$:
\begin{equation}
\begin{tikzpicture}[baseline=-0.25em,scale=0.5]
\begin{pgfonlayer}{nodelayer}
		\node [style={gauge_gp}] (0) at (0, 1) {};
		\node [style={gauge_gp}] (1) at (-1.25, -1) {};
		\node [style={gauge_gp}] (2) at (1.25, -1) {};
		\node [style={flavour_gp}] (3) at (0, 2.75) {};
		\node [style={flavour_gp}] (4) at (-3, -1.75) {};
		\node [style={gauge_gp}] (5) at (3, -1.75) {};
		\node [style={gauge_gp}] (6) at (4.75, -2.5) {};
		\node [style={flavour_gp}] (7) at (8.25, -4) {};
		\node [style=none] (8) at (1, 1) {2};
		\node [style=none] (9) at (1, 2.75) {3};
		\node [style=none] (10) at (-3, -2.75) {1};
		\node [style=none] (11) at (-1.25, -2) {1};
		\node [style=none] (12) at (1.25, -2) {1};
		\node [style=none] (13) at (3, -2.75) {1};
		\node [style=none] (14) at (4.75, -3.5) {1};
		\node [style=none] (15) at (8.25, -5) {1};
		\node [style={gauge_gp}] (16) at (6.5, -3.25) {};
		\node [style=none] (17) at (6.5, -4.25) {1};
	\end{pgfonlayer}
	\begin{pgfonlayer}{edgelayer}
		\draw (2) to (1);
		\draw (1) to (0);
		\draw (2) to (0);
		\draw (3) to (0);
		\draw (4) to (1);
		\draw (6) to (5);
		\draw (5) to (2);
		\draw (6) to (16);
		\draw (16) to (7);
	\end{pgfonlayer}
\end{tikzpicture}
\end{equation}

This quiver can be systematically constructed beginning with the simplest three-loop sunshine quiver, the self-dual loop $[3,\{1,1,1\},\{1,1,1\},\{r_{1,1},r_{2,1},r_{3,1}\}]$. We then introduce a modification by adding ray nodes, as in Eq.\ \eqref{eqn:add_ray}. This process corresponds to introducing a higher-multiplicity edge in the mirror quiver. Following this, we can apply the gluing method to introduce a non-Abelian gauge node, such as $\U(2)$, into the quiver's loop, as follows: 

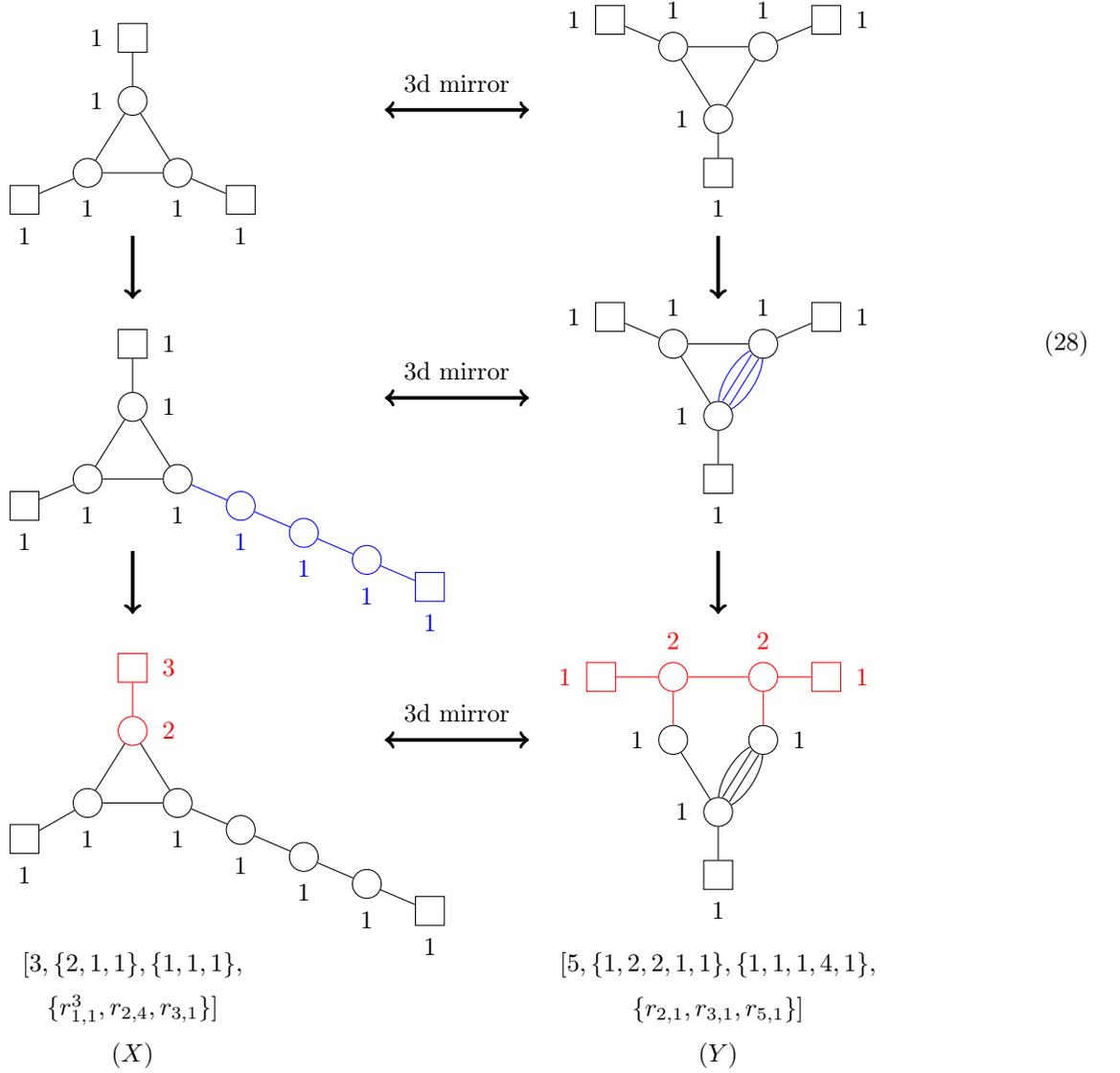
\begin{equation} \label{eqn:nonabelianexample}
\begin{tikzpicture}[baseline=-0.25em,scale=0.5]
	\begin{pgfonlayer}{nodelayer}
		\node [style={gauge_gp}] (1) at (-3, -12.25) {};
		\node [style={gauge_gp}] (2) at (-5.5, -12.25) {};
		\node [style={flavour_gp}] (4) at (-1.25, -13.25) {};
		\node [style={gauge_gp}] (5) at (-6.5, -13.5) {};
		\node [style={gauge_gp}] (6) at (-7.5, -14.75) {};
		\node [style={flavour_gp}] (7) at (-9.5, -17.25) {};
		\node [style=none] (8) at (-5.25, -10.25) {2};
		\node [style=none] (9) at (-5.25, -8.5) {3};
		\node [style=none] (10) at (-1.25, -14.25) {1};
		\node [style=none] (11) at (-3, -13.25) {1};
		\node [style=none] (12) at (-6.25, -12.25) {1};
		\node [style=none] (13) at (-8.25, -14.75) {1};
		\node [style=none] (14) at (-7.25, -13.5) {1};
		\node [style=none] (15) at (-10.25, -17.25) {1};
		\node [style={gauge_gp}] (16) at (-4.25, -1.25) {};
		\node [style={gauge_gp}] (17) at (-3, -3.25) {};
		\node [style={gauge_gp}] (18) at (-5.5, -3.25) {};
		\node [style={flavour_gp}] (19) at (-4.25, 0.5) {};
		\node [style={flavour_gp}] (20) at (-1.25, -4) {};
		\node [style=none] (24) at (-5.25, -1.25) {1};
		\node [style=none] (25) at (-5.25, 0.5) {1};
		\node [style=none] (26) at (-1.25, -5) {1};
		\node [style=none] (27) at (-3, -4.25) {1};
		\node [style=none] (28) at (-5.5, -4.25) {1};
		\node [style=none] (29) at (-7.25, -4.5) {1};
		\node [style=none] (30) at (-8.25, -5.75) {1};
		\node [style=none] (31) at (-10.25, -8.25) {1};
		\node [style={gauge_gp}] (32) at (-4.25, 7.25) {};
		\node [style={gauge_gp}] (33) at (-5.5, 5.25) {};
		\node [style={gauge_gp}] (34) at (-3, 5.25) {};
		\node [style={flavour_gp}] (35) at (-4.25, 9) {};
		\node [style={flavour_gp}] (36) at (-7.25, 4.5) {};
		\node [style={flavour_gp}] (39) at (-1.25, 4.5) {};
		\node [style=none] (40) at (-5.25, 7.25) {1};
		\node [style=none] (41) at (-5.25, 9) {1};
		\node [style=none] (42) at (-7.25, 3.5) {1};
		\node [style=none] (43) at (-5.5, 4.25) {1};
		\node [style=none] (44) at (-3, 4.25) {1};
		\node [style=none] (47) at (-1.25, 3.5) {1};
		\node [style={flavour_gp}] (48) at (-9.5, -8.25) {};
		\node [style={gauge_gp}] (49) at (-7.5, -5.75) {};
		\node [style={gauge_gp}] (50) at (-6.5, -4.5) {};
		\node [style={gauge_gp}] (51) at (5.5, 6.25) {};
		\node [style={gauge_gp}] (52) at (4.25, 8.25) {};
		\node [style={gauge_gp}] (53) at (6.75, 8.25) {};
		\node [style={flavour_gp}] (54) at (5.5, 4.75) {};
		\node [style={flavour_gp}] (55) at (2.5, 9) {};
		\node [style={flavour_gp}] (56) at (8.5, 9) {};
		\node [style=none] (57) at (4.5, 6.25) {1};
		\node [style=none] (58) at (5.5, 3.75) {1};
		\node [style=none] (59) at (1.5, 9) {1};
		\node [style=none] (60) at (4.25, 9.25) {1};
		\node [style=none] (61) at (6.75, 9.25) {1};
		\node [style=none] (62) at (9.5, 9) {1};
		\node [style={gauge_gp}] (63) at (5.25, -2) {};
		\node [style={gauge_gp}] (64) at (6.5, 0) {};
		\node [style={gauge_gp}] (65) at (4, 0) {};
		\node [style={flavour_gp}] (66) at (5.25, -3.75) {};
		\node [style={flavour_gp}] (67) at (8.25, 0.75) {};
		\node [style={flavour_gp}] (68) at (2.25, 0.75) {};
		\node [style=none] (69) at (6.25, -2) {1};
		\node [style=none] (70) at (5.25, -4.75) {1};
		\node [style=none] (71) at (9.25, 0.75) {1};
		\node [style=none] (72) at (6.5, 1) {1};
		\node [style=none] (73) at (4, 1) {1};
		\node [style=none] (74) at (1.25, 0.75) {1};
		\node [style={gauge_gp}] (75) at (5.25, -13) {};
		\node [style={gauge_gp}] (76) at (6.5, -11) {};
		\node [style={gauge_gp}] (77) at (4, -11) {};
		\node [style={flavour_gp}] (78) at (5.25, -14.75) {};
		\node [style=none] (81) at (6.25, -13) {1};
		\node [style=none] (82) at (5.25, -15.75) {1};
		\node [style=none] (84) at (7.5, -11) {1};
		\node [style=none] (85) at (3, -11) {1};
		\node [style={gauge_gp}] (87) at (6.5, -9.25) {};
		\node [style={gauge_gp}] (88) at (4, -9.25) {};
		\node [style={flavour_gp}] (89) at (8.25, -9.25) {};
		\node [style={flavour_gp}] (90) at (2.25, -9.25) {};
		\node [style=none] (91) at (9.25, -9.25) {1};
		\node [style=none] (92) at (1.25, -9.25) {1};
		\node [style=none] (93) at (4, -8.25) {2};
		\node [style=none] (94) at (6.5, -8.25) {2};
		\node [style={flavour_gp}] (95) at (-4.25, -8.5) {};
		\node [style={gauge_gp}] (96) at (-4.25, -10.25) {};
		\node [style=none] (97) at (-4.25, 3.5) {};
		\node [style=none] (98) at (-4.25, 1.75) {};
		\node [style=none] (99) at (5.5, 3) {};
		\node [style=none] (100) at (5.5, 1.25) {};
		\node [style=none] (101) at (-4.25, -5.25) {};
		\node [style=none] (102) at (-4.25, -7) {};
		\node [style=none] (103) at (5.25, -5.75) {};
		\node [style=none] (104) at (5.25, -7.5) {};
		\node [style=none] (105) at (2, 6.5) {};
		\node [style=none] (106) at (-2, 6.5) {};
		\node [style=none] (107) at (0, 7.25) {3D mirror};
		\node [style=none] (108) at (2, -1.5) {};
		\node [style=none] (109) at (-2, -1.5) {};
		\node [style=none] (110) at (0, -0.75) {3D mirror};
		\node [style=none] (111) at (2, -11) {};
		\node [style=none] (112) at (-2, -11) {};
		\node [style=none] (113) at (0, -10.25) {3D mirror};
		\node [style=none] (114) at (5.25, -19.25) {($Y$)};
		\node [style=none] (115) at (-4.25, -19.25) {($X$)};
		\node [style=none] (116) at (-4.25, -16.75) {$[3,\{2,1,1\},\{1,1,1\},$};
		\node [style=none] (117) at (5.25, -16.75) {$[5,\{1,2,2,1,1\},\{1,1,1,4,1\},$};
		\node [style=none] (118) at (-4.25, -18) {$\{r_{1,1}^3,r_{2,4},r_{3,1}\}]$};
		\node [style=none] (119) at (5.25, -18) {$\{r_{2,1},r_{3,1},r_{5,1}\}]$};
		\node [style=none] (120) at (-9.25, -7) {1};
		\node [style={gauge_gp}] (121) at (-8.5, -7) {};
		\node [style={gauge_gp}] (122) at (-8.5, -16) {};
		\node [style=none] (123) at (-9.25, -16) {1};
		\node [style=none] (124) at (0, -21) {$\HS_\mathcal{C}^X(t) = 1 + 18t^2 + 2t^3 + 146t^4 + 56t^5 + 813t^6 + 564t^7 + 3621t^8 + 3516t^9 + 13952t^{10} + \mathcal{O}(t^{11}) = \HS_\mathcal{H}^Y(t)$};
		\node [style=none] (125) at (0, -22.25) {$\HS_\mathcal{H}^X(t) = 1 + 11t^2 + 8t^3 + 80t^4 + 104t^5 + 458t^6 + 742t^7 + 2232t^8 + 3872t^9 + 9430t^{10} + \mathcal{O}(t^{11}) = \HS_\mathcal{C}^Y(t)$};
	\end{pgfonlayer}
	\begin{pgfonlayer}{edgelayer}
		\draw (2) to (1);
		\draw (4) to (1);
		\draw (6) to (5);
		\draw (5) to (2);
		\draw (18) to (17);
		\draw (17) to (16);
		\draw (18) to (16);
		\draw (19) to (16);
		\draw (20) to (17);
		\draw (34) to (33);
		\draw (33) to (32);
		\draw (34) to (32);
		\draw (35) to (32);
		\draw (36) to (33);
		\draw (39) to (34);
		\draw (53) to (52);
		\draw (52) to (51);
		\draw (53) to (51);
		\draw (54) to (51);
		\draw (55) to (52);
		\draw (56) to (53);
		\draw (49) to (50);
		\draw (18) to (50);
		\draw (65) to (64);
		\draw (64) to (63);
		\draw (66) to (63);
		\draw (67) to (64);
		\draw (68) to (65);
		\draw [style=doubleline] (63) to (65);
		\draw [bend left, looseness=0.75] (63) to (65);
		\draw [bend left, looseness=0.75] (65) to (63);
		\draw (76) to (75);
		\draw (78) to (75);
		\draw (87) to (76);
		\draw (88) to (77);
		\draw (88) to (87);
		\draw (89) to (87);
		\draw (90) to (88);
		\draw (95) to (96);
		\draw (96) to (2);
		\draw (96) to (1);
		\draw [style=arrow] (97.center) to (98.center);
		\draw [style=arrow] (99.center) to (100.center);
		\draw [style=arrow] (101.center) to (102.center);
		\draw [style=arrow] (103.center) to (104.center);
		\draw [style=double arrow] (106.center) to (105.center);
		\draw [style=double arrow] (109.center) to (108.center);
		\draw [style=double arrow] (112.center) to (111.center);
		\draw [bend left, looseness=0.75] (75) to (77);
		\draw [bend left, looseness=0.75] (77) to (75);
		\draw [style=doubleline] (77) to (75);
		\draw (49) to (121);
		\draw (121) to (48);
		\draw (122) to (6);
		\draw (122) to (7);
	\end{pgfonlayer}
\end{tikzpicture}
\end{equation}       

\end{widetext}

This systematic construction process, which builds a complex theory from simpler, well-understood components, allows us to propose and verify new 3D mirror pairs that would otherwise be difficult to discover or analyze.

\subsection{Beyond unitary quivers and the breakdown of the gluing method}
Our findings on gluing together linear unitary quivers to create nonlinear sunshine quivers beg the question: is it possible to start with a nonlinear unitary quiver with a nonunitary (orthosymplectic or special unitary) 3D mirror, and use the gluing method to build a larger quiver? It turns out that this is not always true.

A counterexample, involving a special unitary gauge node, is as follows. Consider $\U(2)$ and $\SU(2)$ theories, each with five flavors. Their mirror theories are

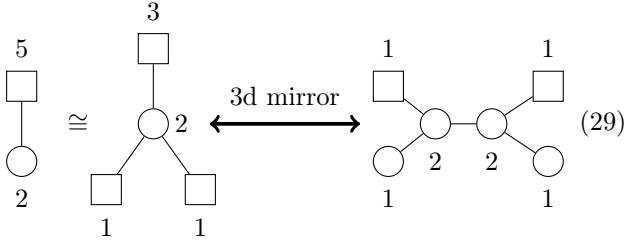
\begin{equation}
    \begin{tikzpicture}[baseline=-0.25em,scale=0.5]
	\begin{pgfonlayer}{nodelayer}
		\node [style={gauge_gp}] (0) at (-3.5, 0) {};
		\node [style={flavour_gp}] (1) at (-3.5, 2) {};
		\node [style={flavour_gp}] (2) at (-2.25, -1.75) {};
		\node [style={flavour_gp}] (3) at (-4.75, -1.75) {};
		\node [style=none] (4) at (2, 0) {};
		\node [style=none] (5) at (-2, 0) {};
		\node [style=none] (6) at (0, 0.75) {3D mirror};
		\node [style={gauge_gp}] (14) at (2.75, -1) {};
		\node [style={gauge_gp}] (15) at (4, 0) {};
		\node [style={gauge_gp}] (16) at (5.5, 0) {};
		\node [style={flavour_gp}] (17) at (2.75, 1) {};
		\node [style=none] (41) at (-2.75, 0) {2};
		\node [style=none] (42) at (-3.5, 3) {3};
		\node [style=none] (43) at (-4.75, -2.75) {1};
		\node [style=none] (44) at (-2.25, -2.75) {1};
		\node [style=none] (49) at (2.75, 2) {1};
		\node [style=none] (50) at (4, -1) {2};
		\node [style=none] (51) at (2.75, -2) {1};
		\node [style=none] (52) at (7, -2) {1};
		\node [style={gauge_gp}] (53) at (7, -1) {};
		\node [style=none] (54) at (5.5, -1) {2};
		\node [style={flavour_gp}] (55) at (7, 1) {};
		\node [style=none] (56) at (7, 2) {1};
		\node [style=none] (81) at (-5.5, 0) {$\cong$};
		\node [style={flavour_gp}] (82) at (-7, 1) {};
		\node [style={gauge_gp}] (83) at (-7, -1) {};
		\node [style=none] (84) at (-7, 2) {5};
		\node [style=none] (85) at (-7, -2) {2};
		\node [style=none] (86) at (-7.75, 0) {};
	\end{pgfonlayer}
	\begin{pgfonlayer}{edgelayer}
		\draw (1) to (0);
		\draw (3) to (0);
		\draw (0) to (2);
		\draw [style=double arrow] (5.center) to (4.center);
		\draw (15) to (14);
		\draw (15) to (16);
		\draw (17) to (15);
		\draw (53) to (16);
		\draw (55) to (16);
		\draw (82) to (83);
	\end{pgfonlayer}
\end{tikzpicture}
\end{equation}
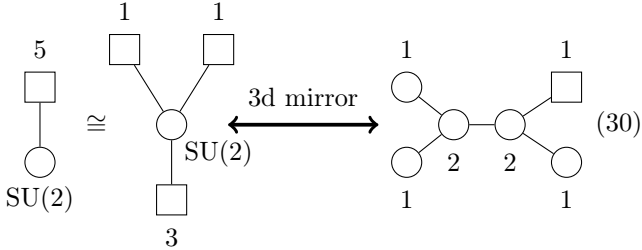
\begin{equation}
    \begin{tikzpicture}[baseline=-0.25em,scale=0.5]
    \begin{pgfonlayer}{nodelayer}
		\node [style={gauge_gp}] (57) at (-3.5, 0) {};
		\node [style={flavour_gp}] (58) at (-3.5, -2) {};
		\node [style={flavour_gp}] (59) at (-2.25, 2) {};
		\node [style={flavour_gp}] (60) at (-4.75, 2) {};
		\node [style=none] (61) at (2, 0) {};
		\node [style=none] (62) at (-2, 0) {};
		\node [style=none] (63) at (0, 0.75) {3D mirror};
		\node [style=none] (68) at (-2.25, -0.75) {SU(2)};
		\node [style=none] (69) at (-3.5, -3) {3};
		\node [style=none] (70) at (-4.75, 3) {1};
		\node [style=none] (71) at (-2.25, 3) {1};
		\node [style={gauge_gp}] (112) at (2.75, -1) {};
		\node [style={gauge_gp}] (113) at (4, 0) {};
		\node [style={gauge_gp}] (114) at (5.5, 0) {};
		\node [style=none] (120) at (2.75, 2) {1};
		\node [style=none] (121) at (4, -1) {2};
		\node [style=none] (122) at (2.75, -2) {1};
		\node [style=none] (123) at (7, -2) {1};
		\node [style={gauge_gp}] (124) at (7, -1) {};
		\node [style=none] (125) at (5.5, -1) {2};
		\node [style={flavour_gp}] (126) at (7, 1) {};
		\node [style=none] (127) at (7, 2) {1};
		\node [style=none] (128) at (-5.5, 0) {$\cong$};
		\node [style={flavour_gp}] (129) at (-7, 1) {};
		\node [style={gauge_gp}] (130) at (-7, -1) {};
		\node [style=none] (131) at (-7, 2) {5};
		\node [style=none] (132) at (-7, -2) {SU(2)};
		\node [style={gauge_gp}] (133) at (2.75, 1) {};
		\node [style=none] (134) at (-7.75, 0) {};
	\end{pgfonlayer}
	\begin{pgfonlayer}{edgelayer}
		\draw (58) to (57);
		\draw (60) to (57);
		\draw (57) to (59);
		\draw [style=double arrow] (62.center) to (61.center);
		\draw (113) to (112);
		\draw (113) to (114);
		\draw (124) to (114);
		\draw (126) to (114);
		\draw (129) to (130);
		\draw (133) to (113);
	\end{pgfonlayer}
\end{tikzpicture}
\end{equation}

They can be glued together by the following:
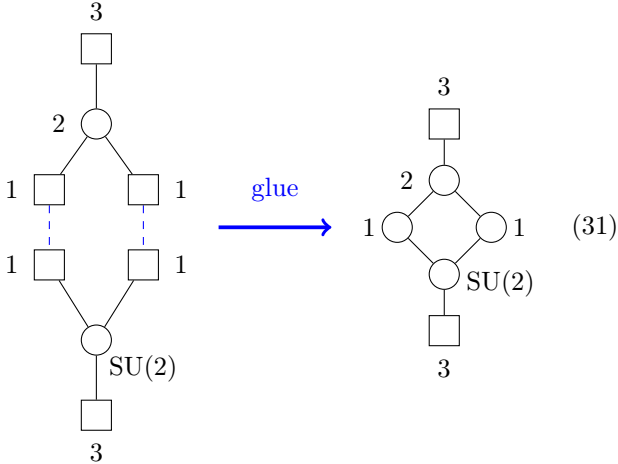
\begin{equation} \label{eqn:SU_glue}
    \begin{tikzpicture}[baseline=-0.25em,scale=0.5]
	\begin{pgfonlayer}{nodelayer}
		\node [style={gauge_gp}] (57) at (-4.75, -3) {};
		\node [style={flavour_gp}] (58) at (-4.75, -5) {};
		\node [style={flavour_gp}] (59) at (-3.5, -1) {};
		\node [style={flavour_gp}] (60) at (-6, -1) {};
		\node [style=none] (68) at (-3.5, -3.75) {SU(2)};
		\node [style=none] (69) at (-4.75, -6) {3};
		\node [style=none] (70) at (-7, -1) {1};
		\node [style=none] (71) at (-2.5, -1) {1};
		\node [style={gauge_gp}] (105) at (-4.75, 2.75) {};
		\node [style={flavour_gp}] (106) at (-4.75, 4.75) {};
		\node [style={flavour_gp}] (107) at (-3.5, 1) {};
		\node [style={flavour_gp}] (108) at (-6, 1) {};
		\node [style=none] (109) at (-5.75, 2.75) {2};
		\node [style=none] (110) at (-4.75, 5.75) {3};
		\node [style=none] (111) at (-7, 1) {1};
		\node [style=none] (112) at (-2.5, 1) {1};
		\node [style={gauge_gp}] (113) at (4.5, 1.25) {};
		\node [style={flavour_gp}] (114) at (4.5, 2.75) {};
		\node [style={gauge_gp}] (115) at (3.25, 0) {};
		\node [style={gauge_gp}] (116) at (5.75, 0) {};
		\node [style={gauge_gp}] (117) at (4.5, -1.25) {};
		\node [style={flavour_gp}] (118) at (4.5, -2.75) {};
		\node [style=none] (119) at (3.5, 1.25) {2};
		\node [style=none] (120) at (4.5, 3.75) {3};
		\node [style=none] (121) at (2.5, 0) {1};
		\node [style=none] (122) at (6.5, 0) {1};
		\node [style=none] (124) at (4.5, -3.75) {3};
		\node [style=none] (125) at (-1.5, 0) {};
		\node [style=none] (126) at (1.5, 0) {};
		\node [style=none] (139) at (0, 0.5) {glue};
		\node [style=none] (140) at (6, -1.5) {SU(2)};
	\end{pgfonlayer}
	\begin{pgfonlayer}{edgelayer}
		\draw (58) to (57);
		\draw (60) to (57);
		\draw (57) to (59);
		\draw (106) to (105);
		\draw (108) to (105);
		\draw (105) to (107);
		\draw [style=dashed] (107) to (59);
		\draw [style=dashed] (108) to (60);
		\draw (114) to (113);
		\draw (118) to (117);
		\draw (115) to (113);
		\draw (113) to (116);
		\draw (117) to (115);
		\draw (117) to (116);
		\draw [style=arrow] (125.center) to (126.center);
	\end{pgfonlayer}
\end{tikzpicture}
\end{equation}
\begin{eqnarray*}
\HS_\mathcal{C}(t)=&& 1 + 3t^2 + 2t^3 + 12t^4 + 12t^5 + 32t^6 + 44t^7 + \\
&& 98t^8 + 136t^9 + 256t^{10} + \mathcal{O}(t^{11})
\end{eqnarray*}

There are two ways to glue their mirror theories together:
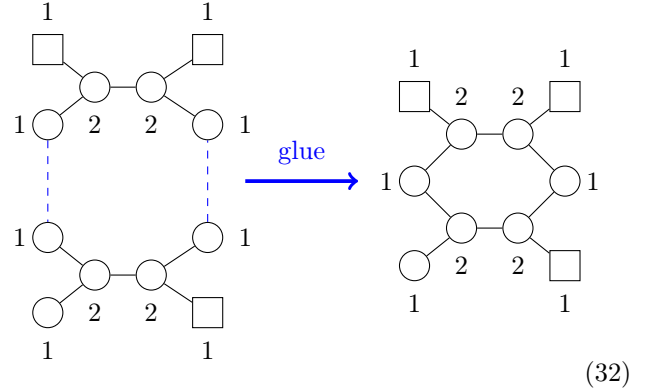
\begin{equation}
    \begin{tikzpicture}[baseline=-0.25em,scale=0.5]
	\begin{pgfonlayer}{nodelayer}
		\node [style=none] (125) at (-1.5, 0) {};
		\node [style=none] (126) at (1.5, 0) {};
		\node [style=none] (139) at (0, 0.5) {glue};
		\node [style={gauge_gp}] (141) at (-6.75, 1.5) {};
		\node [style={gauge_gp}] (142) at (-5.5, 2.5) {};
		\node [style={gauge_gp}] (143) at (-4, 2.5) {};
		\node [style={flavour_gp}] (144) at (-6.75, 3.5) {};
		\node [style=none] (145) at (-6.75, 4.5) {1};
		\node [style=none] (146) at (-5.5, 1.5) {2};
		\node [style=none] (147) at (-7.5, 1.5) {1};
		\node [style=none] (148) at (-1.5, 1.5) {1};
		\node [style={gauge_gp}] (149) at (-2.5, 1.5) {};
		\node [style=none] (150) at (-4, 1.5) {2};
		\node [style={flavour_gp}] (151) at (-2.5, 3.5) {};
		\node [style=none] (152) at (-2.5, 4.5) {1};
		\node [style={gauge_gp}] (154) at (-6.75, -3.5) {};
		\node [style={gauge_gp}] (155) at (-5.5, -2.5) {};
		\node [style={gauge_gp}] (156) at (-4, -2.5) {};
		\node [style=none] (157) at (-7.5, -1.5) {1};
		\node [style=none] (158) at (-5.5, -3.5) {2};
		\node [style=none] (159) at (-6.75, -4.5) {1};
		\node [style=none] (160) at (-2.5, -4.5) {1};
		\node [style={gauge_gp}] (161) at (-2.5, -1.5) {};
		\node [style=none] (162) at (-4, -3.5) {2};
		\node [style={flavour_gp}] (163) at (-2.5, -3.5) {};
		\node [style=none] (164) at (-1.5, -1.5) {1};
		\node [style={gauge_gp}] (165) at (-6.75, -1.5) {};
		\node [style={gauge_gp}] (166) at (3, 0) {};
		\node [style={gauge_gp}] (167) at (4.25, 1.25) {};
		\node [style={gauge_gp}] (168) at (5.75, 1.25) {};
		\node [style={flavour_gp}] (169) at (3, 2.25) {};
		\node [style=none] (170) at (3, 3.25) {1};
		\node [style=none] (171) at (4.25, 2.25) {2};
		\node [style={gauge_gp}] (174) at (7, 0) {};
		\node [style=none] (175) at (5.75, 2.25) {2};
		\node [style={flavour_gp}] (176) at (7, 2.25) {};
		\node [style=none] (177) at (7, 3.25) {1};
		\node [style={gauge_gp}] (178) at (3, -2.25) {};
		\node [style={gauge_gp}] (179) at (4.25, -1.25) {};
		\node [style={gauge_gp}] (180) at (5.75, -1.25) {};
		\node [style=none] (182) at (4.25, -2.25) {2};
		\node [style=none] (183) at (3, -3.25) {1};
		\node [style=none] (184) at (7, -3.25) {1};
		\node [style=none] (186) at (5.75, -2.25) {2};
		\node [style={flavour_gp}] (187) at (7, -2.25) {};
		\node [style=none] (188) at (7.75, 0) {1};
		\node [style=none] (189) at (2.25, 0) {1};
	\end{pgfonlayer}
	\begin{pgfonlayer}{edgelayer}
		\draw [style=arrow] (125.center) to (126.center);
		\draw (142) to (141);
		\draw (142) to (143);
		\draw (144) to (142);
		\draw (149) to (143);
		\draw (151) to (143);
		\draw (155) to (154);
		\draw (155) to (156);
		\draw (161) to (156);
		\draw (163) to (156);
		\draw (165) to (155);
		\draw [style=dashed] (141) to (165);
		\draw [style=dashed] (161) to (149);
		\draw (167) to (166);
		\draw (167) to (168);
		\draw (169) to (167);
		\draw (174) to (168);
		\draw (176) to (168);
		\draw (179) to (178);
		\draw (179) to (180);
		\draw (187) to (180);
		\draw (174) to (180);
		\draw (179) to (166);
	\end{pgfonlayer}
\end{tikzpicture}
\end{equation}
\begin{eqnarray*}
\HS_\mathcal{H}(t) =&& 1 + 3t^2 + 2t^3 + 10t^4 + 12t^5 + 30t^6 + 42t^7 + \\
&& 92t^8 + 130t^9 + 240t^{10} + \mathcal{O}(t^{11})
\end{eqnarray*}

\newpage

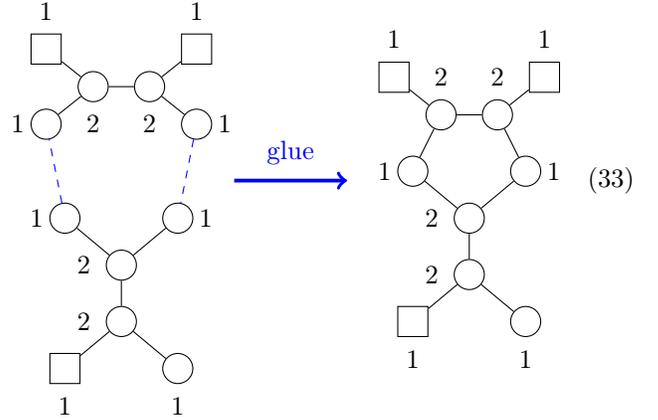
\begin{equation}
    \begin{tikzpicture}[baseline=-0.25em,scale=0.5]
	\begin{pgfonlayer}{nodelayer}
		\node [style=none] (125) at (-1.5, 0) {};
		\node [style=none] (126) at (1.5, 0) {};
		\node [style=none] (139) at (0, 0.5) {glue};
		\node [style={gauge_gp}] (141) at (-6.5, 1.5) {};
		\node [style={gauge_gp}] (142) at (-5.25, 2.5) {};
		\node [style={gauge_gp}] (143) at (-3.75, 2.5) {};
		\node [style={flavour_gp}] (144) at (-6.5, 3.5) {};
		\node [style=none] (145) at (-6.5, 4.5) {1};
		\node [style=none] (146) at (-5.25, 1.5) {2};
		\node [style=none] (147) at (-7.25, 1.5) {1};
		\node [style=none] (148) at (-1.75, 1.5) {1};
		\node [style={gauge_gp}] (149) at (-2.5, 1.5) {};
		\node [style=none] (150) at (-3.75, 1.5) {2};
		\node [style={flavour_gp}] (151) at (-2.5, 3.5) {};
		\node [style=none] (152) at (-2.5, 4.5) {1};
		\node [style={gauge_gp}] (154) at (-6, -1) {};
		\node [style={gauge_gp}] (155) at (-4.5, -2.25) {};
		\node [style={gauge_gp}] (156) at (-4.5, -3.75) {};
		\node [style=none] (157) at (-2.25, -1) {1};
		\node [style=none] (158) at (-5.5, -2.25) {2};
		\node [style=none] (159) at (-6.75, -1) {1};
		\node [style=none] (160) at (-6, -6) {1};
		\node [style={gauge_gp}] (161) at (-3, -5) {};
		\node [style=none] (162) at (-5.5, -3.75) {2};
		\node [style={flavour_gp}] (163) at (-6, -5) {};
		\node [style=none] (164) at (-3, -6) {1};
		\node [style={gauge_gp}] (165) at (-3, -1) {};
		\node [style={gauge_gp}] (190) at (3.25, 0.25) {};
		\node [style={gauge_gp}] (191) at (4, 1.75) {};
		\node [style={gauge_gp}] (192) at (5.5, 1.75) {};
		\node [style={flavour_gp}] (193) at (2.75, 2.75) {};
		\node [style=none] (194) at (2.75, 3.75) {1};
		\node [style=none] (195) at (4, 2.75) {2};
		\node [style=none] (196) at (2.5, 0.25) {1};
		\node [style=none] (197) at (7, 0.25) {1};
		\node [style={gauge_gp}] (198) at (6.25, 0.25) {};
		\node [style=none] (199) at (5.5, 2.75) {2};
		\node [style={flavour_gp}] (200) at (6.75, 2.75) {};
		\node [style=none] (201) at (6.75, 3.75) {1};
		\node [style={gauge_gp}] (203) at (4.75, -1) {};
		\node [style={gauge_gp}] (204) at (4.75, -2.5) {};
		\node [style=none] (206) at (3.75, -1) {2};
		\node [style=none] (208) at (3.25, -4.75) {1};
		\node [style={gauge_gp}] (209) at (6.25, -3.75) {};
		\node [style=none] (210) at (3.75, -2.5) {2};
		\node [style={flavour_gp}] (211) at (3.25, -3.75) {};
		\node [style=none] (212) at (6.25, -4.75) {1};
	\end{pgfonlayer}
	\begin{pgfonlayer}{edgelayer}
		\draw [style=arrow] (125.center) to (126.center);
		\draw (142) to (141);
		\draw (142) to (143);
		\draw (144) to (142);
		\draw (149) to (143);
		\draw (151) to (143);
		\draw (155) to (154);
		\draw (155) to (156);
		\draw (161) to (156);
		\draw (163) to (156);
		\draw (165) to (155);
		\draw [style=dashed] (149) to (165);
		\draw [style=dashed] (154) to (141);
		\draw (191) to (190);
		\draw (191) to (192);
		\draw (193) to (191);
		\draw (198) to (192);
		\draw (200) to (192);
		\draw (203) to (204);
		\draw (209) to (204);
		\draw (211) to (204);
		\draw (198) to (203);
		\draw (203) to (190);
	\end{pgfonlayer}
\end{tikzpicture}
\end{equation}
\begin{eqnarray*}
\HS_\mathcal{H}(t) =&& 1 + 3t^2 + 2t^3 + 8t^4 + 14t^5 + 26t^6 + 46t^7 + \\
&& 82t^8 + 132t^9 + 228t^{10} + \mathcal{O}(t^{11})
\end{eqnarray*}

Neither Higgs branch Hilbert series matches the Coulomb branch Hilbert series of the quiver in Eq.\ \eqref{eqn:SU_glue}, and hence, neither of them is its 3D mirror. The breakdown is likely due to the presence of special unitary gauge groups in the proposed mirror, which is to be expected due to the global symmetries. This ties in to the observation in Ref.\ \cite{Zhong:2024jxw} that we are unable to consistently find 3D mirrors of nonlinear quivers involving special unitary gauge groups. One possibility is that the decoupling of a $\U(1)$, which turns a unitary into a special unitary gauge group, may act nontrivially and on several gauge groups simultaneously. Further experimentations using methods from Refs.\ \cite{Dey:2020hfe,  Giacomelli:2024laq,  Zhong:2024jxw} may shed light on any extensions to these gluings.

In the next section, we will use the quartet of quiver algorithms to see how even a slight perturbation of the structure of our non-Abelian quivers can result in a non-Lagrangian mirror.

\section{Finding and bootstrapping 3D mirrors}\label{sec:boot}

In this section, the four algorithms presented in Sec.\ \ref{sec:algo} are put into practical use. While they were originally introduced as part of the bootstrapping procedure for generating 3D mirrors, they can equally serve as direct tools for constructing the 3D mirror of a specified quiver. To illustrate this, we analyze the quiver depicted at the bottom right of \eqref{eqn:nonabelianexample} and explicitly derive its 3D mirror step by step.

\vspace{2mm}
\begin{minipage}{\columnwidth} 
\centering
\textbf{\underline{Step 1: Perform Quiver Subtraction}}\\[2pt]
\textbf{\underline{OR Decay and Fission}}
\end{minipage}
\vspace{2mm}

\noindent
We begin by performing a single operation---either Quiver Subtraction \textbf{or} Decay and Fission. The goal of this step is to identify which of the resulting daughter theories possesses a known 3D mirror. In our example, we find that there are three possible subtractions:

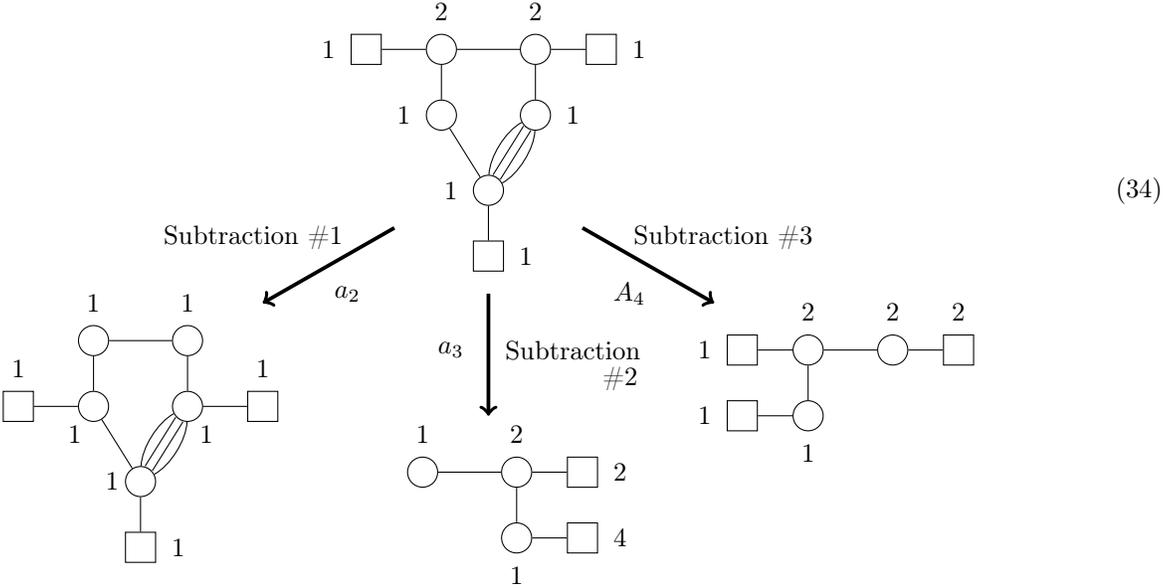
\begin{widetext}
\begin{equation}
    \begin{tikzpicture}[baseline=-0.25em,scale=0.5]
	\begin{pgfonlayer}{nodelayer}
		\node [style={gauge_gp}] (75) at (0, 0) {};
		\node [style={gauge_gp}] (76) at (-1.25, 2) {};
		\node [style={gauge_gp}] (77) at (1.25, 2) {};
		\node [style={flavour_gp}] (78) at (0, -1.75) {};
		\node [style=none] (81) at (-1, 0) {1};
		\node [style=none] (82) at (1, -1.75) {1};
		\node [style=none] (84) at (-2.25, 2) {1};
		\node [style=none] (85) at (2.25, 2) {1};
		\node [style={gauge_gp}] (87) at (-1.25, 3.75) {};
		\node [style={gauge_gp}] (88) at (1.25, 3.75) {};
		\node [style={flavour_gp}] (89) at (-3.25, 3.75) {};
		\node [style={flavour_gp}] (90) at (3, 3.75) {};
		\node [style=none] (91) at (-4.25, 3.75) {1};
		\node [style=none] (92) at (4, 3.75) {1};
		\node [style=none] (93) at (1.25, 4.75) {2};
		\node [style=none] (94) at (-1.25, 4.75) {2};
		\node [style=none] (95) at (0, -2.75) {};
		\node [style=none] (96) at (0, -6) {};
		\node [style=none] (97) at (2.5, -1) {};
		\node [style=none] (98) at (-2.5, -1) {};
		\node [style=none] (99) at (-6, -3) {};
		\node [style=none] (100) at (6, -3) {};
		\node [style=none] (101) at (-6.25, -1.25) {Subtraction \#1};
		\node [style=none] (102) at (6.25, -1.25) {Subtraction \#3};
		\node [style=none] (103) at (2.25, -4.25) {Subtraction};
		\node [style=none] (104) at (3.5, -5) {\#2};
		\node [style={gauge_gp}] (105) at (-9.25, -7.75) {};
		\node [style={gauge_gp}] (106) at (-10.5, -5.75) {};
		\node [style={gauge_gp}] (107) at (-8, -5.75) {};
		\node [style={flavour_gp}] (108) at (-9.25, -9.5) {};
		\node [style=none] (109) at (-10, -7.75) {1};
		\node [style=none] (110) at (-8.25, -9.5) {1};
		\node [style=none] (111) at (-12.5, -4.75) {1};
		\node [style=none] (112) at (-7.5, -6.5) {1};
		\node [style={gauge_gp}] (113) at (-10.5, -4) {};
		\node [style={gauge_gp}] (114) at (-8, -4) {};
		\node [style={flavour_gp}] (115) at (-12.5, -5.75) {};
		\node [style={flavour_gp}] (116) at (-6, -5.75) {};
		\node [style=none] (117) at (-11, -6.5) {1};
		\node [style=none] (118) at (-6, -4.75) {1};
		\node [style=none] (119) at (-8, -3) {1};
		\node [style=none] (120) at (-10.5, -3) {1};
		\node [style=none] (121) at (-3.75, -2.75) {$a_2$};
		\node [style={gauge_gp}] (124) at (0.75, -9.25) {};
		\node [style=none] (129) at (0.75, -10.25) {1};
		\node [style={gauge_gp}] (130) at (-1.75, -7.5) {};
		\node [style={gauge_gp}] (131) at (0.75, -7.5) {};
		\node [style={flavour_gp}] (133) at (2.5, -7.5) {};
		\node [style=none] (134) at (-1.75, -6.5) {1};
		\node [style=none] (135) at (3.5, -7.5) {2};
		\node [style=none] (136) at (0.75, -6.5) {2};
		\node [style={flavour_gp}] (137) at (2.5, -9.25) {};
		\node [style=none] (138) at (3.5, -9.25) {4};
		\node [style={gauge_gp}] (140) at (8.5, -6) {};
		\node [style=none] (145) at (8.5, -7) {1};
		\node [style={gauge_gp}] (147) at (8.5, -4.25) {};
		\node [style={gauge_gp}] (148) at (10.75, -4.25) {};
		\node [style={flavour_gp}] (149) at (6.75, -4.25) {};
		\node [style={flavour_gp}] (150) at (12.5, -4.25) {};
		\node [style=none] (151) at (5.75, -4.25) {1};
		\node [style=none] (152) at (12.5, -3.25) {2};
		\node [style=none] (153) at (10.75, -3.25) {2};
		\node [style=none] (154) at (8.5, -3.25) {2};
		\node [style={flavour_gp}] (155) at (6.75, -6) {};
		\node [style=none] (156) at (5.75, -6) {1};
		\node [style=none] (157) at (-1, -4.25) {$a_3$};
		\node [style=none] (158) at (3.75, -2.75) {$A_4$};
	\end{pgfonlayer}
	\begin{pgfonlayer}{edgelayer}
		\draw (76) to (75);
		\draw (78) to (75);
		\draw [style=none] (87) to (76);
		\draw [style=none] (88) to (77);
		\draw [style=none] (88) to (87);
		\draw [style=none] (89) to (87);
		\draw [style=none] (90) to (88);
		\draw [bend right, looseness=0.75] (75) to (77);
		\draw [bend right, looseness=0.75] (77) to (75);
		\draw [style=doubleline] (77) to (75);
		\draw [style=arrow] (95.center) to (96.center);
		\draw [style=arrow] (98.center) to (99.center);
		\draw [style=arrow] (97.center) to (100.center);
		\draw (106) to (105);
		\draw (108) to (105);
		\draw [style=none] (113) to (106);
		\draw [style=none] (114) to (107);
		\draw [style=none] (114) to (113);
		\draw [bend right, looseness=0.75] (105) to (107);
		\draw [bend right, looseness=0.75] (107) to (105);
		\draw [style=doubleline] (107) to (105);
		\draw (116) to (107);
		\draw (106) to (115);
		\draw [style=none] (131) to (124);
		\draw [style=none] (131) to (130);
		\draw [style=none] (133) to (131);
		\draw (137) to (124);
		\draw [style=none] (147) to (140);
		\draw [style=none] (148) to (147);
		\draw [style=none] (149) to (147);
		\draw [style=none] (150) to (148);
		\draw (155) to (140);
	\end{pgfonlayer}
\end{tikzpicture}
\end{equation}
\vfill
and two decays:
\vfill
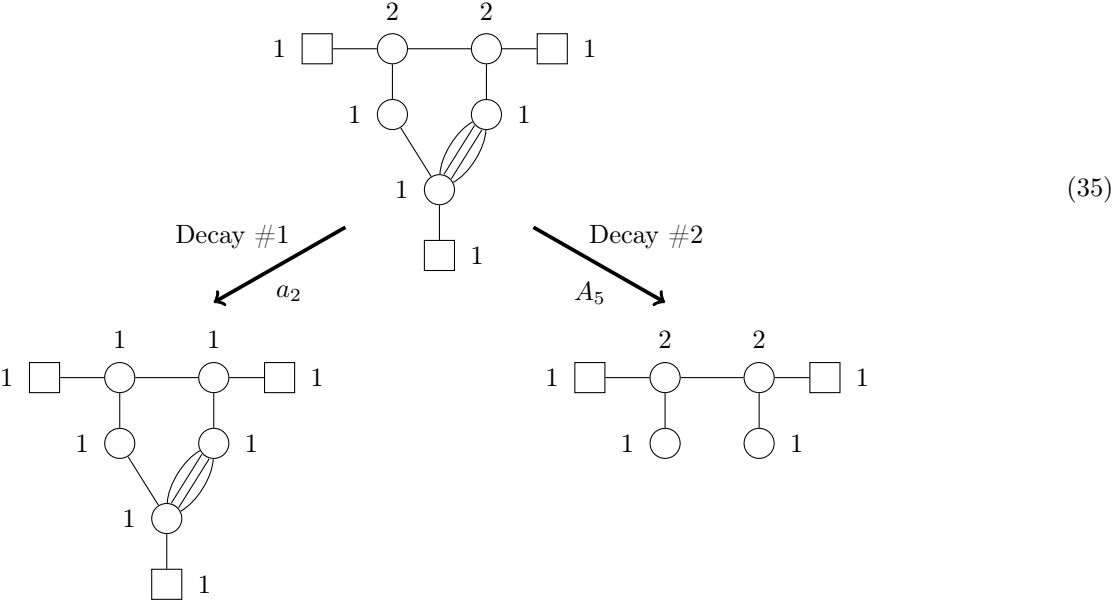
\begin{equation}
    \begin{tikzpicture}[baseline=-0.25em,scale=0.5]
	\begin{pgfonlayer}{nodelayer}
		\node [style={gauge_gp}] (75) at (0, 0) {};
		\node [style={gauge_gp}] (76) at (-1.25, 2) {};
		\node [style={gauge_gp}] (77) at (1.25, 2) {};
		\node [style={flavour_gp}] (78) at (0, -1.75) {};
		\node [style=none] (81) at (-1, 0) {1};
		\node [style=none] (82) at (1, -1.75) {1};
		\node [style=none] (84) at (-2.25, 2) {1};
		\node [style=none] (85) at (2.25, 2) {1};
		\node [style={gauge_gp}] (87) at (-1.25, 3.75) {};
		\node [style={gauge_gp}] (88) at (1.25, 3.75) {};
		\node [style={flavour_gp}] (89) at (-3.25, 3.75) {};
		\node [style={flavour_gp}] (90) at (3, 3.75) {};
		\node [style=none] (91) at (-4.25, 3.75) {1};
		\node [style=none] (92) at (4, 3.75) {1};
		\node [style=none] (93) at (1.25, 4.75) {2};
		\node [style=none] (94) at (-1.25, 4.75) {2};
		\node [style=none] (97) at (2.5, -1) {};
		\node [style=none] (98) at (-2.5, -1) {};
		\node [style=none] (99) at (-6, -3) {};
		\node [style=none] (100) at (6, -3) {};
		\node [style=none] (101) at (-5.5, -1.25) {Decay \#1};
		\node [style=none] (102) at (5.5, -1.25) {Decay \#2};
		\node [style=none] (121) at (-4, -2.75) {$a_2$};
		\node [style=none] (158) at (4, -2.75) {$A_5$};
		\node [style={gauge_gp}] (159) at (-7.25, -8.75) {};
		\node [style={gauge_gp}] (160) at (-8.5, -6.75) {};
		\node [style={gauge_gp}] (161) at (-6, -6.75) {};
		\node [style={flavour_gp}] (162) at (-7.25, -10.5) {};
		\node [style=none] (163) at (-8.25, -8.75) {1};
		\node [style=none] (164) at (-6.25, -10.5) {1};
		\node [style=none] (165) at (-9.5, -6.75) {1};
		\node [style=none] (166) at (-5, -6.75) {1};
		\node [style={gauge_gp}] (167) at (-8.5, -5) {};
		\node [style={gauge_gp}] (168) at (-6, -5) {};
		\node [style={flavour_gp}] (169) at (-10.5, -5) {};
		\node [style={flavour_gp}] (170) at (-4.25, -5) {};
		\node [style=none] (171) at (-11.5, -5) {1};
		\node [style=none] (172) at (-3.25, -5) {1};
		\node [style=none] (173) at (-6, -4) {1};
		\node [style=none] (174) at (-8.5, -4) {1};
		\node [style={gauge_gp}] (176) at (6, -6.75) {};
		\node [style={gauge_gp}] (177) at (8.5, -6.75) {};
		\node [style=none] (181) at (5, -6.75) {1};
		\node [style=none] (182) at (9.5, -6.75) {1};
		\node [style={gauge_gp}] (183) at (6, -5) {};
		\node [style={gauge_gp}] (184) at (8.5, -5) {};
		\node [style={flavour_gp}] (185) at (4, -5) {};
		\node [style={flavour_gp}] (186) at (10.25, -5) {};
		\node [style=none] (187) at (3, -5) {1};
		\node [style=none] (188) at (11.25, -5) {1};
		\node [style=none] (189) at (8.5, -4) {2};
		\node [style=none] (190) at (6, -4) {2};
		\node [style=none] (192) at (12.5, 0) {};
		\node [style=none] (193) at (-12.5, 0) {};
	\end{pgfonlayer}
	\begin{pgfonlayer}{edgelayer}
		\draw (76) to (75);
		\draw (78) to (75);
		\draw [style=none] (87) to (76);
		\draw [style=none] (88) to (77);
		\draw [style=none] (88) to (87);
		\draw [style=none] (89) to (87);
		\draw [style=none] (90) to (88);
		\draw [bend right, looseness=0.75] (75) to (77);
		\draw [bend right, looseness=0.75] (77) to (75);
		\draw [style=doubleline] (77) to (75);
		\draw [style=arrow] (98.center) to (99.center);
		\draw [style=arrow] (97.center) to (100.center);
		\draw (160) to (159);
		\draw (162) to (159);
		\draw [style=none] (167) to (160);
		\draw [style=none] (168) to (161);
		\draw [style=none] (168) to (167);
		\draw [style=none] (169) to (167);
		\draw [style=none] (170) to (168);
		\draw [bend right, looseness=0.75] (159) to (161);
		\draw [bend right, looseness=0.75] (161) to (159);
		\draw [style=doubleline] (161) to (159);
		\draw [style=none] (183) to (176);
		\draw [style=none] (184) to (177);
		\draw [style=none] (184) to (183);
		\draw [style=none] (185) to (183);
		\draw [style=none] (186) to (184);
	\end{pgfonlayer}
\end{tikzpicture}
\end{equation}
\vfill
\clearpage
\end{widetext}

Out of these possibilities, Decay \#2 and Subtraction \#2 and \#3 yield linear quivers, for which the corresponding 3D mirrors can be readily obtained---either through the standard Hanany-Witten D3-D5-NS5 constructions or by using the computational approach described in Ref.\ \cite{USU}. In contrast, Subtraction \#1 and Decay \#1 produce sunshine quivers, whose 3D mirrors can also be identified (see the discussion in the previous section). For clarity and brevity, we proceed with Subtraction \#2, which results in a linear quiver. The 3D mirror of this daughter theory is given by

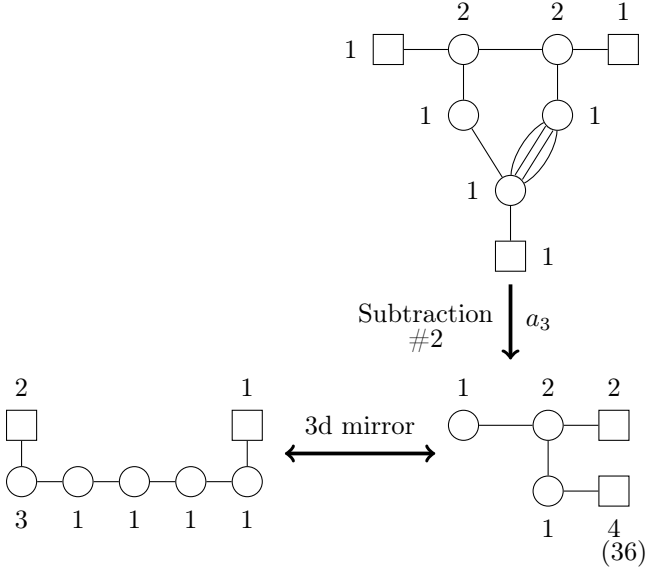
\begin{equation}
    \begin{tikzpicture}[baseline=-0.25em,scale=0.5]
	\begin{pgfonlayer}{nodelayer}
		\node [style={gauge_gp}] (75) at (4.5, 2.25) {};
		\node [style={gauge_gp}] (76) at (3.25, 4.25) {};
		\node [style={gauge_gp}] (77) at (5.75, 4.25) {};
		\node [style={flavour_gp}] (78) at (4.5, 0.5) {};
		\node [style=none] (81) at (3.5, 2.25) {1};
		\node [style=none] (82) at (5.5, 0.5) {1};
		\node [style=none] (84) at (2.25, 4.25) {1};
		\node [style=none] (85) at (6.75, 4.25) {1};
		\node [style={gauge_gp}] (87) at (3.25, 6) {};
		\node [style={gauge_gp}] (88) at (5.75, 6) {};
		\node [style={flavour_gp}] (89) at (1.25, 6) {};
		\node [style={flavour_gp}] (90) at (7.5, 6) {};
		\node [style=none] (91) at (0.25, 6) {1};
		\node [style=none] (92) at (7.5, 7) {1};
		\node [style=none] (93) at (5.75, 7) {2};
		\node [style=none] (94) at (3.25, 7) {2};
		\node [style=none] (95) at (4.5, -0.25) {};
		\node [style=none] (96) at (4.5, -2.25) {};
		\node [style=none] (103) at (2.25, -1) {Subtraction};
		\node [style={gauge_gp}] (124) at (5.5, -5.75) {};
		\node [style=none] (129) at (5.5, -6.75) {1};
		\node [style={gauge_gp}] (130) at (3.25, -4) {};
		\node [style={gauge_gp}] (131) at (5.5, -4) {};
		\node [style={flavour_gp}] (133) at (7.25, -4) {};
		\node [style=none] (134) at (3.25, -3) {1};
		\node [style=none] (135) at (7.25, -3) {2};
		\node [style=none] (136) at (5.5, -3) {2};
		\node [style={flavour_gp}] (137) at (7.25, -5.75) {};
		\node [style=none] (138) at (7.25, -6.75) {4};
		\node [style=none] (157) at (5.25, -1.25) {$a_3$};
		\node [style={gauge_gp}] (158) at (-2.5, -5.5) {};
		\node [style={gauge_gp}] (159) at (-4, -5.5) {};
		\node [style={gauge_gp}] (160) at (-5.5, -5.5) {};
		\node [style={gauge_gp}] (161) at (-7, -5.5) {};
		\node [style={gauge_gp}] (162) at (-8.5, -5.5) {};
		\node [style={flavour_gp}] (163) at (-8.5, -4) {};
		\node [style={flavour_gp}] (164) at (-2.5, -4) {};
		\node [style=none] (165) at (-2.5, -3) {3};
		\node [style=none] (166) at (-8.5, -6.5) {1};
		\node [style=none] (167) at (-8.5, -3) {1};
		\node [style=none] (168) at (-7, -6.5) {1};
		\node [style=none] (169) at (-5.5, -6.5) {1};
		\node [style=none] (170) at (-4, -6.5) {1};
		\node [style=none] (171) at (-2.5, -6.5) {2};
		\node [style=none] (172) at (2.5, -4.75) {};
		\node [style=none] (173) at (-1.5, -4.75) {};
		\node [style=none] (174) at (0.5, -4) {3D mirror};
		\node [style=none] (175) at (2.25, -1.75) { \#2};
		\node [style=none] (177) at (4.5, -1.25) {};
	\end{pgfonlayer}
	\begin{pgfonlayer}{edgelayer}
		\draw (76) to (75);
		\draw (78) to (75);
		\draw [style=none] (87) to (76);
		\draw [style=none] (88) to (77);
		\draw [style=none] (88) to (87);
		\draw [style=none] (89) to (87);
		\draw [style=none] (90) to (88);
		\draw [bend right, looseness=0.75] (75) to (77);
		\draw [bend right, looseness=0.75] (77) to (75);
		\draw [style=doubleline] (77) to (75);
		\draw [style=arrow] (95.center) to (96.center);
		\draw [style=none] (131) to (124);
		\draw [style=none] (131) to (130);
		\draw [style=none] (133) to (131);
		\draw (137) to (124);
		\draw (163) to (162);
		\draw (164) to (158);
		\draw (158) to (159);
		\draw (159) to (160);
		\draw (160) to (161);
		\draw (161) to (162);
		\draw [style=double arrow] (173.center) to (172.center);
	\end{pgfonlayer}
\end{tikzpicture}
\label{36eqn}
\end{equation}

\vspace{2mm}
\begin{minipage}{\columnwidth} 
\centering
\textbf{\underline{Step 2: Perform Quiver Addition}}\\[2pt]
\textbf{\underline{OR Growth and Fusion}}
\end{minipage}
\vspace{2mm}

\noindent
Since we applied quiver subtraction in the previous step, to complete the circle of dualities, we now perform growth and fusion on the mirror linear quiver. As mentioned earlier, growth and fusion offers infinitely many possibilities, so additional information is required to narrow them down. The first observation is that the transverse space after subtraction is $a_3$. Consequently, in the 3D mirror, the transverse space after growth and fusion must be the mirror slice $A_3$. This constraint significantly reduces the possible configurations: if we perform growth on any one of the gauge groups by increasing its rank by one, the resulting slice will not correspond to $A_3$. Therefore, the only consistent possibility is that the growth introduces an additional $\U(1)$ gauge node connected to the rest of the quiver.

The second clue arises from the global symmetries. Determining the exact global symmetry---whether Higgs or Coulomb---is often difficult due to hidden $\U(1)$ factors that frequently appear. However, the non-Abelian subgroups of these symmetries are much easier to identify. From the original quiver, we observe that there exists an $\SU(4) \subset G_{\mathcal{H}}$ subgroup on the Higgs branch and an $\SU(3) \subset G_{\mathcal{C}}$ subgroup on the Coulomb branch. This implies that in the mirror theory on the bottom left of Eq.\ \eqref{36eqn}, the Higgs and Coulomb symmetry must contain an $\SU(3)$ and $\SU(4)$ factor respectively. These non-Abelian symmetries place strong constraints on the possible implementations of the growth algorithm. In particular, to ensure that the additional growth step does not spoil the three balanced nodes responsible for the $\SU(4)$ Coulomb-branch symmetry, we impose the following condition:
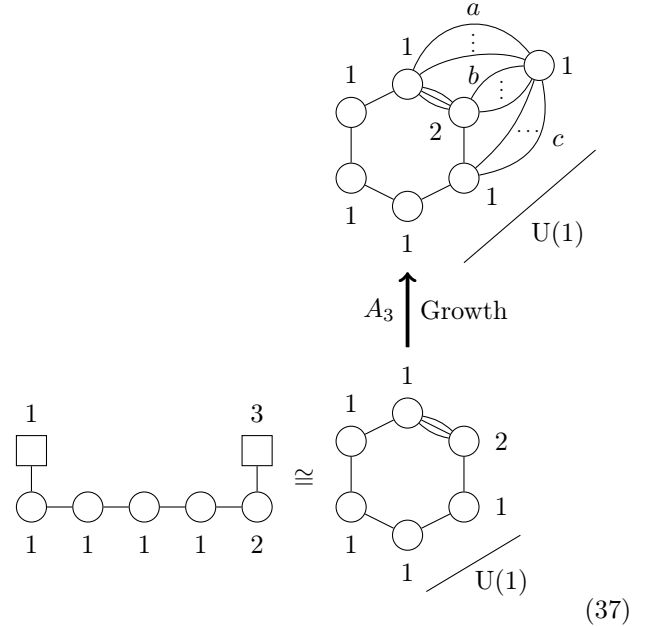
\vfill
\begin{equation}
    \begin{tikzpicture}[baseline=-0.25em,scale=0.5]
	\begin{pgfonlayer}{nodelayer}
		\node [style={gauge_gp}] (0) at (-7.25, -5.5) {};
		\node [style={gauge_gp}] (1) at (-5.75, -5.5) {};
		\node [style={gauge_gp}] (2) at (-4.25, -5.5) {};
		\node [style={gauge_gp}] (3) at (-2.75, -5.5) {};
		\node [style={gauge_gp}] (4) at (-1.25, -5.5) {};
		\node [style={flavour_gp}] (5) at (-1.25, -4) {};
		\node [style={flavour_gp}] (6) at (-7.25, -4) {};
		\node [style=none] (7) at (-7.25, -3) {1};
		\node [style=none] (8) at (-1.25, -3) {3};
		\node [style=none] (9) at (-1.25, -6.5) {2};
		\node [style=none] (10) at (-2.75, -6.5) {1};
		\node [style=none] (11) at (-4.25, -6.5) {1};
		\node [style=none] (12) at (-5.75, -6.5) {1};
		\node [style=none] (13) at (-7.25, -6.5) {1};
		\node [style={gauge_gp}] (14) at (1.25, -3.75) {};
		\node [style={gauge_gp}] (15) at (1.25, -5.5) {};
		\node [style={gauge_gp}] (16) at (2.75, -6.25) {};
		\node [style={gauge_gp}] (17) at (4.25, -5.5) {};
		\node [style={gauge_gp}] (18) at (4.25, -3.75) {};
		\node [style=none] (21) at (2.75, -2) {1};
		\node [style=none] (23) at (5.25, -3.75) {2};
		\node [style=none] (24) at (5.25, -5.5) {1};
		\node [style=none] (25) at (2.75, -7.25) {1};
		\node [style=none] (26) at (1.25, -6.5) {1};
		\node [style=none] (27) at (1.25, -2.75) {1};
		\node [style={gauge_gp}] (28) at (2.75, -3) {};
		\node [style=none] (29) at (0, -4.75) {$\cong$};
		\node [style=none] (30) at (2.75, 0.75) {};
		\node [style=none] (31) at (2.75, -1.25) {};
		\node [style=none] (32) at (4.25, -0.25) {Growth};
		\node [style=none] (33) at (2, -0.25) {$A_3$};
		\node [style={gauge_gp}] (34) at (1.25, 5) {};
		\node [style={gauge_gp}] (35) at (1.25, 3.25) {};
		\node [style={gauge_gp}] (36) at (2.75, 2.5) {};
		\node [style={gauge_gp}] (37) at (4.25, 3.25) {};
		\node [style={gauge_gp}] (38) at (4.25, 5) {};
		\node [style=none] (39) at (2.75, 6.75) {1};
		\node [style=none] (40) at (3.5, 4.5) {2};
		\node [style=none] (41) at (5, 2.75) {1};
		\node [style=none] (42) at (2.75, 1.5) {1};
		\node [style=none] (43) at (1.25, 2.25) {1};
		\node [style=none] (44) at (1.25, 6) {1};
		\node [style={gauge_gp}] (45) at (2.75, 5.75) {};
		\node [style={gauge_gp}] (46) at (6.25, 6.25) {};
		\node [style=none] (47) at (7, 6.25) {1};
		\node [style=none, scale=0.75] (49) at (4.5, 7) {$\vdots$};
		\node [style=none] (50) at (5.25, 5.75) {};
		\node [style=none] (51) at (5.25, 5.75) {};
		\node [style=none, scale=0.75] (52) at (5.25, 5.75) {$\vdots$};
		\node [style=none, scale=0.75] (54) at (6, 4.5) {$\cdots$};
		\node [style=none] (55) at (4.5, 7.75) {$a$};
		\node [style=none] (56) at (6.75, 4.25) {$c$};
		\node [style=none] (57) at (4.5, 6) {$b$};
		\node [style=none] (58) at (3.25, -7.75) {};
		\node [style=none] (59) at (5.75, -6.25) {};
		\node [style=none] (60) at (5.25, -7.5) {U(1)};
		\node [style=none] (61) at (4.25, 1) {};
		\node [style=none] (62) at (7.75, 4) {};
		\node [style=none] (63) at (6.75, 1.75) {U(1)};
	\end{pgfonlayer}
	\begin{pgfonlayer}{edgelayer}
		\draw (5) to (4);
		\draw (6) to (0);
		\draw (0) to (1);
		\draw (1) to (2);
		\draw (2) to (3);
		\draw (3) to (4);
		\draw (14) to (15);
		\draw (15) to (16);
		\draw (16) to (17);
		\draw (17) to (18);
		\draw (28) to (14);
		\draw [bend left=15] (28) to (18);
		\draw [bend right=15] (28) to (18);
		\draw (18) to (28);
		\draw [style=arrow] (31.center) to (30.center);
		\draw (34) to (35);
		\draw (35) to (36);
		\draw (36) to (37);
		\draw (37) to (38);
		\draw (45) to (34);
		\draw [bend left=15] (45) to (38);
		\draw [bend right=15] (45) to (38);
		\draw (38) to (45);
		\draw [bend left] (38) to (46);
		\draw [bend right] (38) to (46);
		\draw [bend left=15] (46) to (37);
		\draw [bend left=45, looseness=1.25] (46) to (37);
		\draw [bend left=60, looseness=1.25] (45) to (46);
		\draw [bend left=330, looseness=0.75] (46) to (45);
		\draw (59.center) to (58.center);
		\draw (62.center) to (61.center);
	\end{pgfonlayer}
\end{tikzpicture}
\end{equation}
\vfill
where we first perform gauging (unframing) to get a flavorless quiver to which the ``grown U(1)'' can connect. Next, we examine the expected Higgs branch symmetry of the mirror we are seeking. We expect the only non-Abelian Higgs branch symmetry to be $\SU(3)$, which already arises from the presence of the triple bond. From the growth algorithm, we know that this triple bond will remain intact regardless of the subsequent operations. This implies that the additional bonds, denoted $a$, $b$, and $c$, must each be single bonds. 

To reproduce the correct $A_3$ symmetry, these bonds must satisfy the condition $a + 2b + c = 4$.
This follows from the fact that the slice $A_n$ associated with a $\U(1)$ node is determined by $\text{balance} + 1$, where the balance of a node is given by the number of hypers minus twice the gauge group rank \cite{Gaiotto:2008ak}. Consequently, the only consistent choice is $a = b = c = 1$. The resulting quiver is

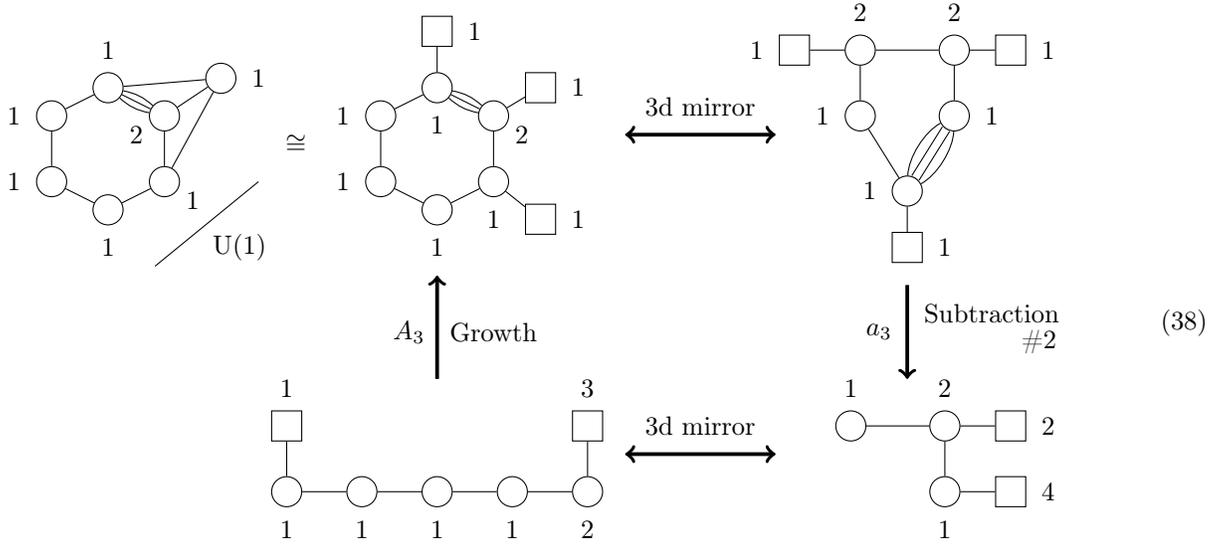
\begin{equation}
	\scalebox{0.9}{
    \begin{tikzpicture}[baseline=-0.25em,scale=0.5]
	\begin{pgfonlayer}{nodelayer}
		\node [style=none] (32) at (-4, 0) {Growth};
		\node [style=none] (33) at (-6.25, 0) {$A_3$};
		\node [style={gauge_gp}] (34) at (-7, 13.75) {};
		\node [style={gauge_gp}] (35) at (-7, 12) {};
		\node [style={gauge_gp}] (36) at (-5.5, 11.25) {};
		\node [style={gauge_gp}] (37) at (-4, 12) {};
		\node [style={gauge_gp}] (38) at (-4, 13.75) {};
		\node [style=none] (39) at (-5.5, 15.5) {1};
		\node [style=none] (40) at (-4.75, 13.25) {2};
		\node [style=none] (41) at (-3.25, 11.5) {1};
		\node [style=none] (42) at (-5.5, 10.25) {1};
		\node [style=none] (43) at (-8, 12) {1};
		\node [style=none] (44) at (-8, 13.75) {1};
		\node [style={gauge_gp}] (45) at (-5.5, 14.5) {};
		\node [style={gauge_gp}] (46) at (-2.5, 14.75) {};
		\node [style=none] (47) at (-1.5, 14.75) {1};
		\node [style=none] (61) at (-3.5, 10) {};
		\node [style=none] (62) at (-2, 12) {};
		\node [style=none] (63) at (-2, 10.5) {U(1)};
		\node [style=none] (64) at (-5.5, -1.25) {};
		\node [style=none] (65) at (-5.5, 1.5) {};
		\node [style={gauge_gp}] (66) at (-7, 5.75) {};
		\node [style={gauge_gp}] (67) at (-7, 4) {};
		\node [style={gauge_gp}] (68) at (-5.5, 3.25) {};
		\node [style={gauge_gp}] (69) at (-4, 4) {};
		\node [style={gauge_gp}] (70) at (-4, 5.75) {};
		\node [style=none] (71) at (-4.5, 8) {1};
		\node [style=none] (72) at (-3.25, 5.25) {2};
		\node [style=none] (73) at (-4, 3) {1};
		\node [style=none] (74) at (-5.5, 2.25) {1};
		\node [style=none] (75) at (-8, 4) {1};
		\node [style=none] (76) at (-8, 5.75) {1};
		\node [style={gauge_gp}] (77) at (-5.5, 6.5) {};
		\node [style={flavour_gp}] (78) at (-2.75, 6.5) {};
		\node [style={flavour_gp}] (79) at (-5.5, 8) {};
		\node [style={flavour_gp}] (80) at (-2.75, 3) {};
		\node [style=none] (81) at (-1.75, 6.5) {1};
		\node [style=none] (82) at (-1.75, 3) {1};
		\node [style=none] (83) at (-5.5, 9.25) {\rotatebox{270}{$\cong$}};
		\node [style={gauge_gp}] (86) at (6, 3.5) {};
		\node [style={gauge_gp}] (87) at (4.75, 5.5) {};
		\node [style={gauge_gp}] (88) at (7.25, 5.5) {};
		\node [style={flavour_gp}] (89) at (6, 2) {};
		\node [style=none] (90) at (5, 3.5) {1};
		\node [style=none] (91) at (7, 2) {1};
		\node [style=none] (92) at (3.75, 5.5) {1};
		\node [style=none] (93) at (8.25, 5.5) {1};
		\node [style={gauge_gp}] (94) at (4.75, 7.25) {};
		\node [style={gauge_gp}] (95) at (7.25, 7.25) {};
		\node [style={flavour_gp}] (96) at (3, 7.25) {};
		\node [style={flavour_gp}] (97) at (8.75, 7.25) {};
		\node [style=none] (98) at (2, 7.25) {1};
		\node [style=none] (99) at (9.75, 7.25) {1};
		\node [style=none] (100) at (7.25, 8.25) {2};
		\node [style=none] (101) at (4.75, 8.25) {2};
		\node [style=none] (102) at (6, 1) {};
		\node [style=none] (103) at (6, -1.5) {};
		\node [style=none] (104) at (8.25, 0.25) {Subtraction};
		\node [style={gauge_gp}] (105) at (7, -4.5) {};
		\node [style=none] (106) at (7, -5.5) {1};
		\node [style={gauge_gp}] (107) at (4.5, -2.75) {};
		\node [style={gauge_gp}] (108) at (7, -2.75) {};
		\node [style={flavour_gp}] (109) at (8.75, -2.75) {};
		\node [style=none] (110) at (4.5, -1.75) {1};
		\node [style=none] (111) at (9.75, -2.75) {2};
		\node [style=none] (112) at (7, -1.75) {2};
		\node [style={flavour_gp}] (113) at (8.75, -4.5) {};
		\node [style=none] (114) at (9.75, -4.5) {4};
		\node [style=none] (115) at (5.25, -0.25) {$a_3$};
		\node [style=none] (130) at (3, -3.5) {};
		\node [style=none] (131) at (-1, -3.5) {};
		\node [style=none] (132) at (1, -2.75) {3D mirror};
		\node [style=none] (133) at (3, 5) {};
		\node [style=none] (134) at (-1, 5) {};
		\node [style=none] (135) at (1, 5.75) {3D mirror};
		\node [style=none] (136) at (-5.5, 5.5) {1};
		\node [style=none] (137) at (9.5, -0.5) {\#2};
		\node [style={gauge_gp}] (138) at (-2.5, -3.75) {};
		\node [style={gauge_gp}] (139) at (-4, -3.75) {};
		\node [style={gauge_gp}] (140) at (-5.5, -3.75) {};
		\node [style={gauge_gp}] (141) at (-7, -3.75) {};
		\node [style={gauge_gp}] (142) at (-8.5, -3.75) {};
		\node [style={flavour_gp}] (143) at (-8.5, -2.25) {};
		\node [style={flavour_gp}] (144) at (-2.5, -2.25) {};
		\node [style=none] (145) at (-2.5, -1.25) {3};
		\node [style=none] (146) at (-8.5, -4.75) {1};
		\node [style=none] (147) at (-8.5, -1.25) {1};
		\node [style=none] (148) at (-7, -4.75) {1};
		\node [style=none] (149) at (-5.5, -4.75) {1};
		\node [style=none] (150) at (-4, -4.75) {1};
		\node [style=none] (151) at (-2.5, -4.75) {2};
	\end{pgfonlayer}
	\begin{pgfonlayer}{edgelayer}
		\draw (34) to (35);
		\draw (35) to (36);
		\draw (36) to (37);
		\draw (37) to (38);
		\draw (45) to (34);
		\draw [bend left=15] (45) to (38);
		\draw [bend right=15] (45) to (38);
		\draw (38) to (45);
		\draw (62.center) to (61.center);
		\draw [style=arrow] (64.center) to (65.center);
		\draw (45) to (46);
		\draw (46) to (38);
		\draw (46) to (37);
		\draw (66) to (67);
		\draw (67) to (68);
		\draw (68) to (69);
		\draw (69) to (70);
		\draw (77) to (66);
		\draw [bend left=15] (77) to (70);
		\draw [bend right=15] (77) to (70);
		\draw (70) to (77);
		\draw (78) to (70);
		\draw (79) to (77);
		\draw (80) to (69);
		\draw (87) to (86);
		\draw (89) to (86);
		\draw [style=none] (94) to (87);
		\draw [style=none] (95) to (88);
		\draw [style=none] (95) to (94);
		\draw [style=none] (96) to (94);
		\draw [style=none] (97) to (95);
		\draw [bend right, looseness=0.75] (86) to (88);
		\draw [bend right, looseness=0.75] (88) to (86);
		\draw [style=doubleline] (88) to (86);
		\draw [style=arrow] (102.center) to (103.center);
		\draw [style=none] (108) to (105);
		\draw [style=none] (108) to (107);
		\draw [style=none] (109) to (108);
		\draw (113) to (105);
		\draw [style=double arrow] (131.center) to (130.center);
		\draw [style=double arrow] (134.center) to (133.center);
		\draw (143) to (142);
		\draw (144) to (138);
		\draw (138) to (139);
		\draw (139) to (140);
		\draw (140) to (141);
		\draw (141) to (142);
	\end{pgfonlayer}
\end{tikzpicture}
	}
\end{equation}
which is the same quiver as in Eq.\ \eqref{eqn:nonabelianexample}.

\vspace{4mm}
\begin{minipage}{\columnwidth} 
\centering
\textbf{\underline{Step 3: A Consistency Check}}
\end{minipage}
\vspace{2mm}

\noindent
Once a mirror pair is constructed, one can perform a quick consistency check by computing all quivers that can be subtracted from the original quiver and comparing these to all possible decays of the corresponding newly obtained mirror quiver. The latter can be efficiently obtained using the \texttt{Mathematica} code described in Ref.\ \cite{Decay}. The reverse can also be done: we compute all quivers that can be subtracted from the mirror quiver and compare them to the decays of the original quiver. In all cases, the results must match. This provides a rapid and reliable method for verifying 3D mirror pairs without the need to explicitly compute Hilbert series.

\subsection{Bootstrapping mirror pairs}
In Section~\ref{subsec:abelian_mirror_pairs}, we provided a general prescription for abelian sunshine quivers and their 3D mirrors. Observing the emerging pattern, we attempted to apply the same reasoning in Sec.\ \ref{subsec:ab&non-ab} to non-Abelian sunshine quivers, but we were unable to obtain a general formula. Using the bootstrapping mirror-pair method introduced earlier, we make it easier to move forward. 

Starting again from our original quiver, we start bootstrapping by choosing the following quiver addition:

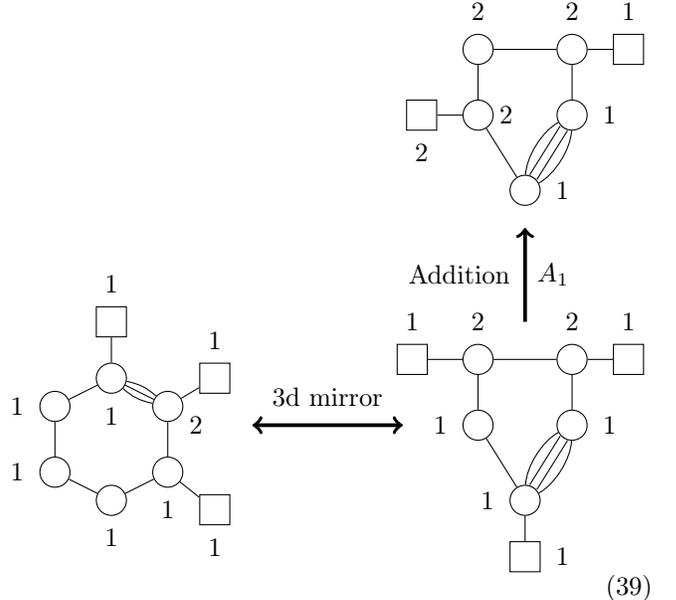
\begin{equation}
    \begin{tikzpicture}[baseline=-0.25em,scale=0.5]
	\begin{pgfonlayer}{nodelayer}
		\node [style={gauge_gp}] (66) at (-7.25, -3.5) {};
		\node [style={gauge_gp}] (67) at (-7.25, -5.25) {};
		\node [style={gauge_gp}] (68) at (-5.75, -6) {};
		\node [style={gauge_gp}] (69) at (-4.25, -5.25) {};
		\node [style={gauge_gp}] (70) at (-4.25, -3.5) {};
		\node [style=none] (71) at (-5.75, -0.25) {1};
		\node [style=none] (72) at (-3.5, -4) {2};
		\node [style=none] (73) at (-4.25, -6.25) {1};
		\node [style=none] (74) at (-5.75, -7) {1};
		\node [style=none] (75) at (-8.25, -5.25) {1};
		\node [style=none] (76) at (-8.25, -3.5) {1};
		\node [style={gauge_gp}] (77) at (-5.75, -2.75) {};
		\node [style={flavour_gp}] (78) at (-3, -2.75) {};
		\node [style={flavour_gp}] (79) at (-5.75, -1.25) {};
		\node [style={flavour_gp}] (80) at (-3, -6.25) {};
		\node [style=none] (81) at (-3, -1.75) {1};
		\node [style=none] (82) at (-3, -7.25) {1};
		\node [style={gauge_gp}] (86) at (5.25, -6) {};
		\node [style={gauge_gp}] (87) at (4, -4) {};
		\node [style={gauge_gp}] (88) at (6.5, -4) {};
		\node [style={flavour_gp}] (89) at (5.25, -7.5) {};
		\node [style=none] (90) at (4.25, -6) {1};
		\node [style=none] (91) at (6.25, -7.5) {1};
		\node [style=none] (92) at (3, -4) {1};
		\node [style=none] (93) at (7.5, -4) {1};
		\node [style={gauge_gp}] (94) at (4, -2.25) {};
		\node [style={gauge_gp}] (95) at (6.5, -2.25) {};
		\node [style={flavour_gp}] (96) at (2.25, -2.25) {};
		\node [style={flavour_gp}] (97) at (8, -2.25) {};
		\node [style=none] (98) at (2.25, -1.25) {1};
		\node [style=none] (99) at (8, -1.25) {1};
		\node [style=none] (100) at (6.5, -1.25) {2};
		\node [style=none] (101) at (4, -1.25) {2};
		\node [style=none] (133) at (2, -4) {};
		\node [style=none] (134) at (-2, -4) {};
		\node [style=none] (135) at (0, -3.25) {3D mirror};
		\node [style=none] (136) at (-5.75, -3.75) {1};
		\node [style=none] (137) at (5.25, -1.25) {};
		\node [style=none] (138) at (5.25, 1.25) {};
		\node [style=none] (139) at (6, 0) {$A_1$};
		\node [style={gauge_gp}] (140) at (5.25, 2.25) {};
		\node [style={gauge_gp}] (141) at (4, 4.25) {};
		\node [style={gauge_gp}] (142) at (6.5, 4.25) {};
		\node [style=none] (144) at (6.25, 2.25) {1};
		\node [style=none] (146) at (4.75, 4.25) {2};
		\node [style=none] (147) at (7.5, 4.25) {1};
		\node [style={gauge_gp}] (148) at (4, 6) {};
		\node [style={gauge_gp}] (149) at (6.5, 6) {};
		\node [style={flavour_gp}] (150) at (8, 6) {};
		\node [style=none] (151) at (8, 7) {1};
		\node [style=none] (152) at (6.5, 7) {2};
		\node [style=none] (153) at (4, 7) {2};
		\node [style={flavour_gp}] (154) at (2.5, 4.25) {};
		\node [style=none] (155) at (2.5, 3.25) {2};
		\node [style=none] (157) at (3.5, 0) {Addition};
	\end{pgfonlayer}
	\begin{pgfonlayer}{edgelayer}
		\draw (66) to (67);
		\draw (67) to (68);
		\draw (68) to (69);
		\draw (69) to (70);
		\draw (77) to (66);
		\draw [bend left=15] (77) to (70);
		\draw [bend right=15] (77) to (70);
		\draw (70) to (77);
		\draw (78) to (70);
		\draw (79) to (77);
		\draw (80) to (69);
		\draw (87) to (86);
		\draw (89) to (86);
		\draw [style=none] (94) to (87);
		\draw [style=none] (95) to (88);
		\draw [style=none] (95) to (94);
		\draw [style=none] (96) to (94);
		\draw [style=none] (97) to (95);
		\draw [bend right, looseness=0.75] (86) to (88);
		\draw [bend right, looseness=0.75] (88) to (86);
		\draw [style=doubleline] (88) to (86);
		\draw [style=double arrow] (134.center) to (133.center);
		\draw [style=arrow] (137.center) to (138.center);
		\draw (141) to (140);
		\draw [style=none] (148) to (141);
		\draw [style=none] (149) to (142);
		\draw [style=none] (149) to (148);
		\draw [style=none] (150) to (149);
		\draw [bend right, looseness=0.75] (140) to (142);
		\draw [bend right, looseness=0.75] (142) to (140);
		\draw [style=doubleline] (142) to (140);
		\draw (154) to (141);
	\end{pgfonlayer}
\end{tikzpicture}
\end{equation}

To find its 3D mirror, we repeat our algorithm by performing a decay or quiver subtraction that reduces it to a linear quiver. The simplest thing is to do a decay on the U(1) node at the bottom. This is an $A_5$ decay and leads to a linear quiver, as follows:
\begin{equation}
    \begin{tikzpicture}[baseline=-0.25em,scale=0.5]
	\begin{pgfonlayer}{nodelayer}
		\node [style=none] (133) at (2.25, -3.5) {};
		\node [style=none] (134) at (-2.25, -3.5) {};
		\node [style=none] (135) at (0, -2.75) {3D mirror};
		\node [style={gauge_gp}] (140) at (6.5, 1.75) {};
		\node [style={gauge_gp}] (141) at (5.25, 3.75) {};
		\node [style={gauge_gp}] (142) at (7.75, 3.75) {};
		\node [style=none] (144) at (7.5, 1.75) {1};
		\node [style=none] (146) at (6, 3.75) {2};
		\node [style=none] (147) at (8.75, 3.75) {1};
		\node [style={gauge_gp}] (148) at (5.25, 5.5) {};
		\node [style={gauge_gp}] (149) at (7.75, 5.5) {};
		\node [style={flavour_gp}] (150) at (9.25, 5.5) {};
		\node [style=none] (151) at (9.25, 6.5) {1};
		\node [style=none] (152) at (7.75, 6.5) {2};
		\node [style=none] (153) at (5.25, 6.5) {2};
		\node [style={flavour_gp}] (154) at (3.75, 3.75) {};
		\node [style=none] (155) at (3.75, 4.75) {2};
		\node [style=none] (160) at (6.5, -1.5) {};
		\node [style=none] (161) at (6.5, 1) {};
		\node [style=none] (162) at (7.25, 0) {$A_5$};
		\node [style=none] (163) at (5.25, 0) {Decay};
		\node [style={gauge_gp}] (169) at (-5.75, -4.5) {};
		\node [style={gauge_gp}] (170) at (-3.5, -4.5) {};
		\node [style={flavour_gp}] (171) at (-3.5, -2.5) {};
		\node [style=none] (172) at (-5.75, -5.5) {1};
		\node [style=none] (173) at (-3.5, -5.5) {2};
		\node [style=none] (174) at (-3.5, -1.5) {5};
		\node [style={gauge_gp}] (176) at (5.25, -4.5) {};
		\node [style={gauge_gp}] (177) at (7.75, -4.5) {};
		\node [style=none] (179) at (5.25, -5.5) {2};
		\node [style=none] (180) at (7.75, -5.5) {1};
		\node [style={gauge_gp}] (181) at (5.25, -2.5) {};
		\node [style={gauge_gp}] (182) at (7.75, -2.5) {};
		\node [style={flavour_gp}] (183) at (9.75, -2.5) {};
		\node [style=none] (184) at (9.75, -1.5) {1};
		\node [style=none] (185) at (7.75, -1.5) {2};
		\node [style=none] (186) at (5.25, -1.5) {2};
		\node [style={flavour_gp}] (187) at (3.25, -4.5) {};
		\node [style=none] (188) at (3.25, -5.5) {2};
	\end{pgfonlayer}
	\begin{pgfonlayer}{edgelayer}
		\draw [style=double arrow] (134.center) to (133.center);
		\draw (141) to (140);
		\draw [style=none] (148) to (141);
		\draw [style=none] (149) to (142);
		\draw [style=none] (149) to (148);
		\draw [style=none] (150) to (149);
		\draw [bend right, looseness=0.75] (140) to (142);
		\draw [bend right, looseness=0.75] (142) to (140);
		\draw [style=doubleline] (142) to (140);
		\draw (154) to (141);
		\draw [style=arrow] (161.center) to (160.center);
		\draw (171) to (170);
		\draw (169) to (170);
		\draw [style=none] (181) to (176);
		\draw [style=none] (182) to (177);
		\draw [style=none] (182) to (181);
		\draw [style=none] (183) to (182);
		\draw (187) to (176);
	\end{pgfonlayer}
\end{tikzpicture}
\end{equation}

Once we obtained the linear quiver, we easily derived its 3D mirror, shown on the bottom left. The resulting mirror is the linear $(1)-(2)-[5]$ theory. Having only two gauge nodes already imposes strong constraints on the possible quiver additions. From the $A_5$ slice that we decayed, we are now seeking an $a_5$-type quiver addition of the bottom left quiver. Although several such additions appear possible at first, the global symmetry restrictions quickly narrow the viable options. 

The quiver of interest on the top right exhibits an $\SU(4)\times \SU(2) \subset G_{\mathcal{H}}$ subgroup on the Higgs branch and an $\SU(3) \subset G_{\mathcal{C}}$ subgroup on the Coulomb branch. This implies that the $a_5$ addition on the $(1)-(2)-[5]$ quiver must reduce the flavor symmetry from $\SU(5)\rightarrow \SU(3)$ without introducing any additional non-Abelian Higgs branch flavor symmetry. After narrowing down the possibilities, we quickly find the correct $a_5$ to add is: 

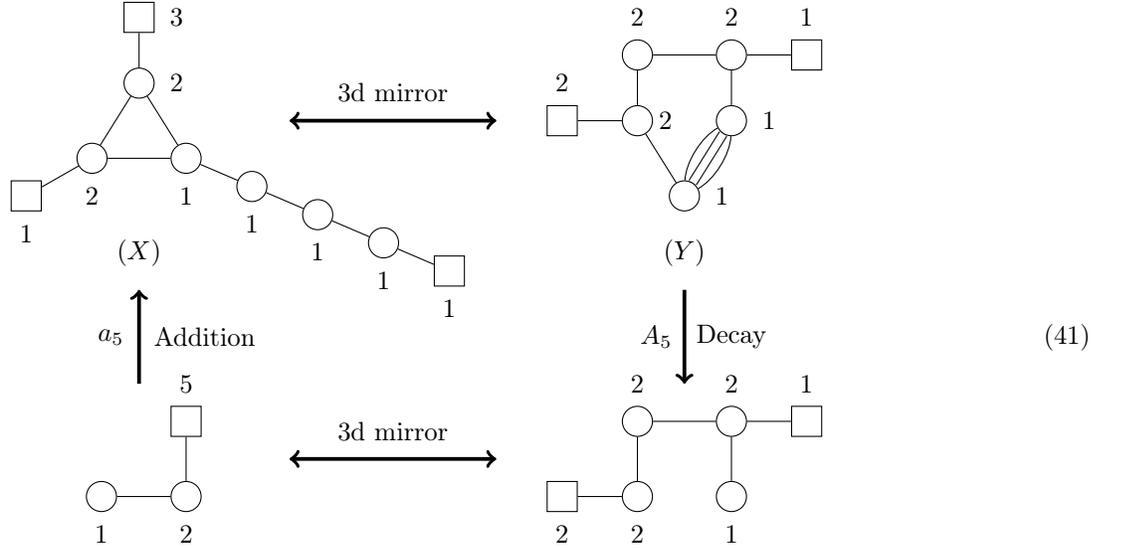
\begin{equation}
	\scalebox{0.93}{
\begin{tikzpicture}[baseline=-0.25em,scale=0.5]
	\begin{pgfonlayer}{nodelayer}
		\node [style=none] (133) at (1.75, -3.25) {};
		\node [style=none] (134) at (-1.75, -3.25) {};
		\node [style=none] (135) at (0, -2.5) {3D mirror};
		\node [style={gauge_gp}] (140) at (6.25, 3.75) {};
		\node [style={gauge_gp}] (141) at (5, 5.75) {};
		\node [style={gauge_gp}] (142) at (7.5, 5.75) {};
		\node [style=none] (144) at (7.25, 3.75) {1};
		\node [style=none] (146) at (5.75, 5.75) {2};
		\node [style=none] (147) at (8.5, 5.75) {1};
		\node [style={gauge_gp}] (148) at (5, 7.5) {};
		\node [style={gauge_gp}] (149) at (7.5, 7.5) {};
		\node [style={flavour_gp}] (150) at (9.5, 7.5) {};
		\node [style=none] (151) at (9.5, 8.5) {1};
		\node [style=none] (152) at (7.5, 8.5) {2};
		\node [style=none] (153) at (5, 8.5) {2};
		\node [style={flavour_gp}] (154) at (3, 5.75) {};
		\node [style=none] (155) at (3, 6.75) {2};
		\node [style=none] (157) at (1.75, 5.75) {};
		\node [style=none] (158) at (-1.75, 5.75) {};
		\node [style=none] (159) at (0, 6.5) {3D mirror};
		\node [style=none] (160) at (6.25, -1.25) {};
		\node [style=none] (161) at (6.25, 1.25) {};
		\node [style=none] (162) at (5.5, 0) {$A_5$};
		\node [style=none] (163) at (7.5, 0) {Decay};
		\node [style=none] (164) at (-4.75, -1.25) {};
		\node [style=none] (165) at (-4.75, 1.25) {};
		\node [style=none] (166) at (-5.5, 0) {$a_5$};
		\node [style=none] (167) at (-3, 0) {Addition};
		\node [style={gauge_gp}] (169) at (-5.75, -4.25) {};
		\node [style={gauge_gp}] (170) at (-3.5, -4.25) {};
		\node [style={flavour_gp}] (171) at (-3.5, -2.25) {};
		\node [style=none] (172) at (-6.25, -5.25) {1};
		\node [style=none] (173) at (-4, -5.25) {2};
		\node [style=none] (174) at (-3.5, -1.25) {5};
		\node [style={gauge_gp}] (176) at (5, -4.25) {};
		\node [style={gauge_gp}] (177) at (7.5, -4.25) {};
		\node [style=none] (179) at (5, -5.25) {2};
		\node [style=none] (180) at (7.5, -5.25) {1};
		\node [style={gauge_gp}] (181) at (5, -2.25) {};
		\node [style={gauge_gp}] (182) at (7.5, -2.25) {};
		\node [style={flavour_gp}] (183) at (9.5, -2.25) {};
		\node [style=none] (184) at (9.5, -1.25) {1};
		\node [style=none] (185) at (7.5, -1.25) {2};
		\node [style=none] (186) at (5, -1.25) {2};
		\node [style={flavour_gp}] (187) at (3, -4.25) {};
		\node [style=none] (188) at (3, -5.25) {2};
		\node [style={gauge_gp}] (189) at (-6, 4.75) {};
		\node [style={gauge_gp}] (190) at (-3.5, 4.75) {};
		\node [style={flavour_gp}] (191) at (-7.75, 3.75) {};
		\node [style={gauge_gp}] (192) at (-1.75, 4) {};
		\node [style={gauge_gp}] (193) at (0, 3.25) {};
		\node [style={flavour_gp}] (194) at (3.75, 1.75) {};
		\node [style=none] (195) at (-3.75, 6.75) {2};
		\node [style=none] (196) at (-3.75, 8.5) {3};
		\node [style=none] (197) at (-7.75, 2.75) {1};
		\node [style=none] (198) at (-6, 3.75) {2};
		\node [style=none] (199) at (-3.5, 3.75) {1};
		\node [style=none] (200) at (-1.75, 3) {1};
		\node [style=none] (201) at (0, 2.25) {1};
		\node [style=none] (202) at (3.75, 0.75) {1};
		\node [style={flavour_gp}] (203) at (-4.75, 8.5) {};
		\node [style={gauge_gp}] (204) at (-4.75, 6.75) {};
		\node [style={gauge_gp}] (205) at (2, 2.5) {};
		\node [style=none] (206) at (2, 1.5) {1};
		\node [style=none] (207) at (-4.75, 2.25) {($X$)};
		\node [style=none] (208) at (6.25, 2.25) {($Y$)};
	\end{pgfonlayer}
	\begin{pgfonlayer}{edgelayer}
		\draw [style=double arrow] (134.center) to (133.center);
		\draw (141) to (140);
		\draw [style=none] (148) to (141);
		\draw [style=none] (149) to (142);
		\draw [style=none] (149) to (148);
		\draw [style=none] (150) to (149);
		\draw [bend right, looseness=0.75] (140) to (142);
		\draw [bend right, looseness=0.75] (142) to (140);
		\draw [style=doubleline] (142) to (140);
		\draw (154) to (141);
		\draw [style=double arrow] (158.center) to (157.center);
		\draw [style=arrow] (161.center) to (160.center);
		\draw [style=arrow] (164.center) to (165.center);
		\draw (171) to (170);
		\draw (169) to (170);
		\draw [style=none] (181) to (176);
		\draw [style=none] (182) to (177);
		\draw [style=none] (182) to (181);
		\draw [style=none] (183) to (182);
		\draw (187) to (176);
		\draw (190) to (189);
		\draw (191) to (189);
		\draw (193) to (192);
		\draw (192) to (190);
		\draw (203) to (204);
		\draw (204) to (190);
		\draw (204) to (189);
		\draw (205) to (193);
		\draw (205) to (194);
	\end{pgfonlayer}
\end{tikzpicture}
	}
\end{equation}

\begin{eqnarray*}
\HS_\mathcal{C}^X(t) =&& \HS_\mathcal{H}^Y(t) \\
=&& 1 + 20t^2 + 182t^4 + 24t^5 + 1143t^6 + 368t^7 + \\
&& 5663t^8 + 3000t^9 + 23769t^{10} + \mathcal{O}(t^{11})
\end{eqnarray*}
\begin{eqnarray*}
\HS_\mathcal{H}^X(t) =&& \HS_\mathcal{C}^Y(t) \\
=&& 1 + 11t^2 + 8t^3 + 81t^4 + 112t^5 + 490t^6 + \\
&& 4822t^9 + 12124t^{10} + \mathcal{O}(t^{11})
\end{eqnarray*}

This demonstrates the power of the bootstrapping algorithm in generating new mirror pairs on top of old mirror pairs. It does not rely on brane configurations and can operate on any quiver gauge theory.  

This is a quick proof-of-concept exercise, and we leave the large systematic production of new mirror pairs for the next installment.

\section{Conclusion and Outlook}
In this paper, we developed an algorithm capable of bootstrapping new 3D mirror pairs. To achieve this, we completed the set of four quiver Higgsing algorithms by introducing the \textbf{Growth and Fusion} algorithm. The strength of the four quiver algorithms lies in their ability to determine 3D mirror pairs without relying on brane constructions, which are rarely applicable beyond linear quivers. Prior to this framework, the standard approach was to conjecture the 3D mirror based solely on the global symmetry structure---a method that is both inefficient and almost impossible in practice. The new bootstrapping framework overcomes this limitation and can be applied to any quiver gauge theory composed of unitary gauge groups, enabling one to (a) determine the 3D mirror of a given unitary quiver, or (b) construct new mirror pairs from a known 3D mirror pair. To demonstrate the utility of this approach, we examined a new class of 3D mirror pairs associated with circular quivers. We began by taking inspiration from Abelian 3D mirror pairs involving circular quivers---dubbed as sunshine quivers---and used these as motivation to construct non-Abelian 3D mirror pairs. We then used our quartet of quiver algorithms to verify these results as well as use them as a basis to generate new mirror pairs.

\subsubsection{Outlook}

This paper is intended as a first step toward discovering all of the landscape of 3D mirror pairs in which both theories are Lagrangian/quiver gauge theories. 
\begin{itemize}
    \item \underline{Computer code:} Thus far, the bootstrapping technique has been carried out with pen and paper, but it can be implemented on a computer in a straightforward manner. One simply needs to encode quivers as adjacency matrices and generate all possible growths and quiver additions. For quiver addition, the only requirement is that the resulting quiver must have nodes with the same balances, which reduces the number of outputs to a finite set. Coding the growth algorithm is slightly more involved, as it requires imposing additional constraints---such as insisting on the correct non-Abelian global symmetries---to similarly restrict the space of solutions. Doable but requires a bit more coding finesse. 
    
    \item \underline{Orthosymplectic quivers:} To complete the framework, it is necessary to extend the algorithm to include $\SU$, $\SO$, and $\Sp$ gauge groups within all four quiver operations. For $\SU$ nodes, such an extension appears to be within reach. For orthosymplectic gauge groups ($\SO$ and $\Sp$), the situation is considerably more challenging. Partial progress has been made for the orthosymplectic decay and fission algorithm in Ref.\ \cite{orthosymplecticdecay}, but the remaining algorithms will require deeper investigation.
    
    \item \underline{Other considerations:} When quivers involve features such as adjoint hypermultiplets, additional subtleties arise in the quiver subtraction algorithm in relation to Higgs branch Higgsings, as discussed in Ref.\ \cite{Bennett:2024loi}. Fortunately, it is not necessary to touch the gauge nodes with adjoint hypers when performing our algorithm. As demonstrated in the previous section, identifying a suitable subset of possible subtractions or decays already provides sufficient information to construct the mirror. Hence, in practice, such complications may be circumvented. Furthermore, more obscure kinds of matter content (beyond bifundamental hypers) were encountered in places like Refs.\ \cite{o5,exo2,exo3,product,Bourget:2024asp,Liu} as well as discrete factors of gauge groups such as Refs.\ \cite{Nawata:2023rdx,Harding:2025vov}. The presence of these objects can allow for more possibilities, but really strains the definition of what we call a Lagrangian/quiver gauge theory. For instance, many of these theories have nonsimply laced edges, which we do not know how to interpret as a Lagrangian theory, nor can we study their Higgs branch. 

\item \underline{Other mirror programs}: In this paper, we looked at the quiver algorithm in generating new 3D mirror pairs. But we also used a gluing technique for circular quivers. There are other techniques in the literature, such as the program in Refs.\ \cite{Dey:2014tka, Dey:2020hfe,Dey:2021rxw, Dey:2023vrt}. It will be interesting to compare our results and see if we can use mirror pairs in those papers as a starting point for generating new mirror pairs as well. Furthermore, there is a program for a 4D analogue of 3D mirror symmetry \cite{Hwang:2020wpd,Hwang:2021ulb}, and it will be interesting if any of our results may have implications for understanding 4D theories. 

\end{itemize}

\begin{acknowledgments}
Parts of this work first appeared in the master's dissertations of L.J., J.E.Z.O.\ and R.S.\ at the University of Oxford. Z.Z.\ was supported by the ERC Consolidator Grant \#864828 ``Algebraic Foundations of Supersymmetric Quantum Field Theory'' (SCFTAlg) and SIMIS starting grant.
All authors contributed equally to this work. 
\end{acknowledgments}

\appendix
\section{Additional background} 
\label{appx:bg}
\subsection{Coulomb Branch Global Symmetry}
The global symmetry of the Coulomb branch (denoted $G_{\mathcal{C}}$) can be deduced from the quiver's structure, through the concept of gauge node \emph{balance}.

The balance $b\in\mathbb{Z}$ of a $\U(k)$ gauge node is defined as the total number of hypermultiplets connected to it minus twice its rank $k$ \cite{Gaiotto:2008ak}. The node is classified as: balanced if $b = 0$; overbalanced if $b > 0$; and underbalanced if $b < 0$.
Each overbalanced node contributes a factor of $\U(1)$, and each connected component of balanced nodes contributes a group corresponding to the Dynkin diagram formed. Flavor nodes do not have a balance value as they do not directly contribute to the global symmetry of the Coulomb branch. 

As an example, consider the following quiver:
\begin{equation}
\label{eqn:bal_ex1}
    \begin{tikzpicture}[baseline=-0.25em,scale=0.5]
        \begin{pgfonlayer}{nodelayer}
		\node [style={flavour_gp}] (0) at (2, 0) {};
		\node [style={gauge_gp}] (1) at (0, 0) {};
		\node [style=none] (3) at (2, -1) {3};
		\node [style={gauge_gp}] (4) at (-2, 0) {};
		\node [style=none] (6) at (0, -1) {2};
		\node [style=none] (7) at (-2, -1) {1};
	\end{pgfonlayer}
	\begin{pgfonlayer}{edgelayer}
		\draw (0) to (1);
		\draw (4) to (1);
	\end{pgfonlayer}
\end{tikzpicture}
\end{equation}
in which both gauge nodes are balanced ($b=0$): the $\U(1)$ node has two edges, and the $\U(2)$ node has four edges (one from $\U(1)$ and three from the rank-3 flavor node). Two connected nodes correspond to the $A_2$ Dynkin diagram, and hence the theory has Coulomb branch global symmetry $G_{\mathcal{C}}=\SU(3)$. 

Now, consider the following two quivers:
\begin{equation}
\label{eqn:bal_ex2}
    \begin{tikzpicture}[baseline=-0.25em,scale=0.5]
    \begin{pgfonlayer}{nodelayer}
		\node [style={flavour_gp}] (0) at (0, 2) {};
		\node [style={gauge_gp}] (1) at (2, 0) {};
		\node [style=none] (2) at (1, 2) {2};
		\node [style=none] (3) at (2, -1) {1};
		\node [style={gauge_gp}] (4) at (0, 0) {};
		\node [style={gauge_gp}] (5) at (-2, 0) {};
		\node [style=none] (6) at (0, -1) {2};
		\node [style=none] (7) at (-2, -1) {1};
	\end{pgfonlayer}
	\begin{pgfonlayer}{edgelayer}
		\draw (5) to (4);
		\draw (4) to (1);
		\draw (0) to (4);
	\end{pgfonlayer}
\end{tikzpicture}
\end{equation}
\begin{equation}
\label{eqn:bal_ex3}
    \begin{tikzpicture}[baseline=-0.25em,scale=0.5]
	\begin{pgfonlayer}{nodelayer}
		\node [style={flavour_gp}] (0) at (3, 0) {};
		\node [style={gauge_gp}] (1) at (1, 0) {};
		\node [style=none] (2) at (3, -1) {4};
		\node [style=none] (3) at (1, -1) {3};
		\node [style={gauge_gp}] (4) at (-1, 0) {};
		\node [style={gauge_gp}] (5) at (-3, 0) {};
		\node [style=none] (6) at (-1, -1) {2};
		\node [style=none] (7) at (-3, -1) {1};
	\end{pgfonlayer}
	\begin{pgfonlayer}{edgelayer}
		\draw (0) to (1);
		\draw (5) to (4);
		\draw (4) to (1);
	\end{pgfonlayer}
\end{tikzpicture}
\end{equation}

In both \eqref{eqn:bal_ex2} and \eqref{eqn:bal_ex3}, all gauge nodes are balanced ($b=0$). A linear chain of three nodes corresponds to the Dynkin diagram $A_3$, and thus they share the Coulomb branch global symmetry $G_\mathcal{C}=\SU(4)$.

Now, consider the quiver corresponding to supersymmetric QED with three massless electrons:
\begin{equation}
\label{eqn:bal_ex4}
    \begin{tikzpicture}[baseline=-0.25em,scale=0.5]
    \begin{pgfonlayer}{nodelayer}
		\node [style={flavour_gp}] (0) at (1, 0) {};
		\node [style={gauge_gp}] (4) at (-1, 0) {};
		\node [style=none] (6) at (1, -1) {3};
		\node [style=none] (7) at (-1, -1) {1};
	\end{pgfonlayer}
	\begin{pgfonlayer}{edgelayer}
		\draw (0) to (4);
	\end{pgfonlayer}
\end{tikzpicture}
\end{equation}

The $\U(1)$ gauge node in \eqref{eqn:bal_ex4} is overbalanced ($b=1$), corresponding to Coulomb branch global symmetry $G_\mathcal{C}=\U(1)$.

\subsection{Hilbert series}
\label{appx:hs}
The Hilbert series is a generating function that counts gauge-invariant chiral operators according to their conformal dimension. For a theory with moduli space $\mathcal{M}$ (associated with either the Higgs branch $\mathcal{H}$ or the Coulomb branch $\mathcal{C}$), the Hilbert series is defined as
\begin{equation}
    \HS(t) = \sum_{d=0}^{\infty} c_d t^d
\end{equation}
Where $c_d$ is the number of independent chiral operators with dimension $d$, and $t$ is a counting variable or fugacity\footnote{This terminology originates from statistical mechanics, where $\log(t)$ can be identified as the chemical potential to conserved charge.} associated with the $R$ symmetry.

The Higgs branch Hilbert series is typically computed using the Molien-Weyl formula \cite{hanany2016quiver}, while the Monopole formula \cite{monopole} is used for the Coulomb branch.

\newpage

\begin{widetext}
\begin{figure*}
\begin{tikzpicture}[baseline=-0.25em,scale=0.5]
	\begin{pgfonlayer}{nodelayer}
		\node [style=none] (24) at (6.75, 1.75) {$\mathrm{I}^Y_{[n',l',p']}$};
		\node [style={gauge_gp}] (26) at (4.75, 0) {};
		\node [style={gauge_gp}] (27) at (7.75, 0) {};
		\node [style=none] (28) at (6.25, 0) {\textcolor{red}{$\vdots$}};
		\node [style={blue_flavour}] (30) at (9.25, 0) {};
		\node [style={purple_flavour}] (32) at (3.25, 0) {};
		\node [style={gauge_gp}] (90) at (4.75, 0) {};
		\node [style={gauge_gp}] (91) at (7.75, 0) {};
		\node [style=none] (92) at (-7, 2.5) {I$^X_{[n',l',p']}$};
		\node [style={flavour_gp}] (93) at (-11, 0) {};
		\node [style={gauge_gp}] (94) at (-8.75, -1.25) {};
		\node [style={gauge_gp}] (95) at (-8.75, 1.25) {};
		\node [style={gauge_gp}] (96) at (-5.75, 1.25) {};
		\node [style={gauge_gp}] (97) at (-5.75, -1.25) {};
		\node [style={gauge_gp}] (98) at (-9.5, 0) {};
		\node [style={gauge_gp}] (99) at (-4.75, 0) {};
		\node [style=none] (100) at (-8, 1.25) {};
		\node [style=none] (101) at (-6.5, 1.25) {};
		\node [style=none] (102) at (-7.5, 1.25) {\textcolor{blue}{$\cdots$}};
		\node [style=none] (103) at (-8, -1.25) {};
		\node [style=none] (104) at (-6.5, -1.25) {};
		\node [style=none] (105) at (-7.25, -1.25) {\textcolor{rgb,255: red,128; green,0; blue,128}{$\cdots$}};
		\node [style=none] (155) at (5.5, -2.75) {\textcolor{blue}{$l'-1$}};
		\node [style=none] (156) at (2, -2.75) {\textcolor{rgb,255: red,128; green,0; blue,128}{$n'-l'+1$}};
		\node [style=none] (157) at (8.25, -2.75) {\textcolor{red}{$p'-1$}};
		\node [style=none] (158) at (10.75, -2.75) {\textcolor{orange}{$1$}};
		\node [style=none] (159) at (-1.75, 0) {};
		\node [style=none] (160) at (1.75, 0) {};
		\node [style=none] (161) at (0, 0.75) {3D mirror};
		\node [style=none] (190) at (-3.25, -1) {};
		\node [style={flavour_gp}] (191) at (-3.25, -4) {};
		\node [style=none] (192) at (-3.25, -1.25) {\textcolor{red}{$\vdots$}};
		\node [style={gauge_gp}] (193) at (-3.25, 0) {};
		\node [style={gauge_gp}] (194) at (-3.25, -2.75) {};
		\node [style=none] (195) at (-3.25, -2) {};
	\end{pgfonlayer}
	\begin{pgfonlayer}{edgelayer}
		\draw [style=orange, bend left=45] (26) to (27);
		\draw [style=red, bend left, looseness=0.75] (26) to (27);
		\draw [style=red, bend left=45] (27) to (26);
		\draw (30) to (91);
		\draw (32) to (90);
		\draw [style=blue] (100.center) to (95);
		\draw [style=blue] (95) to (98);
		\draw [style=blue] (101.center) to (96);
		\draw [style=purple] (98) to (94);
		\draw [style=purple] (99) to (97);
		\draw [style=purple] (94) to (103.center);
		\draw [style=purple] (104.center) to (97);
		\draw [style=blue] (96) to (99);
		\draw [style=orange] (93) to (98);
		\draw [style=double arrow] (159.center) to (160.center);
		\draw [style=red] (190.center) to (193);
		\draw [style=red] (194) to (191);
		\draw [style=red] (195.center) to (194);
		\draw [style=red] (193) to (99);
	\end{pgfonlayer}
\end{tikzpicture}
\caption{The family of 3D mirror pairs denoted as Family I in \cite{Dey:2020hfe}.}
\label{fig:famI}
\end{figure*}
\end{widetext}

\subsection{Quiver algorithms}
\label{appx:algo}
Here, we give a quick summary of the other three quiver algorithms: quiver subtraction, quiver addition, and decay and fission. 

\begin{enumerate}[label=(\roman*)]

\item Quiver Subtraction, introduced and detailed in Ref.\ \cite{Bourget:2019aer}, involves taking a quiver and subtracting an \emph{elementary slice}. The elementary slice is another 3D $\mathcal{N}=4$ quiver gauge theory whose Coulomb branch is an \emph{isolated symplectic singularity} (the singular locus is a point). A list of such singularities is known to be incomplete in the mathematics literature, and the most up-to-date list of singularities that are Coulomb branches of quivers made of only unitary gauge groups is given in Ref.\ \cite{Bourget:2025wsp}. The algorithm simply subtracts the elementary slice quiver from the original quiver, node by node, and then adds flavors to the remaining gauge groups to ensure balance per node remains unchanged. 

\item
The quiver addition algorithm, introduced in Ref.\ \cite{Bourget:2021siw}, is the reversal of the subtraction algorithm. Starting with a quiver, the algorithm adds elementary slice quivers to it, node by node. Such a process is allowed as long as all gauge groups before and after such an addition have the same balance. Note that rebalancing using flavors is not part of quiver addition, and hence not all quivers can be added. 

\item
Decay and fission \cite{Decay, Fission} is a much simpler algorithm, and does not involve subtracting or adding elementary slice quivers. Instead, one takes a quiver, reducing ranks of gauge groups one by one, until one arrives at a quiver where all gauge groups are once again good. Each such process that can be found is a valid decay. A fission is where one can split a quiver into two quivers in a way that all the gauge groups in the fission product are good.

\end{enumerate}

\newpage

\section{A consistency check}
\label{appx:check}

\subsection{Abelian: Families I, II and IIa}
In this appendix, we demonstrate that three Abelian families discussed in Ref.\ \cite{Dey:2020hfe} are simply specific cases of our aforementioned sunshine quivers, and the 3D mirror pairs found align with our findings.
Colored numbers denote the number of edges or the number of flavors of a colored square node (uncolored implies one flavor). All indices from Ref.\ \cite{Dey:2020hfe} are primed to avoid confusion.

\vspace{-1em}
\subsubsection{Family I}
\vspace{-1em}

$\mathrm{I}^X_{[n',l',p']}$ is equivalent to an $n'$-sized loop with two rays of length \textcolor{red}{$p'-1$} and \textcolor{orange}{$1$} respectively, separated by \textcolor{blue}{$l'-1$} edges, as seen in Fig.\ \ref{fig:famI}.

This is an exceptional case, since there are only two rays (see Appendix \ref{appx:no-ray}). Hence, $\mathrm{I}^Y_{n',l',p'}$ theory only has one bond, of multiplicity $p'=\textcolor{red}{p'-1}+\textcolor{orange}{1}$.

\vspace{-1em}
\subsubsection{Family II}
\vspace{-1em}

$X$ quivers of Family $\mathrm{II}$ are described in \cite{Dey:2020hfe} as `two closed loops attached to a linear quiver tail'. By alternate choice of ungauging scheme, $\mathrm{II}_{[n',l',l'_1,l'_2,p'_1,p'_2]}$ is equivalent to a loop of size $n'+l'_1-l'_2+p'_2$, with three rays of lengths \textcolor{orange}{$l'_2-l'$}, \textcolor[HTML]{800080}{$l'-l'_1$} and \textcolor{blue}{$p$}\textcolor{cyan}{$'_1$} respectively, separated by \textcolor{red}{$p'_2$} and \textcolor[HTML]{00c800}{$l'_1$} edges, as seen in Fig.\ \ref{fig:famII}. The node of $\mathrm{II}^X$ ungauged to obtain sunshine form from two-loop form (as seen in Ref.\ \cite{Dey:2020hfe}) is denoted with ``$\mathrm{x}$''.

\vspace{-1em}
\subsubsection{Family IIa}
\vspace{-1em}

\noindent
This family, which we call IIa, is not formally defined in Ref.\ \cite{Dey:2020hfe}, appearing only during construction of Family $\mathrm{II}$; simply denoted by $X^{(p'_1-1,p'_2-2)}$ and $Y^{(p'_1-1,p'_2-2)}$ in Fig.\ 22 of Ref.\ \cite{Dey:2020hfe}. $\mathrm{IIa}^X_{[n',l',l'_1,l'_2,p'_1,p'_2]}$ is equivalent to a loop of size $n'$, with four rays of lengths \textcolor{blue}{$p'_1-1$}, \textcolor{red}{$p'_2-1$}, \textcolor{cyan}{$1$} and \textcolor[HTML]{a8a800}{$1$}, separated by \textcolor[HTML]{00c800}{$l'_1$}, \textcolor[HTML]{800080}{$l'-l'_1$} and \textcolor{orange}{$l'_2-l'$} edges, as seen in Fig.\ \ref{fig:famIIa}.

\begin{figure*}
    \begin{tikzpicture}[baseline=-0.25em,scale=0.5]
	\begin{pgfonlayer}{nodelayer}
		\node [style=none] (0) at (-6.25, 12.75) {$\mathrm{II}^X_{[n',l',l'_1,l'_2,p'_1,p'_2]}$};
		\node [style=none] (1) at (-9, 2.5) {\textcolor{blue}{$\cdots$}};
		\node [style=none] (3) at (-5.5, 9.25) {\textcolor{orange}{$\vdots$}};
		\node [style=none] (4) at (3.75, 7) {$\cong$};
		\node [style=none] (5) at (2, 4.25) {};
		\node [style=none] (6) at (0.5, 1.5) {};
		\node [style=none] (7) at (2, 2) {U(1)};
		\node [style={flavour_gp}] (8) at (-7.75, 11.5) {};
		\node [style={gauge_gp}] (9) at (-7.75, 10) {};
		\node [style={gauge_gp}] (10) at (-4.75, 7) {};
		\node [style={gauge_gp}] (11) at (-7.75, 4) {};
		\node [style={gauge_gp}] (12) at (-10.75, 7) {};
		\node [style={gauge_gp}] (13) at (-6.25, 7) {};
		\node [style={gauge_gp}] (14) at (-9.25, 7) {};
		\node [style=none] (15) at (-8.25, 7) {};
		\node [style=none] (16) at (-7.25, 7) {};
		\node [style=none] (17) at (-7.75, 7) {\textcolor{red}{\footnotesize $\cdots$}};
		\node [style={gauge_gp}] (18) at (-6.25, 10) {};
		\node [style={gauge_gp}] (19) at (-4.75, 8.5) {};
		\node [style=none] (20) at (-5.75, 9.5) {};
		\node [style=none] (21) at (-5.25, 9) {};
		\node [style=none] (22) at (-5.25, 5) {};
		\node [style=none] (23) at (-5.75, 4.5) {};
		\node [style={gauge_gp}] (24) at (-4.75, 5.5) {};
		\node [style={gauge_gp}] (25) at (-6.25, 4) {};
		\node [style={gauge_gp}] (26) at (-9.25, 10) {};
		\node [style={gauge_gp}] (27) at (-10.75, 8.5) {};
		\node [style={gauge_gp}] (28) at (-10.75, 5.5) {};
		\node [style={gauge_gp}] (29) at (-9.25, 4) {};
		\node [style=none] (30) at (-10.25, 5) {};
		\node [style=none] (31) at (-9.75, 4.5) {};
		\node [style=none] (32) at (-9.75, 9.5) {};
		\node [style=none] (33) at (-10.25, 9) {};
		\node [style={flavour_gp}] (34) at (-11.75, 2.5) {};
		\node [style=none] (35) at (-8.5, 2.5) {};
		\node [style=none] (37) at (-9.5, 2.5) {};
		\node [style=none] (38) at (-10, 4.75) {\textcolor{rgb,255: red,0; green,200; blue,0}{$\vdots$}};
		\node [style=none] (39) at (-10, 9.25) {\textcolor{rgb,255: red,128; green,0; blue,128}{$\vdots$}};
		\node [style=none] (40) at (-5.5, 4.75) {\textcolor{magenta}{$\vdots$}};
		\node [style=none] (41) at (0, 10) {\textcolor{red}{$\cdots$}};
		\node [style=none] (42) at (-1.5, 9) {\textcolor{rgb,255: red,128; green,0; blue,}{$\vdots$}};
		\node [style={gauge_gp}] (44) at (0, 5.75) {};
		\node [style={gauge_gp}] (45) at (2.25, 9.25) {};
		\node [style=none] (47) at (0, 4.25) {};
		\node [style=none] (48) at (0, 5) {};
		\node [style=none] (49) at (0, 4.75) {\textcolor{blue}{\footnotesize \vdots}};
		\node [style=none] (50) at (-1.75, 9) {};
		\node [style=none] (51) at (-1.25, 8.5) {};
		\node [style=none] (52) at (1.25, 8.5) {};
		\node [style=none] (53) at (1.75, 9) {};
		\node [style={gauge_gp}] (54) at (-2.5, 8.25) {};
		\node [style={gauge_gp}] (55) at (-1, 4.25) {};
		\node [style={gauge_gp}] (56) at (1, 4.25) {};
		\node [style={gauge_gp}] (57) at (2.5, 8.25) {};
		\node [style=none] (58) at (1.5, 5.5) {};
		\node [style=none] (59) at (2, 6.75) {};
		\node [style=none] (60) at (-2, 6.75) {};
		\node [style=none] (61) at (-1.5, 5.5) {};
		\node [style=none] (62) at (-1.75, 6.25) {\textcolor{rgb,255: red,0; green,200; blue,0}{\footnotesize $\vdots$}};
		\node [style=none] (63) at (1.75, 6.25) {\textcolor{magenta}{$\vdots$}};
		\node [style=none] (66) at (-0.75, 10) {};
		\node [style={gauge_gp}] (67) at (-1.75, 10) {};
		\node [style={gauge_gp}] (68) at (1.75, 10) {};
		\node [style=none] (70) at (0.75, 10) {};
		\node [style=none] (86) at (-3.5, 7) {$\cong$};
		\node [style=none] (87) at (8.25, 9.5) {\textcolor{red}{$\cdots$}};
		\node [style={gauge_gp}] (88) at (5.75, 9.5) {};
		\node [style={gauge_gp}] (89) at (10.75, 9.5) {};
		\node [style={gauge_gp}] (90) at (8.25, 4.25) {};
		\node [style=none] (91) at (10, 3) {};
		\node [style=none] (92) at (9, 3) {};
		\node [style=none] (93) at (9.5, 3) {\textcolor{blue}{\footnotesize $\cdots$}};
		\node [style={gauge_gp}] (98) at (6.25, 8.25) {};
		\node [style={gauge_gp}] (99) at (7.5, 5.5) {};
		\node [style={gauge_gp}] (100) at (9, 5.5) {};
		\node [style={gauge_gp}] (101) at (10.25, 8.25) {};
		\node [style=none] (102) at (9.5, 6.5) {};
		\node [style=none] (103) at (10, 7.25) {};
		\node [style=none] (104) at (6.5, 7.25) {};
		\node [style=none] (105) at (7, 6.5) {};
		\node [style=none] (106) at (5.25, 7.25) {\textcolor{rgb,255: red,128; green,0; blue,128}{\footnotesize $\vdots$}};
		\node [style=none] (107) at (11.25, 7.25) {\textcolor{orange}{$\vdots$}};
		\node [style=none] (108) at (7, 9.5) {};
		\node [style=none] (109) at (9.5, 9.5) {};
		\node [style=none] (117) at (6.75, 7) {\textcolor{rgb,255: red,0; green,200; blue,0}{$\vdots$}};
		\node [style={flavour_gp}] (118) at (5.25, 4.25) {};
		\node [style={flavour_gp}] (120) at (10.75, 1.5) {};
		\node [style={gauge_gp}] (121) at (6.75, -7.5) {};
		\node [style={gauge_gp}] (122) at (9.75, -7.5) {};
		\node [style={gauge_gp}] (123) at (8.25, -4.5) {};
		\node [style={gauge_gp}] (125) at (10.75, 3) {};
		\node [style=none] (126) at (9.75, 7) {\textcolor{magenta}{$\vdots$}};
		\node [style={red_flavour}] (127) at (8.25, -2.75) {};
		\node [style={pink_flavour}] (129) at (11, -8.75) {};
		\node [style={green_flavour}] (130) at (5.5, -8.75) {};
		\node [style=none, scale=0.75] (133) at (7.5, -6) {\textcolor{rgb,255: red,128; green,0; blue,128}{\footnotesize $\cdots$}};
		\node [style=none, scale=0.75] (134) at (9, -6) {\textcolor{orange}{\footnotesize $\cdots$}};
		\node [style=none, scale=0.75] (135) at (8.25, -8) {\textcolor{blue}{\footnotesize $\vdots$}};
		\node [style={gauge_gp}] (161) at (-4.75, 7) {};
		\node [style={gauge_gp}] (162) at (-10.75, 7) {};
		\node [style={gauge_gp}] (163) at (-7.75, 4) {};
		\node [style={gauge_gp}] (164) at (-7.75, 10) {};
		\node [style={gauge_gp}] (165) at (0, 7.25) {};
		\node [style={gauge_gp}] (166) at (3.25, 10) {};
		\node [style={gauge_gp}] (167) at (-3.25, 10) {};
		\node [style=none] (168) at (0, 7.25) {x};
		\node [style={gauge_gp}] (169) at (0, 2) {};
		\node [style={gauge_gp}] (170) at (5.75, 9.5) {};
		\node [style={gauge_gp}] (171) at (10.75, 9.5) {};
		\node [style={gauge_gp}] (173) at (8.25, 4.25) {};
		\node [style={gauge_gp}] (178) at (9.75, -7.5) {};
		\node [style={gauge_gp}] (179) at (6.75, -7.5) {};
		\node [style={gauge_gp}] (180) at (8.25, -4.5) {};
		\node [style={gauge_gp}] (181) at (5.25, 8.25) {};
		\node [style=none] (182) at (5.25, 7.5) {};
		\node [style={gauge_gp}] (183) at (5.25, 5.75) {};
		\node [style=none] (184) at (5.25, 6.5) {};
		\node [style={flavour_gp}] (186) at (11.25, 4.25) {};
		\node [style={gauge_gp}] (187) at (11.25, 8.25) {};
		\node [style=none] (188) at (11.25, 7.5) {};
		\node [style={gauge_gp}] (189) at (11.25, 5.75) {};
		\node [style=none] (190) at (11.25, 6.5) {};
		\node [style={gauge_gp}] (191) at (8.25, 3) {};
		\node [style={gauge_gp}] (192) at (-7.75, 2.5) {};
		\node [style={gauge_gp}] (193) at (-10.25, 2.5) {};
		\node [style=none] (194) at (9, -10.25) {$\mathrm{II}^Y_{[n',l',l'_1,l'_2,p'_1,p'_2]}$};
		\node [style={gauge_gp}] (195) at (-2.25, 9.25) {};
		\node [style={gauge_gp}] (196) at (0.75, 8) {};
		\node [style={gauge_gp}] (197) at (-0.75, 8) {};
		\node [style={gauge_gp}] (198) at (0, 3.5) {};
		\node [style=none] (199) at (1.5, 9) {\textcolor{orange}{$\vdots$}};
		\node [style=none] (204) at (-3, -4.5) {\textcolor{magenta}{$n'-l'_2$}};
		\node [style=none] (207) at (0, -4.5) {\textcolor{rgb,255: red,0; green,200; blue,0}{$l'_1$}};
		\node [style=none] (208) at (2.25, -4.5) {\textcolor{red}{$p'_2$}};
		\node [style=none] (212) at (0, -7) {\textcolor{cyan}{1}};
		\node [style=none] (213) at (-3, -7) {\textcolor{blue}{$p'_1-1$}};
		\node [style=none] (214) at (-3, -5.75) {\textcolor{orange}{$l'_2-l'$}};
		\node [style=none] (215) at (0, -5.75) {\textcolor{rgb,255: red,128; green,0; blue,128}{$l'-l'_1$}};
		\node [style=none] (216) at (8.25, 1.75) {};
		\node [style=none] (217) at (8.25, -1.5) {};
		\node [style=none] (218) at (7.5, 0.5) {3D};
		\node [style=none] (219) at (-7.75, 1) {(choice of ungauging};
		\node [style=none] (220) at (-7.75, 0.25) { as seen in \cite{Dey:2020hfe} )};
		\node [style=none] (221) at (7, -0.25) {mirror};
	\end{pgfonlayer}
	\begin{pgfonlayer}{edgelayer}
		\draw (5.center) to (6.center);
		\draw [style=red] (10) to (13);
		\draw [style=red] (14) to (12);
		\draw [style=red] (13) to (16.center);
		\draw [style=red] (15.center) to (14);
		\draw [style=orange] (19) to (10);
		\draw [style=orange] (18) to (9);
		\draw [style=orange] (20.center) to (18);
		\draw [style=orange] (21.center) to (19);
		\draw [style=pink] (10) to (24);
		\draw [style=pink] (24) to (22.center);
		\draw [style=pink] (23.center) to (25);
		\draw [style=pink] (25) to (11);
		\draw [style=purple] (9) to (26);
		\draw [style=purple] (26) to (32.center);
		\draw [style=purple] (33.center) to (27);
		\draw [style=purple] (27) to (12);
		\draw [style=green] (12) to (28);
		\draw [style=green] (28) to (30.center);
		\draw [style=green] (31.center) to (29);
		\draw [style=green] (29) to (11);
		\draw [style=cyan] (8) to (9);
		\draw [style=red] (88) to (108.center);
		\draw [style=red] (109.center) to (89);
		\draw [style=green] (98) to (88);
		\draw [style=green] (105.center) to (99);
		\draw [style=green] (99) to (90);
		\draw [style=green] (98) to (104.center);
		\draw [style=pink] (89) to (101);
		\draw [style=pink] (101) to (103.center);
		\draw [style=pink] (102.center) to (100);
		\draw [style=pink] (100) to (90);
		\draw [style=cyan] (125) to (120);
		\draw [style=blue] (91.center) to (125);
		\draw [style=purple, bend right=15] (123) to (121);
		\draw [style=purple, bend right=15] (121) to (123);
		\draw [style=orange, bend right=15] (123) to (122);
		\draw [style=orange, bend right=15] (122) to (123);
		\draw [style=blue, bend right=15, looseness=0.75] (121) to (122);
		\draw [style=blue, bend right=60] (121) to (122);
		\draw [style=cyan, bend left=15, looseness=0.75] (121) to (122);
		\draw (130) to (179);
		\draw (178) to (129);
		\draw (180) to (127);
		\draw [style=purple] (181) to (170);
		\draw [style=purple] (181) to (182.center);
		\draw [style=purple] (183) to (118);
		\draw [style=purple] (183) to (184.center);
		\draw [style=orange] (171) to (187);
		\draw [style=orange] (187) to (188.center);
		\draw [style=orange] (190.center) to (189);
		\draw [style=orange] (189) to (186);
		\draw [style=blue] (191) to (173);
		\draw [style=blue] (191) to (92.center);
		\draw [style=blue] (163) to (192);
		\draw [style=blue] (192) to (35.center);
		\draw [style=blue] (34) to (193);
		\draw [style=blue] (193) to (37.center);
		\draw [style=red] (167) to (67);
		\draw [style=red] (67) to (66.center);
		\draw [style=red] (70.center) to (68);
		\draw [style=red] (68) to (166);
		\draw [style=green] (167) to (54);
		\draw [style=green] (54) to (60.center);
		\draw [style=green] (61.center) to (55);
		\draw [style=green] (169) to (55);
		\draw [style=pink] (169) to (56);
		\draw [style=pink] (56) to (58.center);
		\draw [style=pink] (59.center) to (57);
		\draw [style=pink] (57) to (166);
		\draw [style=purple] (195) to (167);
		\draw [style=purple] (195) to (50.center);
		\draw [style=orange] (45) to (166);
		\draw [style=orange] (45) to (53.center);
		\draw [style=orange] (196) to (165);
		\draw [style=orange] (196) to (52.center);
		\draw [style=purple] (51.center) to (197);
		\draw [style=purple] (197) to (165);
		\draw [style=cyan] (165) to (44);
		\draw [style=blue] (47.center) to (198);
		\draw [style=blue] (198) to (169);
		\draw [style=double arrow] (216.center) to (217.center);
		\draw [style=blue] (44) to (49.center);
	\end{pgfonlayer}
\end{tikzpicture}
\caption{The family of 3D mirror pairs denoted as Family II in \cite{Dey:2020hfe}.}
\label{fig:famII}
\end{figure*}

\begin{figure*}
    \begin{tikzpicture}[baseline=-0.25em,scale=0.5]
    \begin{pgfonlayer}{nodelayer}
		\node [style=none] (0) at (4.25, 10.5) {$\mathrm{IIa}^X_{[n',l',l'_1,l'_2,p'_1,p'_2]}$};
		\node [style=none] (1) at (7.25, 5.75) {\textcolor{blue}{$\vdots$}};
		\node [style=none] (2) at (17.75, 8) {$\mathrm{IIa}^Y_{[n',l',l'_1,l'_2,p'_1,p'_2]}$};
		\node [style=none] (3) at (0.5, 6.75) {\textcolor{orange}{$\vdots$}};
		\node [style={flavour_gp}] (8) at (-2, 4.25) {};
		\node [style={gauge_gp}] (9) at (-0.25, 4.25) {};
		\node [style={gauge_gp}] (10) at (2.75, 7.25) {};
		\node [style={gauge_gp}] (11) at (5.75, 4.25) {};
		\node [style={gauge_gp}] (12) at (2.75, 1.25) {};
		\node [style={gauge_gp}] (18) at (-0.25, 5.75) {};
		\node [style={gauge_gp}] (19) at (1.25, 7.25) {};
		\node [style=none] (20) at (0.25, 6.25) {};
		\node [style=none] (21) at (0.75, 6.75) {};
		\node [style=none] (22) at (4.75, 6.75) {};
		\node [style=none] (23) at (5.25, 6.25) {};
		\node [style={gauge_gp}] (24) at (4.25, 7.25) {};
		\node [style={gauge_gp}] (25) at (5.75, 5.75) {};
		\node [style={gauge_gp}] (26) at (-0.25, 2.75) {};
		\node [style={gauge_gp}] (27) at (1.25, 1.25) {};
		\node [style={gauge_gp}] (28) at (4.25, 1.25) {};
		\node [style={gauge_gp}] (29) at (5.75, 2.75) {};
		\node [style=none] (30) at (4.75, 1.75) {};
		\node [style=none] (31) at (5.25, 2.25) {};
		\node [style=none] (32) at (0.25, 2.25) {};
		\node [style=none] (33) at (0.75, 1.75) {};
		\node [style={flavour_gp}] (34) at (7.25, 8.25) {};
		\node [style=none] (35) at (7.25, 5) {};
		\node [style=none] (37) at (7.25, 6) {};
		\node [style=none] (38) at (5, 2.25) {\textcolor{rgb,255: red,0; green,200; blue,0}{$\vdots$}};
		\node [style=none] (39) at (0.5, 2.25) {\textcolor{rgb,255: red,128; green,0; blue,128}{$\vdots$}};
		\node [style=none] (40) at (5, 6.75) {\textcolor{magenta}{$\vdots$}};
		\node [style={gauge_gp}] (161) at (2.75, 7.25) {};
		\node [style={gauge_gp}] (162) at (2.75, 1.25) {};
		\node [style={gauge_gp}] (163) at (5.75, 4.25) {};
		\node [style={gauge_gp}] (164) at (-0.25, 4.25) {};
		\node [style={flavour_gp}] (182) at (2.75, 9) {};
		\node [style=none] (184) at (5, 0) {};
		\node [style={flavour_gp}] (185) at (7.5, 0) {};
		\node [style=none] (186) at (4.25, 0) {\textcolor{red}{$\cdots$}};
		\node [style={gauge_gp}] (237) at (15, 6.25) {};
		\node [style={gauge_gp}] (238) at (19, 6.25) {};
		\node [style={gauge_gp}] (239) at (15, 2.25) {};
		\node [style={gauge_gp}] (240) at (19, 2.25) {};
		\node [style=none] (241) at (17, 2.5) {\textcolor{red}{$\vdots$}};
		\node [style=none] (243) at (19, 4.25) {\textcolor{blue}{$\cdots$}};
		\node [style={orange_flavour}] (258) at (13.5, 6.25) {};
		\node [style={pink_flavour}] (259) at (20.5, 6.25) {};
		\node [style={purple_flavour}] (260) at (13.5, 2.25) {};
		\node [style={green_flavour}] (261) at (20.5, 2.25) {};
		\node [style=none] (316) at (3.5, 0) {};
		\node [style={gauge_gp}] (317) at (6, 0) {};
		\node [style={gauge_gp}] (318) at (2.75, 0) {};
		\node [style={gauge_gp}] (325) at (7.25, 4.25) {};
		\node [style={gauge_gp}] (326) at (7.25, 6.75) {};
		\node [style={gauge_gp}] (327) at (15, 6.25) {};
		\node [style={gauge_gp}] (328) at (19, 6.25) {};
		\node [style={gauge_gp}] (329) at (19, 2.25) {};
		\node [style={gauge_gp}] (330) at (15, 2.25) {};
		\node [style=none] (341) at (10.25, -1.5) {\textcolor{red}{$p'_2-1$}};
		\node [style=none] (347) at (13.5, 0) {\textcolor{magenta}{$n'-l'_2$}};
		\node [style=none] (348) at (10.25, 0) {\textcolor{orange}{$l'_2-l'$}};
		\node [style=none] (349) at (19.25, 0) {\textcolor{rgb,255: red,128; green,0; blue,128}{$l'-l'_1$}};
		\node [style=none] (351) at (16.5, -1.5) {\textcolor{rgb,255: red,168; green,168; blue,0}{1}};
		\node [style=none] (352) at (16.5, 0) {\textcolor{rgb,255: red,0; green,200; blue,0}{$l'_1$}};
		\node [style=none] (353) at (13.5, -1.5) {\textcolor{blue}{$p'_1-1$}};
		\node [style=none] (354) at (19.25, -1.5) {\textcolor{cyan}{1}};
		\node [style=none] (355) at (9, 4.25) {};
		\node [style=none] (356) at (12.5, 4.25) {};
		\node [style=none] (357) at (10.75, 5) {3D mirror};
	\end{pgfonlayer}
	\begin{pgfonlayer}{edgelayer}
		\draw [style=orange] (19) to (10);
		\draw [style=orange] (18) to (9);
		\draw [style=orange] (20.center) to (18);
		\draw [style=orange] (21.center) to (19);
		\draw [style=pink] (10) to (24);
		\draw [style=pink] (24) to (22.center);
		\draw [style=pink] (23.center) to (25);
		\draw [style=pink] (25) to (11);
		\draw [style=purple] (9) to (26);
		\draw [style=purple] (26) to (32.center);
		\draw [style=purple] (33.center) to (27);
		\draw [style=purple] (27) to (12);
		\draw [style=green] (12) to (28);
		\draw [style=green] (28) to (30.center);
		\draw [style=green] (31.center) to (29);
		\draw [style=green] (29) to (11);
		\draw [style=cyan] (8) to (9);
		\draw [style=yellow] (182) to (161);
		\draw [style=yellow] (237) to (238);
		\draw [style=cyan] (239) to (237);
		\draw [style=red, bend left] (239) to (240);
		\draw [style=red, bend right] (239) to (240);
		\draw [style=blue, bend right] (238) to (240);
		\draw [style=blue, bend right] (240) to (238);
		\draw [style=red] (184.center) to (317);
		\draw [style=red] (317) to (185);
		\draw [style=red] (162) to (318);
		\draw [style=red] (318) to (316.center);
		\draw [style=blue] (325) to (35.center);
		\draw [style=blue] (325) to (163);
		\draw [style=blue] (326) to (37.center);
		\draw [style=blue] (34) to (326);
		\draw (258) to (327);
		\draw (330) to (260);
		\draw (329) to (261);
		\draw (328) to (259);
		\draw [style=double arrow] (356.center) to (355.center);
	\end{pgfonlayer}
\end{tikzpicture}
\caption{The family of 3D mirror pairs, which we denote as Family IIa, appearing during construction of Family II in \cite{Dey:2020hfe}.}
\label{fig:famIIa}
\end{figure*}

\clearpage

\subsection{Non-Abelian: Family III and beyond}\label{fam3}
Here we present explicit examples of non-Abelian mirror pairs from Sec.\ \ref{subsec:ab&non-ab}, which extend Family $\mathrm{III}_{[p_1,p_2,p_3]}$ from Ref.\ \cite{Dey:2020hfe}. The simplest example is the pair $\mathrm{III}^X_{[1,1,1]}$ and $\mathrm{III}^Y_{[1,1,1]}$, as seen below: 

\begin{equation}\label{non-abelian_[3]-(2)}
    \begin{tikzpicture}[baseline=-0.25em,scale=0.5]
	\begin{pgfonlayer}{nodelayer}
		\node [style={gauge_gp}] (0) at (-7.5, 0) {};
		\node [style={gauge_gp}] (1) at (-6, 1.5) {};
		\node [style={gauge_gp}] (2) at (-4.5, 0) {};
		\node [style={gauge_gp}] (3) at (-6, -1.5) {};
		\node [style={flavour_gp}] (4) at (-9, 0) {};
		\node [style={flavour_gp}] (5) at (-3, 0) {};
		\node [style=none] (6) at (-9, 1) {3};
		\node [style=none] (7) at (-7.5, 1) {2};
		\node [style=none] (8) at (-6, 2.5) {1};
		\node [style=none] (9) at (-6, -2.5) {1};
		\node [style=none] (10) at (-4.5, 1) {1};
		\node [style=none] (11) at (-3, 1) {1};
		\node [style=none] (12) at (-2, 0) {};
		\node [style=none] (13) at (2, 0) {};
		\node [style=none] (14) at (0, 0.75) {3D mirror};
		\node [style={gauge_gp}] (15) at (4, 1) {};
		\node [style={gauge_gp}] (16) at (4, -1) {};
		\node [style={gauge_gp}] (17) at (6, -1) {};
		\node [style={gauge_gp}] (18) at (6, 1) {};
		\node [style={flavour_gp}] (19) at (2.75, 2.25) {};
		\node [style={flavour_gp}] (20) at (7.25, 2.25) {};
		\node [style={flavour_gp}] (21) at (7.25, -2.25) {};
		\node [style={flavour_gp}] (22) at (2.75, -2.25) {};
		\node [style=none] (23) at (3, -1) {2};
		\node [style=none] (24) at (7, -1) {2};
		\node [style=none] (25) at (1.75, -2.25) {1};
		\node [style=none] (26) at (8.25, -2.25) {1};
		\node [style=none] (27) at (7, 1) {1};
		\node [style=none] (28) at (3, 1) {1};
		\node [style=none] (29) at (1.75, 2.25) {1};
		\node [style=none] (30) at (8.25, 2.25) {1};
		\node [style=none] (33) at (-6, 3.75) {$\mathrm{III}^X_{[1,1,1]}$};
		\node [style=none] (34) at (5, 3.75) {$\mathrm{III}^Y_{[1,1,1]}$};
		\node [style=none] (37) at (-6, -4) {$[4,\{2,1,1,1\},$};
		\node [style=none] (38) at (5, -4) {$[4,\{2,2,1,1\},\{1,1,1,1\},$};
		\node [style=none] (39) at (5, -5) {$\{r_{1,1},r_{2,1},r_{3,1},r_{4,1}\}]$};
		\node [style=none] (40) at (-6, -5) {$\{1,1,1,1\},\{r_{1,1}^3,r_{3,1}\}]$};
		\node [style=none] (41) at (0, -6) {$G_\mathcal{C}^X  \supset \U(1)^4  \subset G_\mathcal{H}^Y$};
		\node [style=none] (42) at (0, -7.25) {$G_\mathcal{H}^X  \supset  \U(1)\times\SU(3) \subset G_\mathcal{C}^Y$};
	\end{pgfonlayer}
	\begin{pgfonlayer}{edgelayer}
		\draw (1) to (0);
		\draw (0) to (3);
		\draw (3) to (2);
		\draw (2) to (1);
		\draw (4) to (0);
		\draw (2) to (5);
		\draw [style=double arrow] (13.center) to (12.center);
		\draw (15) to (16);
		\draw (15) to (18);
		\draw (18) to (17);
		\draw (17) to (16);
		\draw (19) to (15);
		\draw (20) to (18);
		\draw (17) to (21);
		\draw (16) to (22);
	\end{pgfonlayer}
\end{tikzpicture}
\end{equation}

\begin{eqnarray*}
\HS_\mathcal{C}^X(t) =&& \HS_\mathcal{H}^Y(t) \\
=&& 1 + 4t^2 + 8t^3 + 21t^4 + 42t^5 + 95t^6 +  \\
&& 182t^7 + 355t^8 + 636t^9 + 1137t^{10} + \mathcal{O}(t^{11})\\\\
\HS_\mathcal{H}^X(t) =&& \HS_\mathcal{C}^Y(t) \\
=&& 1 + 10t^2 + 77t^4 + 446t^6 + 2098t^8 + \\
&& 8268t^{10} + \mathcal{O}(t^{12})
\end{eqnarray*}

This family can be extended by different types of gluing. For example, we could glue together three (or arbitrarily many) quivers to form the following pair:
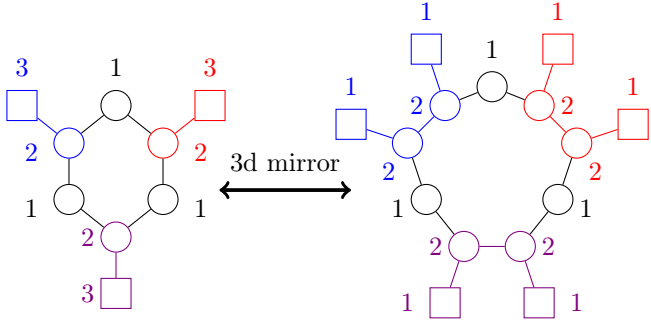
\begin{equation}\label{non-abelian_(2-3)^3}
    \begin{tikzpicture}[baseline=-0.25em,scale=0.5]
		\begin{pgfonlayer}{nodelayer}
		\node [style={gauge_gp}] (1) at (-4.5, 2.25) {};
		\node [style=none] (6) at (-7, 3.25) {3};
		\node [style=none] (7) at (-6.75, 1) {2};
		\node [style=none] (8) at (-4.5, 3.25) {1};
		\node [style=none] (9) at (-5.25, -2.75) {3};
		\node [style=none] (10) at (-2.25, 1) {2};
		\node [style=none] (11) at (-2, 3.25) {3};
		\node [style=none] (12) at (1.75, 0) {};
		\node [style=none] (13) at (-1.75, 0) {};
		\node [style=none] (14) at (0, 0.75) {3D mirror};
		\node [style={gauge_gp}] (31) at (-5.75, -0.25) {};
		\node [style={gauge_gp}] (32) at (-3.25, -0.25) {};
		\node [style=none] (34) at (-5.25, -1.25) {2};
		\node [style=none] (35) at (-6.75, -0.5) {1};
		\node [style=none] (36) at (-2.25, -0.5) {1};
		\node [style={gauge_gp}] (37) at (3.75, -0.25) {};
		\node [style={gauge_gp}] (40) at (7.25, -0.25) {};
		\node [style={gauge_gp}] (41) at (5.5, 2.75) {};
		\node [style=none] (52) at (3.75, 4.75) {1};
		\node [style=none] (53) at (7.25, 4.75) {1};
		\node [style=none] (54) at (9.25, 2.75) {1};
		\node [style=none] (55) at (1.75, 2.75) {1};
		\node [style=none] (56) at (3.25, -3) {1};
		\node [style=none] (57) at (7.75, -3) {1};
		\node [style=none] (58) at (2.75, 0.5) {2};
		\node [style=none] (59) at (8.25, 0.5) {2};
		\node [style=none] (60) at (3.5, 2.25) {2};
		\node [style=none] (61) at (7.5, 2.25) {2};
		\node [style=none] (62) at (4, -1.5) {2};
		\node [style=none] (63) at (7, -1.5) {2};
		\node [style=none] (64) at (3, -0.5) {1};
		\node [style=none] (65) at (5.5, 3.75) {1};
		\node [style=none] (66) at (8, -0.5) {1};
		\node [style=none] (67) at (6, -7.25) {($Y$)};
		\node [style=none] (68) at (-4.5, -7) {($X$)};
		\node [style={gauge_gp}] (69) at (-5.75, 1.25) {};
		\node [style={gauge_gp}] (70) at (4.25, 2.25) {};
		\node [style={flavour_gp}] (72) at (3.75, 3.75) {};
		\node [style={flavour_gp}] (73) at (1.75, 1.75) {};
		\node [style={gauge_gp}] (74) at (6.75, 2.25) {};
		\node [style={gauge_gp}] (75) at (7.75, 1.25) {};
		\node [style={flavour_gp}] (76) at (7.25, 3.75) {};
		\node [style={flavour_gp}] (77) at (9.25, 1.75) {};
		\node [style={gauge_gp}] (78) at (4.75, -1.5) {};
		\node [style={gauge_gp}] (79) at (6.25, -1.5) {};
		\node [style={flavour_gp}] (80) at (4.25, -3) {};
		\node [style={flavour_gp}] (81) at (6.75, -3) {};
		\node [style={gauge_gp}] (82) at (3.25, 1.25) {};
		\node [style={flavour_gp}] (83) at (-7, 2.25) {};
		\node [style={gauge_gp}] (84) at (-3.25, 1.25) {};
		\node [style={flavour_gp}] (85) at (-2, 2.25) {};
		\node [style={gauge_gp}] (86) at (-4.5, -1.25) {};
		\node [style={flavour_gp}] (87) at (-4.5, -2.75) {};
		\node [style=none] (88) at (-4.5, -4.5) {$[6,\{2,1,2,1,2,1\},$};
		\node [style=none] (89) at (-4.25, -5.5) {$\{1^6\}, \{r_{1,1}^3,r_{3,1}^3,r_{5,1}^3\}]$};
		\node [style=none] (90) at (5, -4.5) {$[9,\{2,2,1,2,2,1,2,2,1\},\{1^9\},$};
		\node [style=none] (91) at (5, -5.5) {$\{r_{1,1},r_{2,1},r_{4,1},r_{5,1},r_{7,1},r_{8,1}\}]$};
		\node [style=none] (92) at (0, -8.25) {$G_\mathcal{C}^X \supset \U(1)^6 \subset G_\mathcal{H}^Y$};
		\node [style=none] (93) at (0, -9.5) {$G_\mathcal{H}^X \supset \U(1)^3\times\SU(3)^3 \subset G_\mathcal{C}^Y$};
	\end{pgfonlayer}
	\begin{pgfonlayer}{edgelayer}
		\draw [style=double arrow] (13.center) to (12.center);
		\draw (70) to (41);
		\draw (82) to (37);
		\draw (73) to (82);
		\draw (82) to (70);
		\draw (70) to (72);
		\draw (83) to (69);
		\draw (69) to (31);
		\draw (69) to (1);
		\draw (80) to (78);
		\draw (78) to (79);
		\draw (79) to (81);
		\draw (37) to (78);
		\draw (79) to (40);
		\draw (74) to (76);
		\draw (75) to (77);
		\draw (75) to (74);
		\draw (41) to (74);
		\draw (75) to (40);
		\draw (85) to (84);
		\draw (32) to (84);
		\draw (84) to (1);
		\draw (87) to (86);
		\draw (32) to (86);
		\draw (86) to (31);
	\end{pgfonlayer}
\end{tikzpicture}
\end{equation}

\begin{eqnarray*}
\HS_\mathcal{C}^X(t) =&& \HS_\mathcal{H}^Y(t) \\
=&& 1 + 6t^2 + 6t^3 + 30t^4 + 54t^5 + 137t^6 + 294t^7 + \\
&& 636t^8 + 1306t^9 + 2670t^{10} + \mathcal{O}(t^{11})\\\\
\HS_\mathcal{H}^X(t) =&& \HS_\mathcal{C}^Y(t) \\
=&& 1 + 27t^2 + 456t^4 + 5786t^6 + 59910t^8 +  \\
&& 526458t^{10} + \mathcal{O}(t^{12})
\end{eqnarray*}

Additionally, as a further consistency check, all of these mirror pairs can be generated using the bootstrapping method developed in Sec.\ \ref{sec:boot}.

\section{The degeneration of sunshine quivers} \label{appx:no-ray}

In Section \ref{subsec:abelian_mirror_pairs}, we established a correspondence where distance between rays is dual to the number of flavors in the mirror quiver. We will now examine the systematic degeneration of sunshine quivers by subtracting rays, where the mirror quivers are special cases of the sunshine quiver family.

The equations below illustrate the degeneration process, showing how the mirror quiver transitions from a loop-based structure to a star-based structure as the number of rays is reduced.

\begin{itemize}
    \item \textbf{Two Rays on a Loop:} The quiver on the left features a four-node loop with two rays of flavor ranks $\alpha$ and $\beta$, respectively. The two rays in the original quiver are mirrored to two linear chains of $\alpha - 1$ and $\beta - 1$ $\U(1)$ gauge nodes, respectively, connected between the same two ray-emitting loop nodes, as follows:
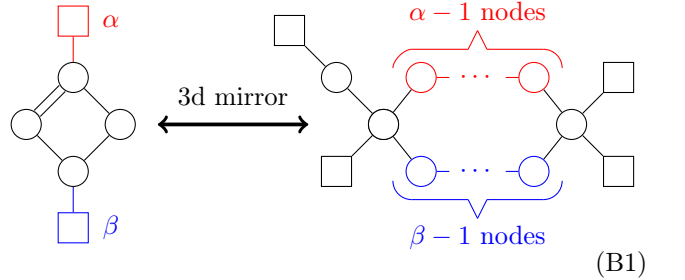
    \begin{equation}
        \begin{tikzpicture}[baseline=-0.25em,scale=0.5]
    	\begin{pgfonlayer}{nodelayer}
    		\node [style=none] (0) at (2, 0) {};
    		\node [style=none] (1) at (-2, 0) {};
    		\node [style=none] (2) at (0, 0.75) {3D mirror};
    		\node [style={gauge_gp}] (3) at (-5.5, 0) {};
    		\node [style={gauge_gp}] (4) at (-4.25, 1.25) {};
    		\node [style={gauge_gp}] (5) at (-3, 0) {};
    		\node [style={gauge_gp}] (6) at (-4.25, -1.25) {};
    		\node [style={flavour_gp}] (7) at (-4.25, 2.75) {};
    		\node [style={flavour_gp}] (8) at (-4.25, -2.75) {};
    		\node [style=none] (9) at (-3.25, -2.75) {$\beta$};
    		\node [style=none] (10) at (-3.25, 2.75) {$\alpha$};
    		\node [style={gauge_gp}] (11) at (4, 0) {};
    		\node [style={gauge_gp}] (12) at (9, 0) {};
    		\node [style={flavour_gp}] (13) at (1.5, 2.5) {};
    		\node [style={gauge_gp}] (15) at (2.75, 1.25) {};
    		\node [style={gauge_gp}] (16) at (4, 0) {};
    		\node [style={flavour_gp}] (17) at (2.75, -1.25) {};
    		\node [style={flavour_gp}] (18) at (10.25, -1.25) {};
    		\node [style={flavour_gp}] (19) at (10.25, 1.25) {};
    		\node [style={gauge_gp}] (20) at (8, -1.25) {};
    		\node [style={gauge_gp}] (21) at (5, -1.25) {};
    		\node [style=none] (22) at (7.25, -1.25) {};
    		\node [style=none] (23) at (5.75, -1.25) {};
    		\node [style=none] (24) at (6.5, -1.25) {$\cdots$};
    		\node [style=none] (25) at (6.5, -3) {$\beta-1$ nodes};
    		\node [style=none] (28) at (4.25, -1.5) {};
    		\node [style=none] (29) at (4.75, -2) {};
    		\node [style=none] (30) at (6.25, -2) {};
    		\node [style=none] (31) at (6.75, -2) {};
    		\node [style=none] (32) at (6.5, -2.5) {};
    		\node [style=none] (33) at (8.75, -1.5) {};
    		\node [style=none] (34) at (8.25, -2) {};
    		\node [style={gauge_gp}] (35) at (5, 1.25) {};
    		\node [style={gauge_gp}] (36) at (8, 1.25) {};
    		\node [style=none] (37) at (7.25, 1.25) {};
    		\node [style=none] (38) at (5.75, 1.25) {};
    		\node [style=none] (39) at (6.5, 1.25) {$\cdots$};
    		\node [style=none] (40) at (4.25, 1.5) {};
    		\node [style=none] (41) at (4.75, 2) {};
    		\node [style=none] (42) at (6.25, 2) {};
    		\node [style=none] (43) at (6.75, 2) {};
    		\node [style=none] (44) at (6.5, 2.5) {};
    		\node [style=none] (45) at (8.75, 1.5) {};
    		\node [style=none] (46) at (8.25, 2) {};
    		\node [style=none] (47) at (6.5, 3) {$\alpha-1$ nodes};
    		\node [style=none] (50) at (0, 0) {};
    	\end{pgfonlayer}
    	\begin{pgfonlayer}{edgelayer}
    		\draw [style=double arrow] (1.center) to (0.center);
    		\draw (5) to (6);
    		\draw (4) to (5);
    		\draw (6) to (3);
    		\draw [style=doubleline] (3) to (4);
    		\draw (7) to (4);
    		\draw (8) to (6);
    		\draw (16) to (15);
    		\draw (15) to (13);
    		\draw (17) to (16);
    		\draw (18) to (12);
    		\draw (19) to (12);
    		\draw (23.center) to (21);
    		\draw (20) to (22.center);
    		\draw (16) to (21);
    		\draw (20) to (12);
    		\draw (32.center) to (30.center);
    		\draw (32.center) to (31.center);
    		\draw [bend left=45] (33.center) to (34.center);
    		\draw (34.center) to (31.center);
    		\draw (30.center) to (29.center);
    		\draw [bend left=45] (29.center) to (28.center);
    		\draw (36) to (12);
    		\draw (35) to (16);
    		\draw (35) to (38.center);
    		\draw (37.center) to (36);
    		\draw [bend right=45, looseness=0.75] (41.center) to (40.center);
    		\draw [bend left=45] (46.center) to (45.center);
    		\draw (46.center) to (43.center);
    		\draw (41.center) to (42.center);
    		\draw (44.center) to (42.center);
    		\draw (44.center) to (43.center);
    	\end{pgfonlayer}
    \end{tikzpicture}
    \end{equation}
    When the four-node loop has two rays of length $\alpha$ and $\beta$, the two rays are mirrored to edges with multiplicity $\alpha$ and $\beta$. However, since they connect the same two gauge nodes here, the two edges merge to form one edge of multiplicity $\alpha+\beta$ in the mirror theory, as follows:
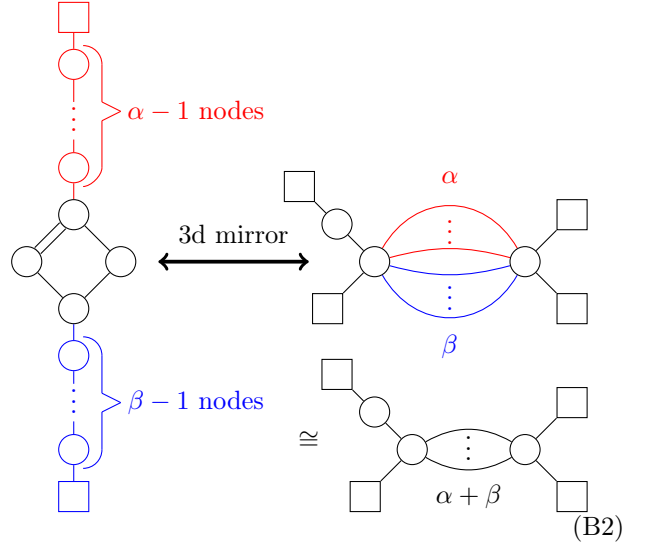
    \begin{equation}
        \begin{tikzpicture}[baseline=-0.25em,scale=0.5]
	\begin{pgfonlayer}{nodelayer}
		\node [style=none] (0) at (2, 0) {};
		\node [style=none] (1) at (-2, 0) {};
		\node [style=none] (2) at (0, 0.75) {3D mirror};
		\node [style={gauge_gp}] (3) at (-5.5, 0) {};
		\node [style={gauge_gp}] (4) at (-4.25, 1.25) {};
		\node [style={gauge_gp}] (5) at (-3, 0) {};
		\node [style={gauge_gp}] (6) at (-4.25, -1.25) {};
		\node [style={flavour_gp}] (7) at (-4.25, 6.5) {};
		\node [style={flavour_gp}] (8) at (-4.25, -6.25) {};
		\node [style={gauge_gp}] (12) at (7.75, 0) {};
		\node [style={flavour_gp}] (13) at (1.75, 2) {};
		\node [style={gauge_gp}] (15) at (2.75, 1) {};
		\node [style={gauge_gp}] (16) at (3.75, 0) {};
		\node [style={flavour_gp}] (17) at (2.5, -1.25) {};
		\node [style={flavour_gp}] (18) at (9, -1.25) {};
		\node [style={flavour_gp}] (19) at (9, 1.25) {};
		\node [style=none] (50) at (0, 0) {};
		\node [style={gauge_gp}] (51) at (-4.25, -2.5) {};
		\node [style={gauge_gp}] (52) at (-4.25, -5) {};
		\node [style=none] (53) at (-4.25, -3.25) {};
		\node [style=none] (54) at (-4.25, -4.25) {};
		\node [style=none] (55) at (-4.25, -3.5) {$\vdots$};
		\node [style=none] (56) at (-1, -3.75) {$\beta-1$ nodes};
		\node [style={gauge_gp}] (64) at (-4.25, 2.5) {};
		\node [style={gauge_gp}] (65) at (-4.25, 5.25) {};
		\node [style=none] (66) at (-4.25, 4.25) {};
		\node [style=none] (67) at (-4.25, 3.25) {};
		\node [style=none] (68) at (-4.25, 4) {$\vdots$};
		\node [style=none] (76) at (-1, 4) {$\alpha-1$ nodes};
		\node [style=none] (77) at (5.75, 1) {$\vdots$};
		\node [style=none] (78) at (5.75, -0.75) {$\vdots$};
		\node [style=none] (79) at (5.75, -2.25) {$\beta$};
		\node [style=none] (80) at (5.75, 2.25) {$\alpha$};
		\node [style={gauge_gp}] (82) at (7.75, -5) {};
		\node [style={flavour_gp}] (83) at (2.75, -3) {};
		\node [style={gauge_gp}] (84) at (3.75, -4) {};
		\node [style={gauge_gp}] (85) at (4.75, -5) {};
		\node [style={flavour_gp}] (86) at (3.5, -6.25) {};
		\node [style={flavour_gp}] (87) at (9, -6.25) {};
		\node [style={flavour_gp}] (88) at (9, -3.75) {};
		\node [style=none] (90) at (6.25, -4.75) {$\vdots$};
		\node [style=none] (91) at (6.25, -6.25) {$\alpha+\beta$};
		\node [style=none] (92) at (2, -4.75) {$\cong$};
		\node [style=none] (101) at (-4, -2) {};
		\node [style=none] (102) at (-3.5, -2.5) {};
		\node [style=none] (103) at (-3.5, -3.5) {};
		\node [style=none] (104) at (-3.5, -4) {};
		\node [style=none] (105) at (-3, -3.75) {};
		\node [style=none] (106) at (-4, -5.5) {};
		\node [style=none] (107) at (-3.5, -5) {};
		\node [style=none] (113) at (-4, 2) {};
		\node [style=none] (114) at (-3.5, 2.5) {};
		\node [style=none] (115) at (-3.5, 3.75) {};
		\node [style=none] (116) at (-3.5, 4.25) {};
		\node [style=none] (117) at (-3, 4) {};
		\node [style=none] (118) at (-4, 5.75) {};
		\node [style=none] (119) at (-3.5, 5.25) {};
	\end{pgfonlayer}
	\begin{pgfonlayer}{edgelayer}
		\draw [style=double arrow] (1.center) to (0.center);
		\draw (5) to (6);
		\draw (4) to (5);
		\draw (6) to (3);
		\draw [style=doubleline] (3) to (4);
		\draw (16) to (15);
		\draw (15) to (13);
		\draw (17) to (16);
		\draw (18) to (12);
		\draw (19) to (12);
		\draw (54.center) to (52);
		\draw (51) to (53.center);
		\draw (51) to (6);
		\draw (52) to (8);
		\draw (64) to (67.center);
		\draw (66.center) to (65);
		\draw (7) to (65);
		\draw (64) to (4);
		\draw [bend right=15] (16) to (12);
		\draw [bend right=60, looseness=1.25] (16) to (12);
		\draw [bend left=60, looseness=1.25] (16) to (12);
		\draw [bend right=15] (12) to (16);
		\draw (85) to (84);
		\draw (84) to (83);
		\draw (86) to (85);
		\draw (87) to (82);
		\draw (88) to (82);
		\draw [bend right] (85) to (82);
		\draw [bend left] (85) to (82);
		\draw (105.center) to (103.center);
		\draw (105.center) to (104.center);
		\draw [bend right=45] (106.center) to (107.center);
		\draw (107.center) to (104.center);
		\draw (103.center) to (102.center);
		\draw [bend right=45] (102.center) to (101.center);
		\draw [bend left=45, looseness=0.75] (114.center) to (113.center);
		\draw [bend right=45] (119.center) to (118.center);
		\draw (119.center) to (116.center);
		\draw (114.center) to (115.center);
		\draw (117.center) to (115.center);
		\draw (117.center) to (116.center);
	\end{pgfonlayer}
\end{tikzpicture}
    \end{equation}
    
    \item \textbf{One Ray on a Loop:} Reducing the quiver further leaves only a single ray with $\alpha$ flavours. The mirror theory now degenerates into a star-like quiver. The ray manifests in the mirror as a loop of size $\alpha$ from the ray-emitting gauge node.
    
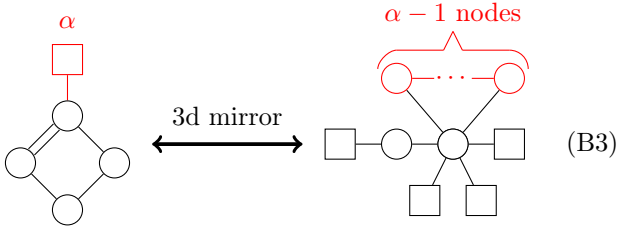
\begin{equation}
        \begin{tikzpicture}[baseline=-0.25em,scale=0.5]
    	\begin{pgfonlayer}{nodelayer}
    		\node [style=none] (0) at (2, 0) {};
    		\node [style=none] (1) at (-2, 0) {};
    		\node [style=none] (2) at (0, 0.75) {3D mirror};
    		\node [style={gauge_gp}] (3) at (-5.5, -0.5) {};
    		\node [style={gauge_gp}] (4) at (-4.25, 0.75) {};
    		\node [style={gauge_gp}] (5) at (-3, -0.5) {};
    		\node [style={gauge_gp}] (6) at (-4.25, -1.75) {};
    		\node [style={flavour_gp}] (7) at (-4.25, 2.25) {};
    		\node [style=none] (10) at (-4.25, 3.25) {$\alpha$};
    		\node [style={gauge_gp}] (11) at (6, 0) {};
    		\node [style={flavour_gp}] (13) at (3, 0) {};
    		\node [style={gauge_gp}] (15) at (4.5, 0) {};
    		\node [style={gauge_gp}] (16) at (6, 0) {};
    		\node [style={flavour_gp}] (17) at (7.5, 0) {};
    		\node [style={flavour_gp}] (18) at (6.75, -1.5) {};
    		\node [style={flavour_gp}] (19) at (5.25, -1.5) {};
    		\node [style={gauge_gp}] (35) at (4.5, 1.75) {};
    		\node [style={gauge_gp}] (36) at (7.5, 1.75) {};
    		\node [style=none] (37) at (6.5, 1.75) {};
    		\node [style=none] (38) at (5.5, 1.75) {};
    		\node [style=none] (39) at (6, 1.75) {$\cdots$};
    		\node [style=none] (40) at (4, 2) {};
    		\node [style=none] (41) at (4.5, 2.5) {};
    		\node [style=none] (42) at (5.75, 2.5) {};
    		\node [style=none] (43) at (6.25, 2.5) {};
    		\node [style=none] (44) at (6, 3) {};
    		\node [style=none] (45) at (8, 2) {};
    		\node [style=none] (46) at (7.5, 2.5) {};
    		\node [style=none] (47) at (6, 3.5) {$\alpha-1$ nodes};
    		\node [style=none] (50) at (0, 0) {};
    	\end{pgfonlayer}
    	\begin{pgfonlayer}{edgelayer}
    		\draw [style=double arrow] (1.center) to (0.center);
    		\draw (5) to (6);
    		\draw (4) to (5);
    		\draw (6) to (3);
    		\draw [style=doubleline] (3) to (4);
    		\draw (7) to (4);
    		\draw (16) to (15);
    		\draw (15) to (13);
    		\draw (17) to (16);
    		\draw (35) to (16);
    		\draw (35) to (38.center);
    		\draw (37.center) to (36);
    		\draw [bend right=45, looseness=0.75] (41.center) to (40.center);
    		\draw [bend left=45] (46.center) to (45.center);
    		\draw (46.center) to (43.center);
    		\draw (41.center) to (42.center);
    		\draw (44.center) to (42.center);
    		\draw (44.center) to (43.center);
    		\draw (19) to (16);
    		\draw (16) to (18);
    		\draw (36) to (16);
    	\end{pgfonlayer}
    \end{tikzpicture}
    \end{equation}
    
    When there is one ray of length $\alpha$, the mirror is again a star-shaped quiver, but this time with a self-loop (a bifundamental hypermultiplet connecting a gauge node back to itself) of multiplicity $\alpha$ on the central gauge node of the mirror. The presence of the self-loop ensures that the twisted free-hypers match the hypers in the mirror.
    
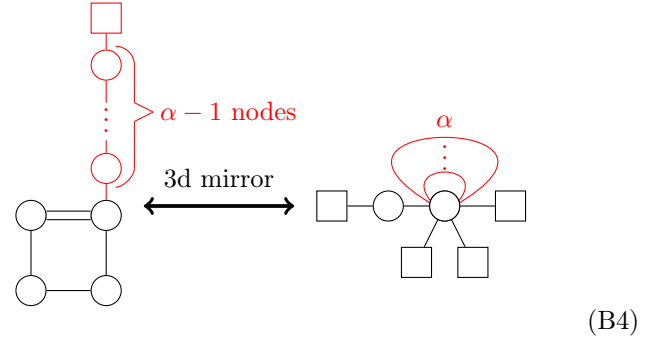
\begin{equation}
        \begin{tikzpicture}[baseline=-0.25em,scale=0.5]
    	\begin{pgfonlayer}{nodelayer}
    		\node [style=none] (0) at (2, 0) {};
    		\node [style=none] (1) at (-2, 0) {};
    		\node [style=none] (2) at (0, 0.75) {3D mirror};
    		\node [style={gauge_gp}] (3) at (-5, -0.25) {};
    		\node [style={gauge_gp}] (4) at (-3, -0.25) {};
    		\node [style={gauge_gp}] (5) at (-3, -2.25) {};
    		\node [style={gauge_gp}] (6) at (-5, -2.25) {};
    		\node [style=none] (50) at (0, 0) {};
    		\node [style={gauge_gp}] (82) at (6, 0) {};
    		\node [style={flavour_gp}] (83) at (3, 0) {};
    		\node [style={gauge_gp}] (84) at (4.5, 0) {};
    		\node [style={gauge_gp}] (85) at (6, 0) {};
    		\node [style={flavour_gp}] (86) at (5.25, -1.5) {};
    		\node [style={flavour_gp}] (87) at (7.75, 0) {};
    		\node [style={flavour_gp}] (88) at (6.75, -1.5) {};
    		\node [style=none] (93) at (6, 1.5) {$\vdots$};
    		\node [style=none] (94) at (6.25, 0) {};
    		\node [style=none] (95) at (5.75, 0) {};
    		\node [style=none] (96) at (6, 2.25) {$\alpha$};
    		\node [style={flavour_gp}] (106) at (-3, 5) {};
    		\node [style={gauge_gp}] (107) at (-3, 1) {};
    		\node [style={gauge_gp}] (108) at (-3, 3.75) {};
    		\node [style=none] (109) at (-3, 2.75) {};
    		\node [style=none] (110) at (-3, 1.75) {};
    		\node [style=none] (111) at (-3, 2.5) {$\vdots$};
    		\node [style=none] (112) at (0.25, 2.5) {$\alpha-1$ nodes};
    		\node [style=none] (113) at (-2.75, 0.5) {};
    		\node [style=none] (114) at (-2.25, 1) {};
    		\node [style=none] (115) at (-2.25, 2.25) {};
    		\node [style=none] (116) at (-2.25, 2.75) {};
    		\node [style=none] (117) at (-1.75, 2.5) {};
    		\node [style=none] (118) at (-2.75, 4.25) {};
    		\node [style=none] (119) at (-2.25, 3.75) {};
    	\end{pgfonlayer}
    	\begin{pgfonlayer}{edgelayer}
    		\draw [style=double arrow] (1.center) to (0.center);
    		\draw (5) to (6);
    		\draw (4) to (5);
    		\draw (6) to (3);
    		\draw [style=doubleline] (3) to (4);
    		\draw (85) to (84);
    		\draw (84) to (83);
    		\draw (86) to (85);
    		\draw (88) to (85);
    		\draw (85) to (87);
    		\draw [bend left=150, looseness=25.00] (95.center) to (94.center);
    		\draw [bend right=135, looseness=8.75] (94.center) to (95.center);
    		\draw (107) to (110.center);
    		\draw (109.center) to (108);
    		\draw (106) to (108);
    		\draw [bend left=45, looseness=0.75] (114.center) to (113.center);
    		\draw [bend right=45] (119.center) to (118.center);
    		\draw (119.center) to (116.center);
    		\draw (114.center) to (115.center);
    		\draw (117.center) to (115.center);
    		\draw (117.center) to (116.center);
    		\draw (107) to (4);
    	\end{pgfonlayer}
    \end{tikzpicture}
    \end{equation}

    \item \textbf{Ray-Free Loop:} The final stage of degeneration is the `ray-free' sunshine quiver, which is simply a loop of $\U(1)$ gauge nodes (with possible multiple bonds, here shown with a double bond). The mirror quiver is a simple star of linear chains, where rays are attached from a central node. The number of rays is equal to the number of loop edges in the mirror theory, while the length of each ray is the corresponding edge multiplicity.
    \begin{equation}
        \begin{tikzpicture}[baseline=-0.25em,scale=0.5]
            	\begin{pgfonlayer}{nodelayer}
    		\node [style=none] (0) at (2, 0) {};
    		\node [style=none] (1) at (-2, 0) {};
    		\node [style=none] (2) at (0, 0.75) {3D mirror};
    		\node [style={gauge_gp}] (3) at (-5, 1) {};
    		\node [style={gauge_gp}] (4) at (-3, 1) {};
    		\node [style={gauge_gp}] (5) at (-3, -1) {};
    		\node [style={gauge_gp}] (6) at (-5, -1) {};
    		\node [style={gauge_gp}] (11) at (4.5, -0.5) {};
    		\node [style={flavour_gp}] (13) at (4.5, 2.5) {};
    		\node [style={gauge_gp}] (15) at (4.5, 1) {};
    		\node [style={gauge_gp}] (16) at (4.5, -0.5) {};
    		\node [style={flavour_gp}] (17) at (6, -0.5) {};
    		\node [style={flavour_gp}] (18) at (4.5, -2) {};
    		\node [style={flavour_gp}] (19) at (3, -0.5) {};
    		\node [style=none] (22) at (0, 0) {};
    	\end{pgfonlayer}
    	\begin{pgfonlayer}{edgelayer}
    		\draw [style=double arrow] (1.center) to (0.center);
    		\draw (5) to (6);
    		\draw (4) to (5);
    		\draw (6) to (3);
    		\draw [style=doubleline] (3) to (4);
    		\draw (16) to (15);
    		\draw (15) to (13);
    		\draw (17) to (16);
    		\draw (19) to (16);
    		\draw (16) to (18);
    	\end{pgfonlayer}
    \end{tikzpicture}
    \end{equation}
\end{itemize}

\vfill\null

\bibliography{bibli.bib}

\end{document}